%% file: sample631.tex
\documentclass[aps,prd,twocolumn,superscriptaddress,nofootinbib,floatfix]{revtex4-2}

\usepackage{graphicx}
\usepackage{bm}  

\newcommand{\nhat}{\bm{\hat{n}}}
\usepackage{hyperref}

\usepackage{amsmath} 
\usepackage{float}
\usepackage[section]{placeins} 
\usepackage{color}

\usepackage{bm}
\newcommand{\bold}[1]{\bm{#1}}
\usepackage{amssymb}

\usepackage[utf8]{inputenc}

\definecolor{linkcolor}{rgb}{0.6,0,0}
\definecolor{citecolor}{rgb}{0,0,0.75}
\definecolor{urlcolor}{rgb}{0.12,0.46,0.7}
\hypersetup{colorlinks,linkcolor=magenta,citecolor=magenta,urlcolor=magenta}

\usepackage{savesym}
\usepackage{siunitx}
\usepackage[mathlines]{lineno}

\begin{document}

\title{Precision Kinematic Sunyaev--Zel'dovich Measurements Across Halo Mass and Redshift with DESI DR2 and ACT DR6: Part I. Luminous Red Galaxies}

\input{authors.tex}


\begin{abstract}

We present the most precise measurements of the kinetic Sunyaev-Zel'dovich (kSZ) effect around luminous red galaxies to date, detecting the signal at $18\sigma$ significance in both harmonic and configuration space. Our analysis cross-correlates 2.4 million spectroscopic LRGs from the Dark Energy Spectroscopic Instrument (DESI) DR2 sample with Data Release 6 (DR6) of the Atacama Cosmology Telescope (ACT). We develop a novel harmonic-space cross-correlation approach using momentum-weighted kSZ templates, yielding nearly uncorrelated bandpowers within a framework consistent with other large-scale structure analyses. By incorporating the LRG halo occupation distribution (HOD) and its uncertainty, we convert measured galaxy gas profiles into halo gas profiles and provide generalized Navarro-Frenk-White (GNFW) fitting profiles, providing empirical targets for tuning feedback efficiency in hydrodynamical simulations and for baryonic modeling in large-scale structure analyses. We find strong evidence that gas profiles do not trace dark matter, providing direct evidence for gas redistribution beyond gravitational collapse. Comparing to hydrodynamical simulations, our measurements favor feedback efficiencies exceeding those in the Battaglia profile, suggesting more efficient gas ejection in group-scale halos than previously predicted. 
Splitting by redshift, we detect the kSZ signal 
at SNR $\approx 5$--$10$
in each of four bins
and find amplitude evolution consistent with the expected decline in mean halo mass at fixed comoving number density. Splitting by stellar mass, we study the scaling of kSZ amplitude with galaxy properties. Together with BGS and ELG measurements in Paper II, these results span $0.1 \lesssim z \lesssim 1.6$ across three galaxy populations, demonstrating the potential of spectroscopic kSZ to map circumgalactic gas and constrain baryonic feedback.
\end{abstract}
\maketitle

\section{Introduction} \label{sec:intro}

Mapping the distribution of baryons and understanding the feedback processes that govern their evolution remain among the most pressing challenges in modern cosmology and astrophysics. While baryons constitute more than $15\%$ of the universe's total matter content, their spatial distribution relative to dark matter remains poorly constrained. This uncertainty has profound implications for upcoming large-scale structure surveys, including the Vera Rubin Observatory~\cite{lsstdarkenergysciencecollaboration2012largesynopticsurveytelescope, Ivezi__2019}, Euclid~\citep{2022A&A...662A.112E}, and the Nancy Grace Roman Space Telescope~\citep{spergel2015widefieldinfrarredsurveytelescopeastrophysics}, which require sub-percent precision in modeling baryonic effects to achieve their cosmological goals. Additionally, characterizing how baryons populate dark matter halos is essential for understanding the physical mechanisms driving galaxy formation and evolution. The majority of baryons reside not in galaxies themselves but in diffuse reservoirs: the circumgalactic medium (CGM) surrounding individual galaxies and the intracluster medium (ICM) permeating galaxy clusters. The thermodynamic state and spatial extent of these gas phases are regulated by energetic feedback from active galactic nuclei (AGN) and supernovae~\citep{2006ApJ...650..560C}, processes whose efficiency and impact remain incompletely understood.

A powerful observational tool for probing diffuse baryons is the Sunyaev-Zel'dovich (SZ) effect, which arises from inverse Compton scattering of cosmic microwave background (CMB) photons off free electrons. The thermal SZ (tSZ) effect, proportional to the integrated electron pressure, has been widely used to study hot gas in massive clusters. By contrast, the kinetic SZ (kSZ) effect results from the Doppler shift induced by electrons with bulk line-of-sight velocities relative to the CMB rest frame. Unlike the tSZ effect, which depends on the electron temperature, or X-ray observations, which scale with the square of the density and are sensitive to gas metallicity, the kSZ signal depends only on the electron momentum field, making it sensitive to the total ionized baryon distribution regardless of temperature or metallicity. This enables kSZ measurements to directly constrain the abundance and spatial distribution of baryons, including in the diffuse outskirts of halos, a regime which has traditionally been challenging to access (see~\citet{Birkinshaw_1999, Mroczkowski_2019} for reviews).

The kSZ has been measured in the past with various techniques, including pairwise methods that exploit the fact that galaxies in adjacent pairs along the line of sight induce correlated temperature decrements and increments in the CMB due to their opposing bulk velocities in the large-scale flow~\citep{PhysRevLett.109.041101,PhysRevLett.115.191301,Bernardis_2017,Soergel_2016,Sugiyama_2018,PhysRevD.104.043502,li2024detectionpairwisekineticsunyaevzeldovich,Hadzhiyska:2025egz,harscouet2025kszeveryonepseudoclapproach}, projected-field estimators that measure the kSZ-large-scale structure cross-correlation without requiring spectroscopic redshifts for individual galaxies~\citep{PhysRevLett.117.051301,PhysRevD.94.123526,PhysRevD.104.043518,Bolliet_2023}, and velocity reconstruction stacking methods that use the continuity equation to estimate each galaxy's line-of-sight velocity and coherently stack CMB temperature fluctuations weighted by these velocity estimates~\citep{PhysRevD.93.082002,PhysRevD.103.063513,PhysRevD.108.023516,hadzhiyska2025evidencelargebaryonicfeedback,guachalla2025backlightingextendedgashalos} , and quadratic estimator methods that exploit the kSZ-induced bispectrum to reconstruct the large-scale velocity field from CMB temperature and galaxy density maps~\citep{mccarthy2024atacamacosmologytelescopelargescale,mccarthy2025atacamacosmologytelescopecrosscorrelation,hotinli2025velocityreconstructionkszmeasuring,lai2025kszvelocityreconstructionact,lague2024constraintslocalprimordialnongaussianity}.

In this work, we present kSZ stacking and template cross-correlation measurements using high-resolution CMB data from Data Release 6 (DR6) of the Atacama Cosmology Telescope (ACT)~\citep{PhysRevD.109.063530} cross-correlated with Luminous Red Galaxies (LRGs) from the Dark Energy Spectroscopic Instrument (DESI) Data Release 2. With nearly double the sample size of DESI Year 1, we achieve significantly improved constraints on both the combined profile and individual redshift and stellar mass bins. The primary methodological advance of this work is the construction of an explicit momentum-weighted template for the kSZ signal, which we cross-correlate with the CMB map in harmonic space\footnote{A similar harmonic-space methodology was independently developed by \citep{harscouet2025kszeveryonepseudoclapproach}, whose work appeared as this paper was being finalized.}. 
This approach offers several advantages over real-space stacking: it provides clearer scale-dependent information and achieves better decoupling between different angular scales. Furthermore, the harmonic-space formalism facilitates joint analyses with other cross-correlation probes (such as galaxy-galaxy lensing and CMB lensing--galaxy  cross-correlations) since these measurements naturally share the same angular power spectrum framework and can be combined with consistent covariance estimation.

We further interpret these measurements by forward-modeling the kSZ signal under a halo occupation distribution (HOD) framework for the LRGs, allowing us to constrain gas parameters that characterize the thermodynamic profiles of baryons around these galaxies.

Recent efforts have begun to confront kSZ measurements with predictions from state-of-the-art hydrodynamical simulations to constrain baryonic feedback processes \citep{Schneider_2022,hadzhiyska2025evidencelargebaryonicfeedback,Sunseri:2025hhj,Bigwood:2025kur,bigwood2025kineticsunyaevzeldovicheffect,Siegel:2025ivd,salcido2025implicationsfeedbacksolutionss8,mccarthy2025flamingocombiningkineticsz,Lague:2025txe,kovac2025baryonificationiiconstrainingfeedback}. 

These studies have compared observed kSZ profiles across multiple simulation suites with varying feedback implementations, and have combined kSZ observations with complementary probes such as X-ray observations and galaxy-galaxy lensing to disentangle baryons from dark matter and break model degeneracies. 


Central to interpreting these comparisons is the correlation coefficient $r$ between reconstructed velocities from the continuity equation and the true halo velocities. This coefficient directly scales the measured kSZ amplitude and depends on the quality of the velocity reconstruction. For our full LRG sample, light-cone simulations provide calibration of $r$ at the $\sim10\%$ level. However, $r$ may vary with redshift, halo mass, and survey properties, and its calibration for individual subsamples (e.g., specific redshift or mass bins) remains less certain. We therefore interpret trends across subsamples with appropriate caution, noting that improved calibration of $r$ as a function of redshift and halo mass will be needed to fully exploit the constraining power of precision kSZ measurements on feedback physics.

In a companion paper (Paper~II; \citep{HadzhiyskaBGSELG}), we present 
analogous kSZ measurements for the DESI Bright Galaxy Survey (BGS) and 
Emission Line Galaxies (ELGs). The BGS sample, with a mean redshift of 
$\bar{z} \simeq 0.26$, probes typical $L_\star$ galaxies in lower-mass 
halos than the LRGs studied here, providing a low-redshift anchor for 
kSZ-based feedback constraints. The ELG sample extends to $z \simeq 1.6$, 
targeting star-forming galaxies in halos with $M_{\rm halo} \sim 10^{12} M_\odot$, 
where stellar feedback is expected to dominate over AGN feedback. Together 
with the LRG measurements presented here, these companion analyses span 
the redshift range $0.1 \lesssim z \lesssim 1.6$ and halo masses from 
$\sim 10^{12}$ to $\sim 10^{14} M_\odot$, providing a comprehensive view 
of baryon distributions across a wide range of galaxy population accessible to DESI.

This paper is organized as follows. In Section~\ref{data}, we describe the DESI LRG galaxy sample and ACT DR6 CMB maps used in this analysis. Section~\ref{sec.methodology} presents our methodology for measuring the kSZ effect through cross-correlation with the galaxy density field. Section~\ref{sec.harmonic} presents our main kSZ measurement in harmonic space, including fits to generalized Navarro-Frenk-White (GNFW) profiles and comparisons to hydrodynamical simulations. In Section~\ref{sec.real_space}, we present complementary real-space measurements using compensated aperture photometry (CAP) filters. Finally, Section~\ref{sec.conclusion} summarizes our findings and discusses implications for baryon feedback models and future kSZ measurements with next-generation CMB experiments.

\section{Datasets}\label{data}

\subsection{DESI DR2 spectroscopic LRG sample} \label{sec.desi_data}

The Dark Energy Spectroscopic Instrument (DESI) is a robotic, fiber-fed, highly multiplexed spectroscopic surveyor operating on the Mayall 4-meter telescope at Kitt Peak National Observatory \citep{DESI2022.KP1.Instr}. The instrument obtains simultaneous spectra for nearly 5000 objects across a $\sim3^\circ$ field of view \citep{Corrector.Miller.2023, FiberSystem.Poppett.2024}. Over its planned eight-year operation, DESI will observe approximately 17,000~deg$^2$ of sky, targeting an estimated 63 million spectroscopically-confirmed galaxies and quasars spanning redshifts $0.1 < z < 1.6$, substantially exceeding the original forecast of 39 million objects \citep{DESI2016b.Instr}. The survey's scale requires sophisticated supporting infrastructure, including dedicated software pipelines for spectroscopic reduction and survey operations \citep{Spectro.Pipeline.Guy.2023, SurveyOps.Schlafly.2023}. Cosmological analyses using the First Data Release DR1 \citep{DESI2024.I.DR1} have yielded results from full-shape modeling of galaxy clustering \citep{DESI2024.VII.KP7B} and baryon acoustic oscillations \citep{DESI2024.III.KP4}, with ongoing work utilizing the forthcoming DR2 dataset \citep{DESI.DR2.DR2}.

The survey employs multiple galaxy tracers optimized for different science goals, with Luminous Red Galaxies (LRGs) being particularly well-suited for kSZ studies. LRGs serve as highly-biased tracers of large-scale structure, residing in massive dark matter halos with typical masses $M_{200c} \sim 10^{13}-10^{14}~M_\odot$ \citep{yuan2023desionepercentsurveyexploring}. Their distinctive spectral features enable robust photometric and spectroscopic selection, making them reliable targets for cross-correlation analyses \citep{2023AJ....165...58Z}. Furthermore, LRGs exhibit low satellite fractions ($\sim 11\%$) compared to other DESI tracers \citep{yuan2023desionepercentsurveyexploring}, minimizing contamination from redshift-space distortions and velocity dispersion effects that would otherwise complicate velocity reconstruction, and reducing miscentering in stacked halo profiles since the majority of LRGs are true halo centrals.

\begin{figure}[!htbp]
\centering
\includegraphics[width=\linewidth]{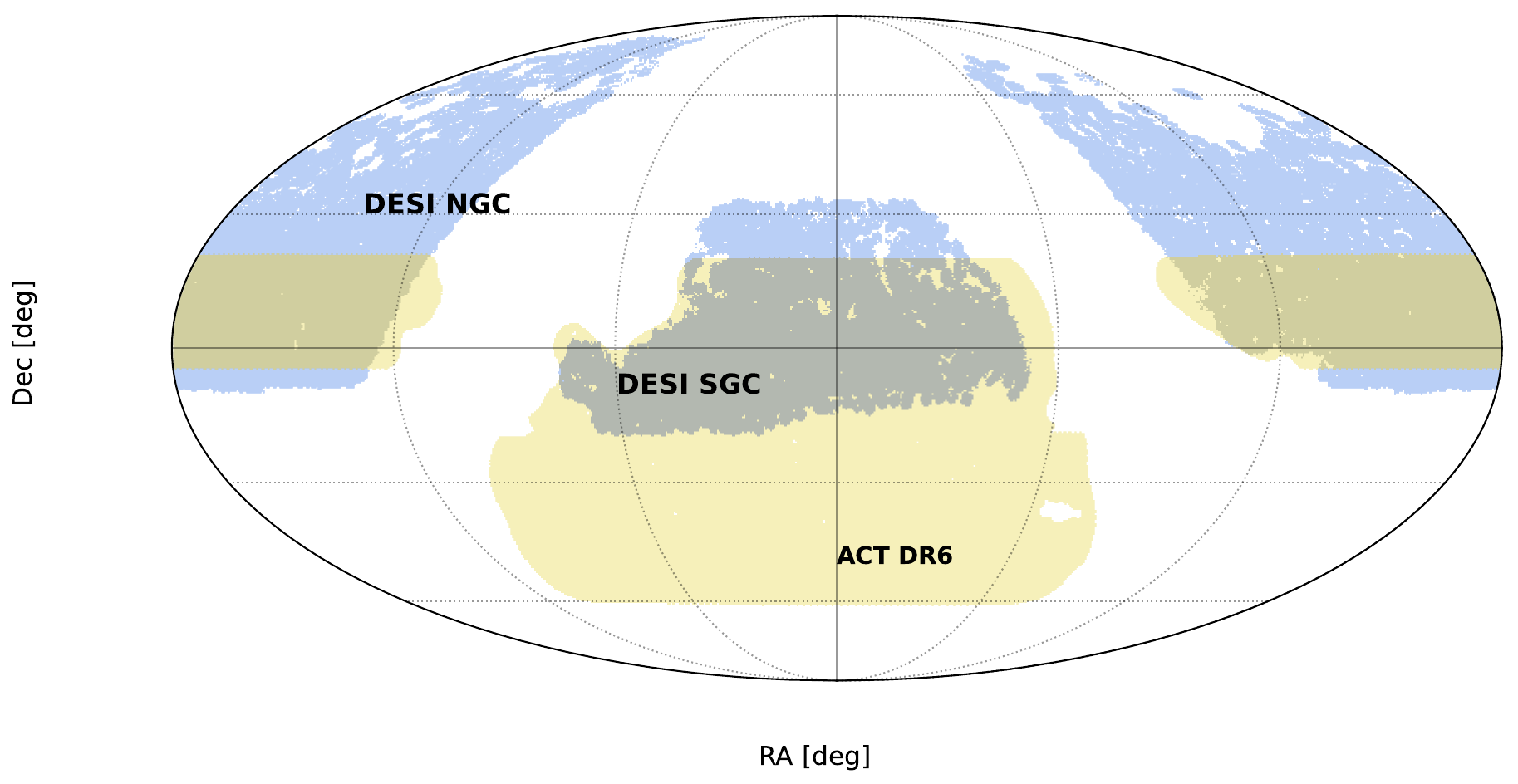}
\caption{Sky coverage of the DESI LRG DR2 sample (blue) and the ACT DR6 CMB survey (yellow) in Mollweide projection. The DESI LRG sample is divided into the North Galactic Cap (DESI NGC) and South Galactic Cap (DESI SGC), containing approximately 2.4 million spectroscopically confirmed galaxies. The ACT DR6 footprint covers roughly 18,000~deg$^2$ of the sky with the Galactic plane masked out. Additionally, the ACT mask excludes regions contaminated by tSZ signals from massive clusters. The overlap between the DESI and ACT footprints defines our analysis region for kSZ measurements.}
\label{fig:footprint}
\end{figure}

The DESI main survey commenced observations on 14 May 2021 following a period of survey validation \citep{DESI2023.KP1.SV}. In this analysis, we use the LRG large-scale structure (LSS) catalogs from Data Assembly~2 (DA2),
based on the \texttt{loa} spectroscopic reduction, constructed following the
methodology of \citep{ross2024constructionlargescalestructurecatalogs} and validated in \citep{andrade2025validationdesidr2measurements}.
These catalogs correspond to the DESI Year~3 spectroscopic data included in
Data Release~2 \citep{Abdul_Karim_2025}. The sample is divided into the North Galactic Cap (NGC) and the South Galactic Cap (SGC), both shown in blue in Figure~\ref{fig:footprint}.

The analysis of the full galaxy catalogue begins with an initial filtering stage, retaining only galaxies that lie within the CMB survey footprint to ensure subsequent measurements are restricted to regions with reliable CMB coverage. A cluster mask is then applied, with holes of 6 arcminute radius around 4,922 regular clusters and 10 arcminute radius around 436 large clusters, thereby reducing contamination from the thermal Sunyaev-Zel'dovich effect in the kSZ signal extraction.  When extracting the kSZ signal, galaxies are  filtered based on this mask: for each galaxy, a cutout of the CMB mask is examined within the $6'$ Compensated Aperture Photometry (CAP) filter aperture. If any masked pixel (mask value equal to zero) falls inside the aperture, the galaxy is excluded from the sample. Finally, galaxies in regions where the CMB map exhibits fluctuations exceeding $\pm 5\sigma$ are discarded. After these cuts, our final sample contains 2,376,871 objects.

\begin{figure}[!htbp]
\centering
\includegraphics[width=\linewidth]{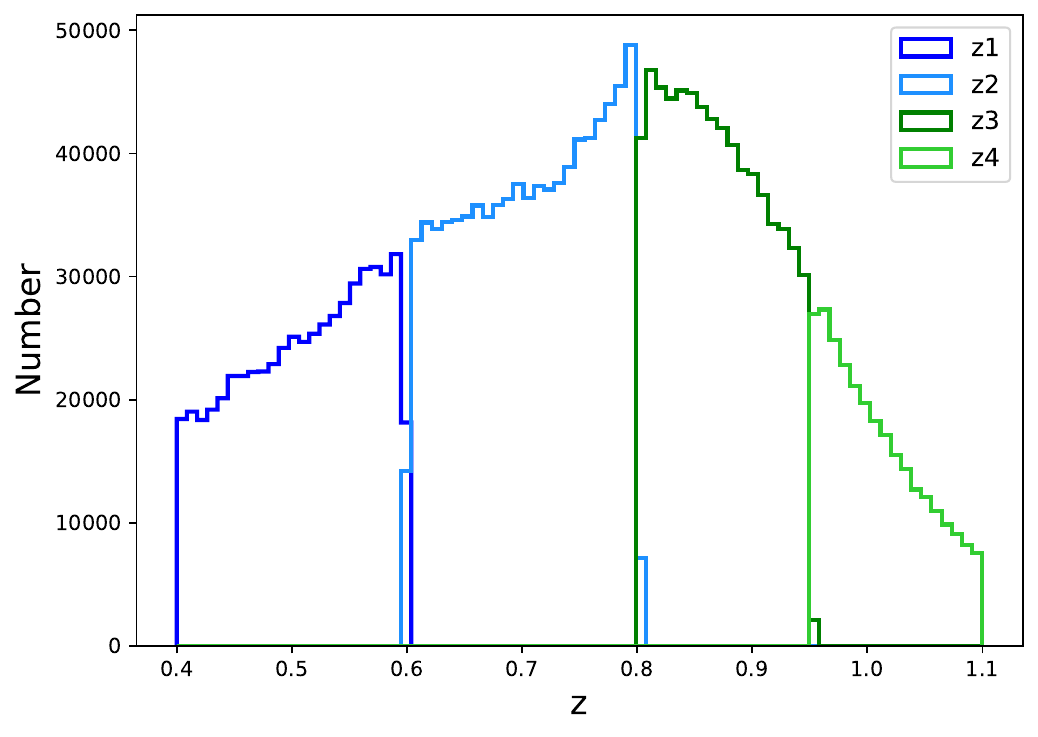}
\caption{Redshift distribution of the final DESI LRG sample entering the stacked kSZ measurement, divided into four redshift bins: z1 ($0.4 < z \leq 0.6$), z2 ($0.6 < z \leq 0.8$), z3 ($0.8 < z \leq 0.95$), and z4 ($0.95 < z \leq 1.1$). For the harmonic-space analysis requiring HOD modeling, we combine bins z3 and z4 into a single high-redshift bin to match with the provided HOD.}
\label{fig:zdist}
\end{figure}

For the redshift-dependent analysis, we divide the LRGs into four bins: z1 ($0.4 < z \leq 0.6$), z2 ($0.6 < z \leq 0.8$), z3 ($0.8 < z \leq 0.95$), and z4 ($0.95 < z \leq 1.1$), as shown in Fig.~\ref{fig:zdist}. To probe the evolution and mass dependence of the kSZ signal, we also analyse the sample in independent bins of stellar mass. For the stellar-mass-dependent analysis, we divide the LRGs into four subsamples according to their stellar mass. Stellar masses are taken from \citet{Zhou_2023}, who applied a random-forest algorithm trained on DESI Legacy Imaging Survey photometry with stellar masses from \citet{Bundy_2015} as the training set. We adopt four bins in $\log_{10}(M_\ast/M_\odot)$: $(10.5,,11.2)$, $(11.2,,11.4)$, $(11.4,,11.6)$, and $(11.6,,12.5)$, which we denote as mass1--mass4, respectively. These bins contain $7.02\times 10^5$, $9.25\times 10^5$, $5.68\times 10^5$, and $1.60\times 10^5$ galaxies of the final filtered sample. This method is well-suited to the LRG population given its close similarity to the CMASS sample used as the training set; since stellar mass serves here as a proxy for halo mass, our results are expected to be largely insensitive to the choice of stellar mass estimation method.

\begin{figure}[!htbp]
\centering
\includegraphics[width=\linewidth]{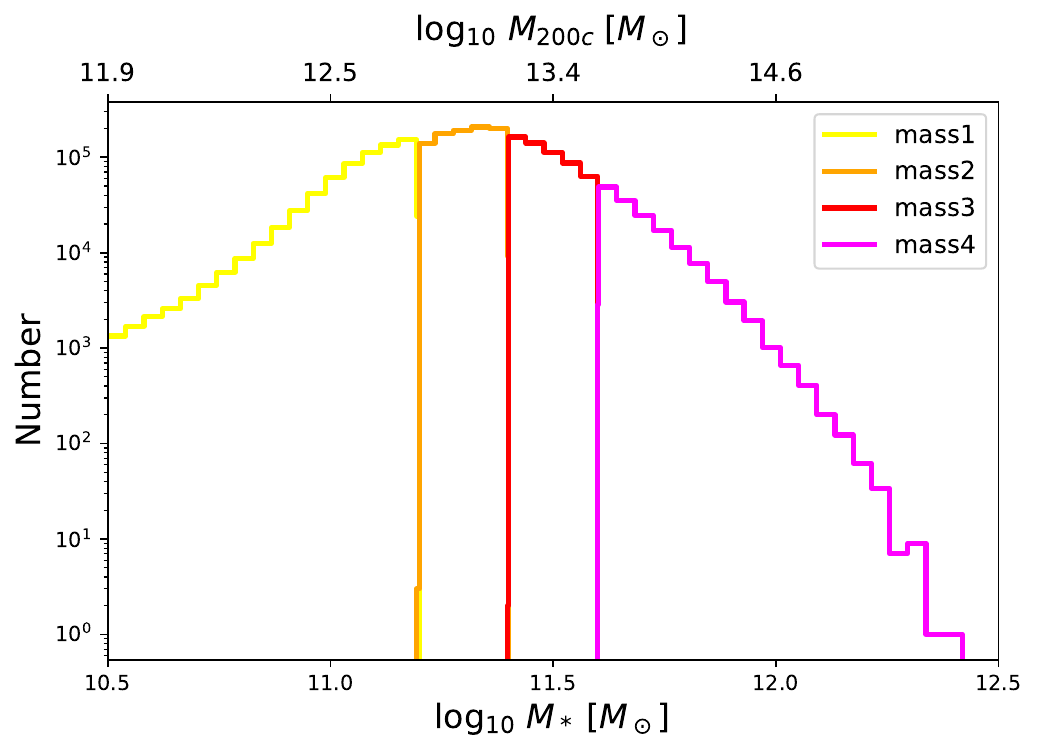}
\caption{Stellar mass distribution of the final DESI LRG sample entering the stacked kSZ measurement (2,376,871 galaxies after the footprint, cluster and outlier cuts described in the text, of which 2,355,530 have a valid stellar mass estimate in the range shown). The histogram shows the distribution of $\log_{10}(M_*/M_\odot)$, with color-coded regions indicating the four stellar mass subsamples (mass1--mass4). The top axis shows approximate halo masses $M_{200c}$ inferred from the \citep{Kravtsov_2018} stellar-to-halo mass relation, converted using the \citep{Duffy_2008} concentration-mass relation at $z=0.7$.}

\label{fig.mass_redshift}
\end{figure}

\subsection{ACT DR6 CMB map}

We use data from the Atacama Cosmology Telescope (ACT)~\citep{2016ApJS..227...21T}, 
a millimeter-wavelength survey that operated from 2007 to 2022 from Cerro Toco 
in the Atacama Desert in northern Chile. ACT surveyed roughly 
$18{,}000\,{\rm deg}^2$ of the sky with the primary goal of measuring the cosmic 
microwave background (CMB) temperature and polarization anisotropies at arcminute 
resolution. In this work we make use of Data Release 6 
(DR6)~\citep{act_0}, which includes observations collected between 2017 and 
2021 in three frequency bands: f090 (77--112\,GHz), f150 (124--172\,GHz), and 
f220 (182--277\,GHz). We focus on the night-time data and employ the 
harmonic-space internal linear combination (hILC) CMB temperature maps (version 
dr6.01)~\citep{PhysRevD.109.063530}, which combine ACT and \textit{Planck} data 
across frequencies. The maps are provided at $0.5'$ pixel resolution and are 
smoothed with a Gaussian beam of FWHM $1.6'$. To mitigate contamination from 
bright point sources and massive clusters, we apply the ACT DR6 survey and 
cluster mask~\citep{hilton2021} that covers $31\%$ of the sky and we further 
exclude regions with temperature fluctuations exceeding $\pm 5\sigma$ relative to 
the mean CMB, following the approach of \citep{PhysRevD.103.063513}. The final ACT 
DR6 CMB footprint used in our analysis is shown in yellow in 
Fig.~\ref{fig:footprint}.

\section{Methodology}\label{sec.methodology}

\subsection{The kSZ effect}

The kSZ effect arises when CMB photons inverse Compton scatter off free electrons that possess a net bulk velocity relative to the CMB rest frame~\citep{1986ApJ...306L..51O}. The temperature perturbation induced by the kSZ effect in direction $\hat{\mathbf{n}}$ on the sky is given by
\begin{equation}\label{eq:deltaTksz}
\frac{\delta T_{\rm kSZ}(\hat{\mathbf{n}})}{T_{\textrm{CMB}}} = -\int d\chi \, a^{-2}(\chi) \, n_e(\chi\hat{\mathbf{n}}, z) \, \sigma_T \, e^{-\tau} \left( \frac{\mathbf{v}_e \cdot \hat{\mathbf{n}}}{c} \right),
\end{equation}
where $\chi$ is the comoving distance, $z$ is the redshift, $a(\chi) = 1/(1+z)$ is the scale factor, $\sigma_T$ is the Thomson scattering cross-section, $c$ is the speed of light, $n_e$ is the comoving free electron number density, $\mathbf{v}_e$ is the electron peculiar velocity, and $\tau$ is the mean integrated optical depth along the line of sight,
\begin{equation}\label{eq:tau}
\tau = \int d\chi \, a^{-2}(\chi) \, n_e(\chi\hat{\mathbf{n}}) \, \sigma_T.
\end{equation}
The optical depth of gaseous halos is typically much smaller than unity ($\tau \ll 1$) for the redshift range considered here, so we can approximate $e^{-\tau} \approx 1$. Under this approximation, the kSZ can be expressed as being proportional to the optical depth and the bulk velocity of the electrons along the line of sight:
\begin{equation}\label{eq:ksz_simple}
\frac{\delta T_{\rm kSZ}(\hat{\mathbf{n}})}{T_{\textrm{CMB}}} = -\tau \left( \frac{v_e^{\rm LOS}}{c} \right),
\end{equation}
where $v_e^{\rm LOS} = \mathbf{v}_e \cdot \hat{\mathbf{n}}$ represents the line-of-sight component of the electron bulk velocity.

A challenge in measuring the kSZ effect is obtaining reliable estimates of the line-of-sight velocities. Unlike galaxy positions and redshifts, which are directly measured by spectroscopic and photometric surveys, the peculiar velocities of individual galaxies are not directly observable. However, the velocity field can be reconstructed from the observed galaxy distribution using the continuity equation that relates density and velocity perturbations in linear theory, as we describe in the following section.

\subsection{Velocity reconstruction}

The linearized continuity equation relates the velocity field of the galaxies $\mathbf{v}$ 
to the galaxy number overdensity $\delta_g$:

\begin{equation} \label{eq. continuity}
    \nabla\cdot{\mathbf{v}}=-aHf\frac{\delta_g}{b},
\end{equation}

where $a$, $H$ and $f$ are the scale factor, the Hubble parameter, and the logarithmic 
growth rate respectively. $f$ is related to the linear growth factor $D(a)$ by 
$f=d\ln{D(a)}/d\ln{a}$. In practice, we evaluate $f$, $H$, and the comoving distances 
assuming the \textit{Planck} 2018 \citep{2020A&A...641A...6P} $\Lambda$CDM cosmology \footnote{The sensitivity of the reconstructed velocities 
to misspecification of these parameters has been quantified in \citep{PhysRevD.109.103533}, 
where it was found that the correlation coefficient $r$ between reconstructed and true halo velocities 
changes at the sub-percent level for cosmological parameters within the Planck $5\sigma$ 
uncertainty, with at most a few-percent bias on the velocity standard deviation. }. 
Redshift space distortions (RSD) add an extra term on the LHS of Eq.~\ref{eq. continuity} 
\citep{1994ApJ...421L...1N}:
\begin{equation} \label{eq. continuity_full}
    \nabla\cdot{\bold{v}}+\frac{f}{b}\nabla\cdot[(\bold{v}\cdot\nhat)\nhat]=-aHf\frac{\delta_g}{b},
\end{equation} 
where $v^\mathrm{LOS} = \bold{v}\cdot\hat{\mathbf{n}} \approx v^\mathrm{LOS}_e$ is the electron bulk velocity of the halo.

Eq.~\ref{eq. continuity_full} can be solved iteratively in Fourier space. In the Zel'dovich approximation, the velocity field is expressed using the Lagrangian displacement field $\boldsymbol{\psi}$, defined via $\delta_m = -\nabla \cdot \boldsymbol{\psi}$:

\begin{equation}\label{eq.zeld}
    \bold{v}(\bold{x})=aHf\boldsymbol{\psi}(\bold{x}),
\end{equation}

where 

\begin{equation}
    \boldsymbol{\psi}(\bold{x})=\bold{x}-\bold{q}.
\end{equation}

Here $\bold{x}$ denotes the final observed galaxy position after being displaced by gravity and $\bold{q}$ denotes the initial positions before the large scale bulk flow shifted the galaxy positions.

We obtain the displacement field using the BAO reconstruction displacements used by the DESI DR2 analysis \cite{Paillas_2025,chen2024extensiveanalysisreconstructionalgorithms}.

\begin{figure}[!htbp]
    \centering
    \includegraphics[width=\linewidth]{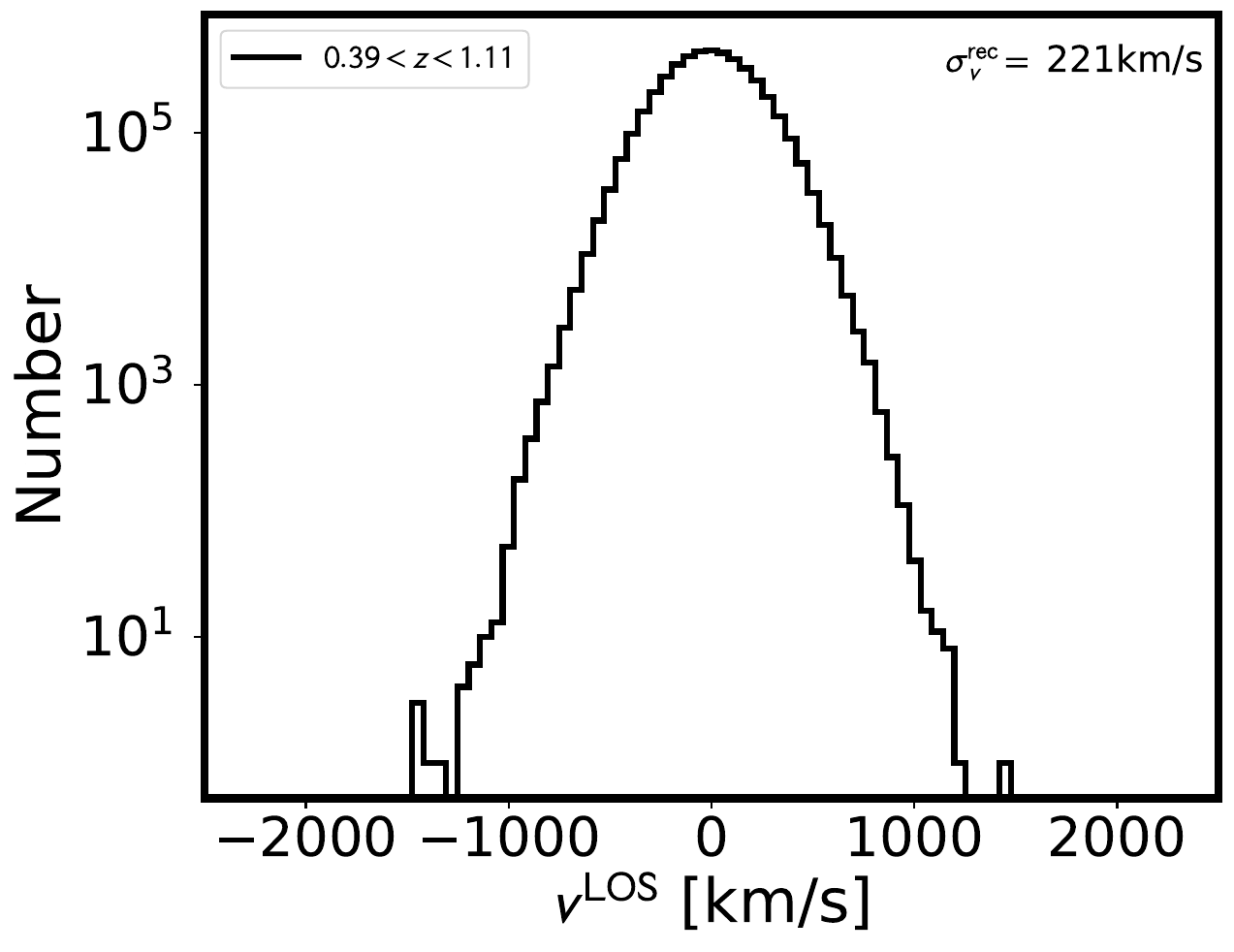} \\ 
    \caption{Distribution of reconstructed line-of-sight velocities for the full DESI DR2 LRG sample ($0.39 < z < 1.11$, ${\sim}4.5$ million galaxies). Of these, ${\sim}2.4$ million fall within the ACT DR6 footprint and are used in the cross-correlation analysis. The quantity $\sigma^{\rm rec}_v$ denotes the standard deviation of the velocity distribution.}
    \label{fig.dist_vel}
\end{figure}

The displacement field is obtained using the 
\texttt{pyrecon}\footnote{\url{https://github.com/cosmodesi/pyrecon}} 
code with the \texttt{RECSYM} convention, where a factor of $(1+f)$ is 
multiplied by the displacement along the line of sight of the galaxies and randoms. For our analysis, we divide by $(1+f_\mathrm{eff})$, where 
$f_\mathrm{eff}$ is the growth rate evaluated at the effective redshift 
$z_\mathrm{eff}=0.79$ for the LRGs. The velocity is then obtained from 
Eq.~(\ref{eq.zeld}), with $a$, $H$, and $f$ evaluated at each galaxy's individual redshift. In Fig.~\ref{fig.dist_vel} we show the reconstructed radial velocity distribution derived for the DESI DR2 data.

\subsubsection{Degeneracy of the correlation coefficient $r$ with the gas amplitude}\label{sec.r}

A consideration when using these reconstructed velocities for kSZ measurements is how well they trace the true gas velocity field. The kSZ stacking estimator measures correlations between CMB temperature fluctuations and galaxy-derived velocity fields. For both real-space and harmonic-space estimators, the kSZ estimator $\hat{p}$ can be expressed as
\begin{equation}
    \hat{p}\sim\langle\delta T^{\rm kSZ} v_{\rm rec}\rangle 
    \sim \langle n_e v_{\rm gas} v_{\rm rec}\rangle 
    = r n_e \sigma_{\rm gas} \sigma_{\rm rec},
\end{equation}
where $n_e$ is the electron density, $\sigma_{\rm gas}$ is the RMS gas peculiar 
velocity, $\sigma_{\rm rec}$ is the RMS reconstructed velocity, and $r$ is the 
correlation coefficient between the reconstructed and true halo velocities \citep{PhysRevD.109.103533}. The correlation coefficient $r$ quantifies how well the galaxy-based velocity reconstruction traces the true gas velocity field.

\begin{equation}
r\equiv\frac{\langle{v_{rec}v_{gas}}\rangle}{\sigma_{rec}\sigma_{gas}}
\end{equation}

Here $\sigma_{\rm gas}$ is the halo RMS velocity and we adopt $r=0.65$ from the light cone analysis of \citep{PhysRevD.109.103534}, which represents the most realistic assessment available for velocity reconstruction performance averaged over the full DESI LRG redshift range. The light cone analysis incorporates redshift evolution, realistic survey geometry, and the effects of varying number density and galaxy properties across the survey volume. This provides the best single estimate of $r$ for our fiducial analysis that combines all redshifts and masses. However, this averaged value may not accurately represent $r$ in individual redshift or mass bins. Below we discuss several effects that might cause variations in $r$ when splitting the sample:

\begin{itemize}
    \item \textbf{Nonlinear structure growth}: The Zel'dovich approximation underlying velocity reconstruction is fundamentally based on linear theory and becomes increasingly inaccurate at lower redshifts where nonlinear gravitational evolution is more important. \citep{PhysRevD.109.103533} showed that velocity distributions develop significant non-Gaussian tails at high velocities due to nonlinear effects, which cannot be captured by linear reconstruction.  This represents a significant expected source of uncertainty in $r$ across our redshift range. We expect $r$ to be  lower in our lowest redshift bins where nonlinear evolution is most pronounced, though quantifying this effect precisely requires dedicated light cone simulations that span the given redshift range.

\item \textbf{Satellite fraction evolution}: Satellite galaxies exhibit larger virial motions that are uncorrelated with the large-scale velocity field relevant for kSZ. These virial motions create Fingers-of-God (FoG) in redshift space, smoothing the observed number density distribution and reducing the available information for velocity reconstruction. The resulting FoG effects, which are not captured in our model, cause the line-of-sight correlation coefficient $r_\parallel$ to degrade mildly as satellite fraction increases \citep{PhysRevD.109.103533}. While the LRG satellite fraction at $z=0.5$ is approximately $f_{\rm sat} = 0.12$ \citep{yuan2023desionepercentsurveyexploring}, this fraction varies with both halo mass and redshift \citep{Baleato_Lizancos_2025}. Higher mass halos typically host more satellites, leading to stronger FoG effects and greater miscentering of the assumed halo position relative to the true center of mass. When splitting our sample by redshift or halo mass, we may therefore be selecting subsamples with different satellite fractions, resulting in varying degrees of velocity decorrelation and halo miscentering. In particular, maintaining constant comoving number density across redshift requires selecting increasingly massive halos at higher redshift, which would have systematically higher satellite fractions and correspondingly different values of $r$.

\item \textbf{Fiber assignment incompleteness}: The fiber assignment algorithm prevents simultaneous observation of galaxies that are close in angular separation, primarily suppressing high $k_\perp$ modes in the observed density field. Since velocity reconstruction is most sensitive to $k_\parallel$ modes, purely angular incompleteness does not directly bias $r$. However, fiber collisions preferentially remove galaxies in angularly dense regions, which tend to correspond to physically overdense environments with higher satellite fractions. This can indirectly modify the effective satellite fraction and velocity distribution of the observed sample relative to the parent population, introducing an additional source of uncertainty in $r$ that is difficult to separate from the intrinsic satellite fraction effect described above.

       \item \textbf{Galaxy bias evolution}: Linear galaxy bias $b$ enters the velocity reconstruction as a conversion factor between the observed galaxy overdensity and the underlying matter overdensity. \citep{PhysRevD.109.103533} show that incorrect assumptions about the linear galaxy bias have minimal impact on the correlation coefficient $r$ in the absence of redshift space distortions (RSD), as the bias only scales the reconstructed velocity as $b^{\rm fid}/b^{\rm true}$. However, more luminous galaxy samples tend to be more highly biased tracers of the underlying dark matter distribution, and at sufficiently high bias, nonlinear bias terms may become important. When splitting our sample by redshift or mass, we are effectively selecting galaxies with different luminosities and therefore different bias values. Higher redshift bins and more massive halos correspond to more luminous, more highly biased LRGs where nonlinear bias corrections could become significant. This could introduce variations in $r$ across our bins that are not captured by the linear bias framework used in the velocity reconstruction method.

          \item \textbf{Halo occupation distribution (HOD) evolution}: The HOD model used in the light cone simulations is calibrated at $z=0.5$ and then applied across the full redshift range with appropriate $N(z)$ downsampling. While this captures the redshift evolution of number density and large-scale clustering, it does not fully account for potential evolution in the detailed relationship between galaxies and their host halos (such as changes in satellite fraction, velocity bias parameters, or the mass-luminosity relation) that could affect velocity reconstruction performance across our redshift and mass bins.

\end{itemize}

Given that our kSZ detection reaches 18$\sigma$ significance and individual redshift bins are also measured with high precision, systematic uncertainties in $r$  could become important for interpreting variations in the inferred gas profiles across redshift and mass. At this level of statistical precision, seemingly subdominant systematic effects warrant careful consideration. Future work with dedicated simulations using evolving HOD models, full light cone geometry matching our survey footprint, and direct calibration to DESI data at multiple redshifts and mass bins will be essential for reducing these systematic uncertainties and fully exploiting the statistical power of our measurement.

\subsection{Building the kSZ template}\label{sec.measurement}

We construct the kSZ template by binning the DESI DR2 LRG catalog onto a CAR (Cylindrical Equal-Area) grid using \texttt{pixell}, with pixel size $\simeq 0.5'$, restricting to galaxies within the ACT footprint (declination $\delta > -20^\circ$). Each galaxy contributes its line-of-sight velocity $v_{r,i}/c$ to the corresponding pixel, normalized by the expected galaxy count per pixel:
\begin{equation}
\hat{\pi}(\boldsymbol{\theta})
= \frac{1}{\Omega_{\mathrm{pix}}\bar{n}_g^{2D}}\sum_{i \in \mathrm{pixel}(\boldsymbol{\theta})} \frac{v_{r,i}}{c},
\end{equation}
where $\bar{n}_g^{2D}$ is the mean galaxy surface density and $\Omega_{\mathrm{pix}}$ is the pixel solid angle. This normalization ensures the template is an intensive quantity independent of pixel size. We construct templates for both the full LRG sample and for subsamples in different redshift bins.






To connect this discrete estimator to the underlying continuous fields, we note that the sum over galaxies approximates a line-of-sight integral weighted by the three-dimensional galaxy density:

\begin{align}
\hat{\pi}(\boldsymbol{\theta})
\nonumber &=  \frac{1}{\Omega_{\mathrm{pix}}\bar{n}_g^{2D}}\int_{\mathrm{l.o.s.}} n_g^{3D}(\chi, \boldsymbol{\theta}) \frac{v_r(\chi, \boldsymbol{\theta})}{c} dV \\
\nonumber &=  \frac{1}{\bar{n}_g^{2D}}\int n_g^{3D}(z, \boldsymbol{\theta}) \frac{v_r(z, \boldsymbol{\theta})}{c} \chi^2(z) \frac{c}{H(z)} dz \\
\nonumber &= \int \frac{\bar{n}_g^{3D}(z)}{\bar{n}_g^{2D}} [1 + \delta_g(z, \boldsymbol{\theta})] \frac{v_r(z, \boldsymbol{\theta})}{c} \chi^2(z) \frac{c}{H(z)} dz \\
 &=  \int_0^{\infty} \frac{dp}{d\chi} [1 + \delta_g(\chi, \boldsymbol{\theta})] \frac{v_r(\chi, \boldsymbol{\theta})}{c} d\chi.
\label{eq:template}
\end{align}

\begin{figure*}
    \centering
    \includegraphics[width=0.9\linewidth]{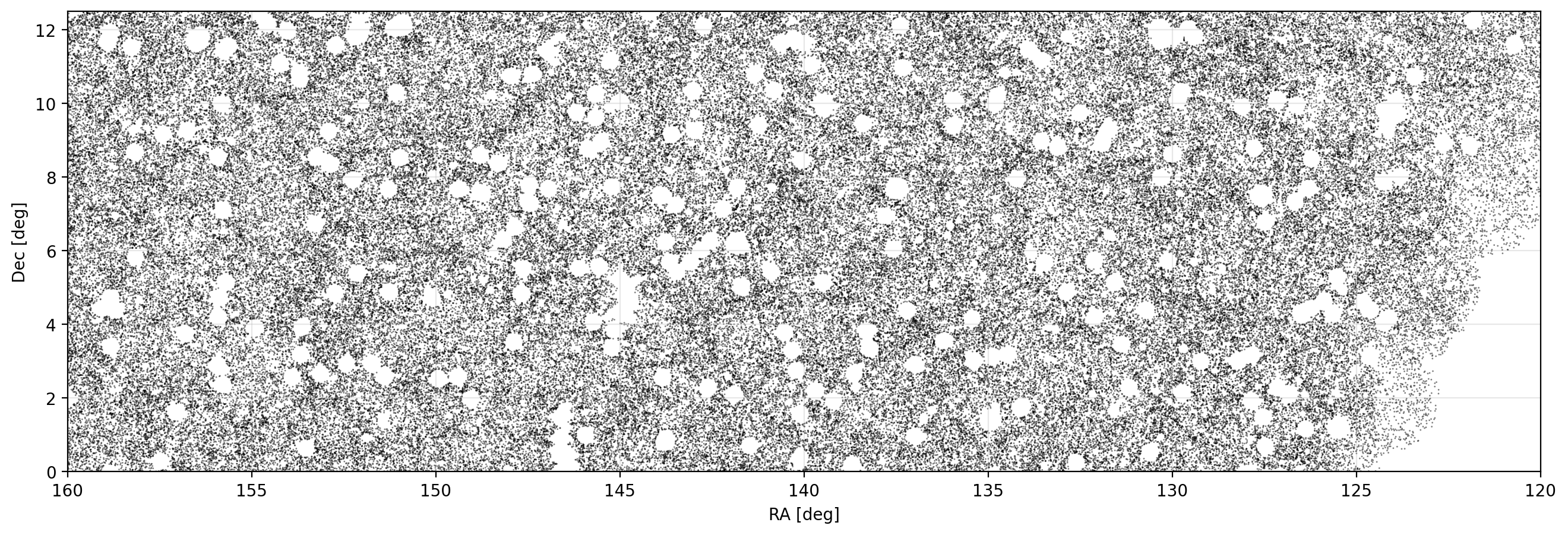}
    \vspace{0.2cm}
    \includegraphics[width=0.9\linewidth]{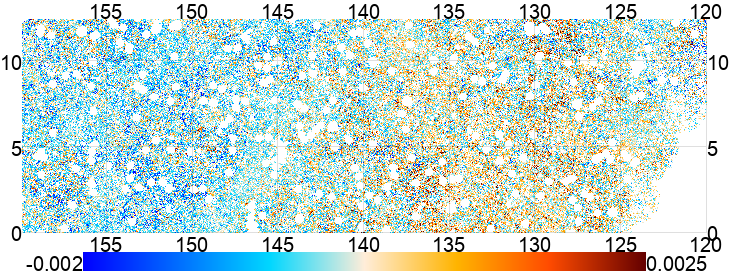}
    \caption{
\textbf{Top:} Projected distribution of DESI LRG galaxies in a representative sky region ($120^\circ < \mathrm{RA} < 160^\circ$, $0^\circ < \mathrm{Dec} < 12.5^\circ$). Each point represents an individual galaxy, with denser regions appearing darker due to overlapping markers. For display purposes, the marker size ($\sim 2\,\mathrm{arcmin}$) is larger than the true angular extent of individual galaxies.  The sparse region in the lower right corresponds to the survey mask. The kSZ signal arises from the modulation of this galaxy distribution by coherent large-scale velocities.
\textbf{Bottom:} Momentum-weighted template showing the projected line-of-sight velocity field used in the kSZ analysis. For visualization purposes, the map has been downsampled by a factor of 7 in resolution. The map exhibits coherent structures on scales of $\gtrsim 10$ degrees, consisting of large regions of positive (orange) and negative (blue) velocities tracing large-scale cosmic flows. The same mask is applied as in the top panel. In both panels, the white circular regions correspond to masked ACT clusters.
    }
    \label{fig:template}
\end{figure*}

where $dp/d\chi \equiv \bar{n}_g^{3D}(z)\chi^2c/[H(z)\bar{n}_g^{2D}]$ is the normalized radial selection function. Figure~\ref{fig:template} illustrates the key ingredients of the kSZ measurement for a representative sky region. The top panel shows the projected distribution of DESI LRG galaxies, which serve as a proxy for the small-scale electron density field. The bottom panel displays the corresponding momentum-weighted velocity template, reconstructed from the same galaxy catalog. The velocity field exhibits coherent structures on scales of $\gtrsim 10$ degrees, with regions of positive (orange) and negative (blue) line-of-sight velocity tracing large-scale cosmic flows. The kSZ signal arises from the coupling between these large-scale bulk velocities and the small-scale electron distribution.

\subsubsection{kSZ temperature power spectrum measurement}

We measure the cross-correlation between the reconstructed momentum field $\hat{\pi}$ and the CMB temperature map $T_{\rm CMB}$ using a pseudo-$C_\ell$ estimator implemented in \textsc{NaMaster} \cite{Alonso_2019}. Expanding both fields in spherical harmonics gives
\begin{equation}
\hat{\pi}(\hat{\boldsymbol{n}}) = \sum_{\ell m} \pi_{\ell m} Y_{\ell m}(\hat{\boldsymbol{n}}),
\qquad
T_{\rm CMB}(\hat{\boldsymbol{n}}) = \sum_{\ell m} \Theta_{\ell m} Y_{\ell m}(\hat{\boldsymbol{n}}),
\end{equation}
from which the pseudo-cross-spectrum estimator is constructed as
\begin{equation}
\tilde{C}_\ell^{\hat{\pi} \times {T_{CMB}}}
= \frac{1}{2\ell+1}\sum_{m=-\ell}^{\ell} \pi_{\ell m}\, \Theta_{\ell m}^\ast .
\end{equation}

The pseudo-$C_\ell$ formalism accounts for the impact of sky masking and pixelization, which induce mode coupling between multipoles. The raw pseudo-spectra $\tilde{C}_\ell^{\hat{\pi}\times{T_{CMB}}}$ differ from the true underlying cross-spectrum $C_\ell^{\hat{\pi}\times{T_{CMB}}}$ because of this effect, but their expectation values can be related through the mode-coupling matrix $M_{\ell\ell'}$:
\begin{equation}
\langle \tilde{C}_\ell^{\hat{\pi}\times{T_{CMB}}} \rangle
= \sum_{\ell'} M_{\ell\ell'} C_{\ell'}^{\hat{\pi}\times{T_{CMB}}} .
\end{equation}
Here $M_{\ell\ell'}$ is determined entirely by the survey mask. We can invert this relationship approximately to extract the true power spectrum by assuming it is piecewise constant across several discrete bins. In practice, rather than inverting this relationship, we forward-model the bandpower window function, computed from $M_{\ell\ell'}$ and the binning scheme, onto the theory predictions and compare directly to the binned pseudo-$C_\ell$ measurements.

\subsubsection{Theoretical modelling}

We use \texttt{class-sz}\citep{Bolliet:2023eob} to compute the cross-power spectrum between the 
electron distribution and galaxies. In the Limber approximation this can be written as
\begin{align}
C_{\ell}^{\tau g}
&=\int_0^\infty \frac{d\chi}{\chi^2}\,
W_\tau(\chi)\,W_g(\chi)\,
P_{eg}\!\left(k_\ell(\chi),z(\chi)\right),
\label{eq:cltaug}
\end{align}
where
\begin{equation}
W_\tau(\chi)=\sigma_T\,a^{-2}(\chi)\,\bar n_e(\chi),
\qquad
W_g(\chi)=\frac{dp}{d\chi},
\end{equation}
and $k_\ell(\chi)=(\ell+1/2)/\chi$. 
Here $P_{eg}(k,z)$ denotes the 3D electron--galaxy cross-power spectrum. 
Within the halo model this splits into one-halo and two-halo terms:
\begin{align}
P_{eg}^{1h}(k,z) &= \int dM\,\frac{dn}{dM}(M,z)\,
\tilde u_{\tau}(k|M,z)\,
\tilde u_{g}(k|M,z), \label{eq:peg1h}\\
P_{eg}^{2h}(k,z) &= 
\left[\int dM\,\frac{dn}{dM}(M,z)\,b(M,z)\,\tilde u_{\tau}(k|M,z)\right]\nonumber\\
&\quad\times
\left[\int dM\,\frac{dn}{dM}(M,z)\,b(M,z)\,\tilde u_{g}(k|M,z)\right]
P_L(k,z),
\label{eq:peg2h}
\end{align}
so that $P_{eg}(k,z)=P_{eg}^{1h}(k,z)+P_{eg}^{2h}(k,z)$. 
The different terms are defined as follows:
\begin{itemize}
    \item $W_g(\chi)=dp/d\chi$ is the galaxy selection function, normalized such that $\int d\chi\,W_g(\chi)=1$. It encodes the redshift distribution and weights used to project the 3D galaxy density into the observed sky field.
    \item $W_\tau(\chi)=\sigma_T a^{-2}(\chi)\bar n_e(\chi)$ is the optical-depth kernel, where $\sigma_T$ is the Thomson cross-section and $\bar n_e$ is the mean electron density.
    \item $\tfrac{dn}{dM}(M,z)$ is the halo mass function.
    \item $\tilde u_{\tau}(k|M,z)$ is the Fourier transform of the electron profile of halos of mass $M$ at redshift $z$ (e.g. the GNFW profile described in the following section).
    \item $\tilde u_{g}(k|M,z)$ is the Fourier transform of the galaxy density profile, determined by the halo occupation distribution (HOD). 
    \item $b(M,z)$ is the halo bias, and $P_L(k,z)$ is the linear matter power spectrum.
\end{itemize}

Having established the expression for $C_\ell^{\tau g}$, we now relate it to the 
cross-correlation measured between the reconstructed momentum template $\hat{\pi}$
and the CMB temperature, $C_\ell^{\pi\times{T_{CMB}}}$. 

In the high-$k$ Ostriker--Vishniac  (OV) limit, the kSZ anisotropy is sourced by the electron density--velocity field as
\begin{equation}
\Theta_{\rm kSZ}(\hat{\mathbf{n}})
= -\,\sigma_T \int_0^\infty d\chi \, a^{-2}(\chi)\,\bar{n}_e(\chi)\,
\delta_e(\chi,\hat{\mathbf{n}})\,\frac{v_r(\chi,\hat{\mathbf{n}})}{c}\,.
\end{equation}

Cross-correlating this with our momentum template $\hat{\pi}$ in Eq. \ref{eq:template} results in

\begin{align}
C_\ell^{\hat{\pi}\times {T_{CMB}}}
&= -\,\frac{\sigma_T}{c^2}
\int d\chi\,\frac{a^{-2}(\chi)}{\chi^2}\,
\bar{n}_e(\chi)\,\frac{dp}{d\chi} \nonumber\\
&\qquad\times
P_{\delta_e v_r,\,\delta_g \hat{v}_r}
\!\left(k=\frac{\ell+1/2}{\chi},\,\chi\right).
\label{eq.total}
\end{align}

Analogous to the treatment in the Ostriker--Vishniac 
limit  for kSZ \cite{PhysRevLett.88.211301,McQuinn_2005,Park_2016}, we can simplify 
$P_{\delta_e v_r,\,\delta_g \hat{v}_r}$ into an expression involving 
$P_{e g}$:

\begin{align}
P_{\delta_e v_r,\,\delta_g \hat{v}_r}(k,\chi)
&\simeq \int \frac{d^3\boldsymbol{k}^\prime}{(2\pi)^3}\,P_{v_r \hat{v}_r}(|\boldsymbol{k}'|)\,P_{e g}(k,\chi)\,
\nonumber\\&= r\sigma_\mathrm{true}\sigma_{\mathrm{rec}}P_{{e{g}}}(k,\chi).
\label{eq:Pegr_factorization}
\end{align}

In Eq.~\ref{eq:Pegr_factorization}, $\sigma_{\rm true}$ denotes the velocity dispersion of the true line-of-sight velocities, which we obtain from the AbacusSummit $N$-body simulation suite \cite{PhysRevD.109.103534}. The quantity $\sigma_{\rm rec}$ corresponds to the dispersion of the reconstructed velocities in our galaxy catalog and  $r$ is the cross-correlation coefficient between the reconstructed and true velocities that we again adopt as $r \simeq 0.65$ as described in Sec. \ref{sec.r}. 

Substituting Eq.~\ref{eq:Pegr_factorization} in Eq.~\ref{eq.total} results in

\begin{equation}
    C_\ell^{\hat{\pi}\times T}=-r\sigma_\mathrm{true}\sigma_{\mathrm{rec}}C^{\tau{g}}_\ell
\label{eq.theory}
\end{equation}


\subsubsection{GNFW gas profile}

We model the gas distribution using the generalized NFW (GNFW) profile, which provides a flexible parametrization of the three-dimensional gas density. The profile is defined as

\begin{align}\label{eq.1h}
    \rho_\mathrm{GNFW}(r) &= \rho_0
    \Bigg(\frac{r}{x_c\,r_{200c}}\Bigg)^{\gamma}
    \Bigg[1+\Bigg(\frac{r}{x_c\,r_{200c}}\Bigg)^\alpha\Bigg]^{-\frac{\beta+\gamma}{\alpha}}, \\
    \rho_\mathrm{gas}(r) &= \rho_\mathrm{GNFW}(r)\,\rho_\mathrm{cr}(z)\,f_b,
\end{align}

where $r_{200c}$ is the characteristic radius enclosing mass $m_{200c}$ at 200 times the critical density, $x_c$ is the dimensionless core scale, and $(\alpha,\beta,\gamma)$ control the slopes at $r\sim x_c r_{200c}$, $r \gg x_c r_{200c}$, and $r \ll x_c r_{200c}$, respectively. Throughout, we fix $\gamma=-0.2$ since our measurements are relatively insensitive to the inner profile shape. 

\begin{figure*}
    \centering
    \includegraphics[width=\linewidth]{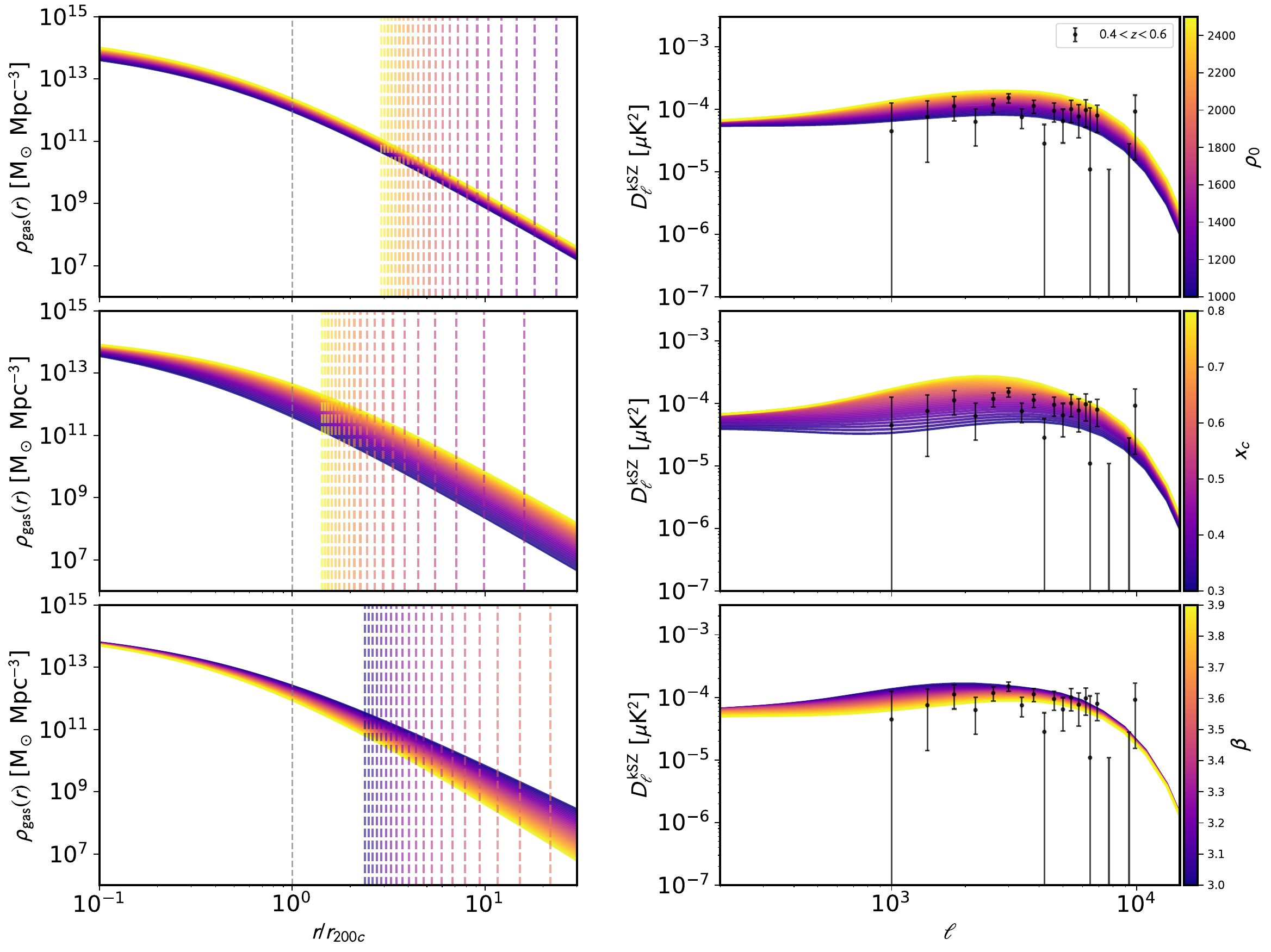} \\ 

    \caption{
Sensitivity of the GNFW gas profile and the kSZ-galaxy correlation to the three sampled shape parameters.
\textbf{Left:} Gas density profiles $\rho_{\rm gas}(r)$ versus $r/r_{200c}$ when varying one GNFW parameter at a time (rows, top to bottom: $\rho_0$, $x_c$, $\beta$); all other parameters are held at their fiducial values. The gray dashed line marks $r=r_{200c}$, while the colored dashed lines indicate the truncation radius $r_{\rm cut}$ for each profile, where the enclosed gas mass equals $f_b M_{200c}$.
\textbf{Right:} Corresponding kSZ angular power spectra $D_\ell^{\rm kSZ} \equiv \ell(\ell+1)C_\ell^{\rm kSZ}/(2\pi)$ computed from the same profiles. The kSZ spectra include both 1-halo and 2-halo contributions, with the 2-halo amplitude $A_{k2h}$ fixed to its best-fit value. Colored curves sweep from low (purple) to high (yellow) values of the parameter indicated by each colorbar; black points show the measurement in the redshift bin $0.4 < z < 0.6$.
Due to strong degeneracies between the five GNFW parameters ($\rho_0$, $x_c$, $\alpha$, $\beta$, $\gamma$), we restrict our analysis to the three parameters shown here, fixing $\alpha = 0.88$ and $\gamma = -0.2$ to their fiducial values from \citet{Battaglia_2016}.
The truncation radius $r_{\rm cut}$ is determined by requiring the enclosed gas mass to equal the cosmic baryon fraction times the halo mass, $M_{\rm gas}(<r_{\rm cut}) = f_b M_{200c}$. This constraint causes higher $\rho_0$ (more gas) to yield smaller $r_{\rm cut}$, while steeper outer slopes ($\beta$) or larger core radii ($x_c$) extend $r_{\rm cut}$ outward. As a result, the kSZ-galaxy power spectra converge on large scales (low $\ell$) where all gas is enclosed within $r_{\rm cut}$.
  }
    \label{fig:gnfw_param_sensitivity}
\end{figure*}

As shown in Fig.~\ref{fig:gnfw_param_sensitivity}, each GNFW parameter has a distinct impact on the profile. Increasing $\rho_0$ rescales the overall amplitude; increasing $x_c$ (i.e.\ shifting the scale radius $r_s = x_c r_{200c}$) moves the turnover to larger radii and broadens the profile; larger $\alpha$ sharpens the transition; and larger $\beta$ steepens the outer slope while leaving the inner slope fixed by $\gamma$. The corresponding effects on the kSZ bandpowers derived from these profiles are shown in the right panels of Fig.~\ref{fig:gnfw_param_sensitivity}. In addition to the one-halo gas profile (Eq.~\ref{eq.1h}), we include a two-halo contribution by allowing a free amplitude $A_{k2h}$ that rescales the two-halo term $P_{eg}^{2h}(k,z)$ in Eq.~\ref{eq:peg2h}, so that the total electron--galaxy power spectrum is $P_{eg}(k,z) = P_{eg}^{1h}(k,z) + A_{k2h}\,P_{eg}^{2h}(k,z)$.

\subsection{Impact of gas profile truncation}

A subtle but important caveat is that the profile normalization is chosen to preserve the total gas mass. Following \citep{Bolliet_2023}, the GNFW profile is truncated at a radius $r_\mathrm{cut}$ such that the enclosed gas mass equals $f_b m_{200c}$, consistent with the NFW case. This guarantees mass conservation and avoids the unphysical accumulation of gas mass at large radii that would arise if the profile were left untruncated.

The choice of truncation radius significantly impacts the predicted kSZ signal and is an important consideration when comparing to simulations. Different truncation schemes yield substantially different predictions at large radii ($r \gtrsim r_{200c}$), even when profiles agree at smaller radii where simulations provide direct constraints. Typical truncation radii lie outside the range where gas profiles can be reliably measured in hydrodynamical simulations due to resolution limits and contamination from neighboring halos, making the choice of $r_{\rm cut}$ somewhat arbitrary. Different studies therefore adopt different prescriptions: for example, \citep{sunseri2025disentanglinghalojointmodel} normalize the gas profile by requiring the baryon-to-total-matter ratio within a specified radius $r_b$ to equal $f_b$, rather than explicitly conserving total halo gas mass as we do here. Because the kSZ signal integrates the gas profile out to radii where simulations cannot provide direct constraints, a fitting function that matches the simulated density profile at small radii can still yield a biased kSZ prediction once combined with a given truncation scheme. Self-consistency therefore requires validating the fit plus truncation scheme against the forward-modelled kSZ signal from the simulation itself, not only against its density profile.

\subsubsection{Covariance matrix estimation}

We estimate the uncertainties in our measurements using the Gaussian covariance formalism implemented in \textsc{NaMaster} \citep{Alonso_2019}, which accounts for mask-induced mode coupling through a precomputed covariance workspace. For clarity, we show here the simplified Knox formula \citep{Knox_1995}, to which the full expression reduces in the limit of a uniform mask:
\begin{equation}
\begin{split}
    \sigma^2(C^{\hat{\pi} \times T_{\mathrm{CMB}}}_{\ell_b}) &= \frac{1}{(2\ell_b+1)\Delta\ell f_{\mathrm{sky}}} \\
    &\quad \times \left[ C^{\hat{\pi}\hat{\pi}}_{\ell_b} C^{T_{\mathrm{CMB}}T_{\mathrm{CMB}}}_{\ell_b} + \left(C^{\hat{\pi} \times T_{\mathrm{CMB}}}_{\ell_b}\right)^2 \right],
\end{split}
\label{eq.cov}
\end{equation}
where $C^{\hat{\pi}\hat{\pi}}_{\ell_b}$ is the auto-spectrum of the momentum template, $C^{T_{\mathrm{CMB}}T_{\mathrm{CMB}}}_{\ell_b}$ is the CMB temperature auto-spectrum, $\Delta\ell$ is the bin width, and $f_{\mathrm{sky}}$ is the sky fraction. In practice, our analysis uses the full  \textsc{NaMaster}  covariance, which incorporates the exact coupling coefficients between bandpowers due to the survey mask geometry.

\subsubsection{Likelihood and Parameter Inference}

For our analysis, we construct a Gaussian likelihood using the measured angular 
power spectrum $C_\ell$ over the multipole range $1000 \leq \ell \leq 7000$ and 
the Gaussian covariance matrix described in Section~\ref{sec.methodology}. The likelihood takes the standard quadratic form,
\begin{equation}
\ln \mathcal{L}(\mathbf{C}_\ell | \boldsymbol{\theta}) \;=\; -\tfrac{1}{2}\, 
\left[ \mathbf{C}_\ell^{\rm obs} - \mathbf{C}_\ell^{\rm th}(\boldsymbol{\theta}) \right]^{\!\rm T} 
\mathbb{C}^{-1}\,
\left[ \mathbf{C}_\ell^{\rm obs} - \mathbf{C}_\ell^{\rm th}(\boldsymbol{\theta}) \right],
\end{equation}
where $\mathbf{C}_\ell^{\rm obs}$ denotes the measured bandpowers, 
$\mathbf{C}_\ell^{\rm th}(\boldsymbol{\theta})$ is the theoretical prediction 
for the GNFW gas profile parameters $\boldsymbol{\theta} = \{\rho_0, x_c, \beta, A_{2h}\}$ 
computed with \texttt{class-sz} (Eq.~\ref{eq.theory}), and $\mathbb{C}$ is the 
covariance matrix (Eq.~\ref{eq.cov}). We adopt uniform priors on the 
four gas profile parameters: $\log_{10}\rho_0 \in [1, 5]$,
$x_{c,k} \in [0.1, 1]$, $\beta_k \in [3.01, 6]$, and
$A_{\rm k2h} \in [0, 5]$, as summarized in Table~\ref{table.prior}. We
additionally impose an analytic prior that excludes GNFW parameter combinations 
for which the profile truncation condition 
$M_{\rm gas}(<r_{\rm cut}) = f_b\,M_{200c}$ has no solution 
(Appendix~\ref{app:zbrent}). We sample the posterior using 
\texttt{Cobaya} \citep{Torrado_2021}, running chains until the Gelman--Rubin 
convergence criterion reaches $R - 1 < 0.01$.

\subsection{Configuration Space estimator }\label{harmonic_CAP}
We employ the CAP filter to measure the CMB temperatures $\mathcal{T}_i(\theta_d)$ around each galaxy $i$. 
The CAP provides a profile-agnostic way to reconstruct spherically averaged profiles and is an effective method for removing the large-scale primary CMB. The CAP filter is defined as
\begin{equation}
\mathcal{T}(\theta_d) = \int d^2\theta \, \delta T(\theta) \, W_{\theta_d}(\theta),
\end{equation}
where $\delta T(\theta)$ are the CMB temperature fluctuations and the filter $W_{\theta_d}$ is constructed as:
\begin{equation}
W_{\theta_d}(\theta) = \begin{cases}
1 & \text{for } \theta < \theta_d, \\
-1 & \text{for } \theta_d \leq \theta \leq \sqrt{2}\theta_d, \\
0 & \text{otherwise}.
\end{cases}
\end{equation}
This filter measures the integrated temperature fluctuation within a disk of radius $\theta_d$ and subtracts a concentric ring of equal area:
\begin{equation}
\mathcal{T}(R) = \int_0^R d^2\theta \, \delta T(\theta) - \int_R^{\sqrt{2}R} d^2\theta \, \delta T(\theta),
\end{equation}
where $R$ is the physical radius corresponding to the angular scale $\theta_d$ at a given redshift. At large radii, the CAP-filtered profiles behave analogously to a cumulative density profile.

We adopt a velocity-weighted, uniform-mean estimator  by weighting each galaxy's CAP measurement according to its reconstructed line-of-sight velocity:
\begin{equation}
\hat{T}_{\rm kSZ}(R) = -\frac{1}{r} \frac{\sigma_v^{\rm rec}}{c} \frac{\sum_i \mathcal{T}_i(R)(v_{\rm rec,i}/c)}{\sum_i (v_{\rm rec,i}/c)^2},
\label{eq.stacking}
\end{equation}
The resulting stacked kSZ profile obtained from this procedure is shown in Figure~\ref{fig.profile_real}.

\subsubsection{Covariance matrix estimation}

\begin{figure}[tbp]
    \centering
    \includegraphics[width=\linewidth]{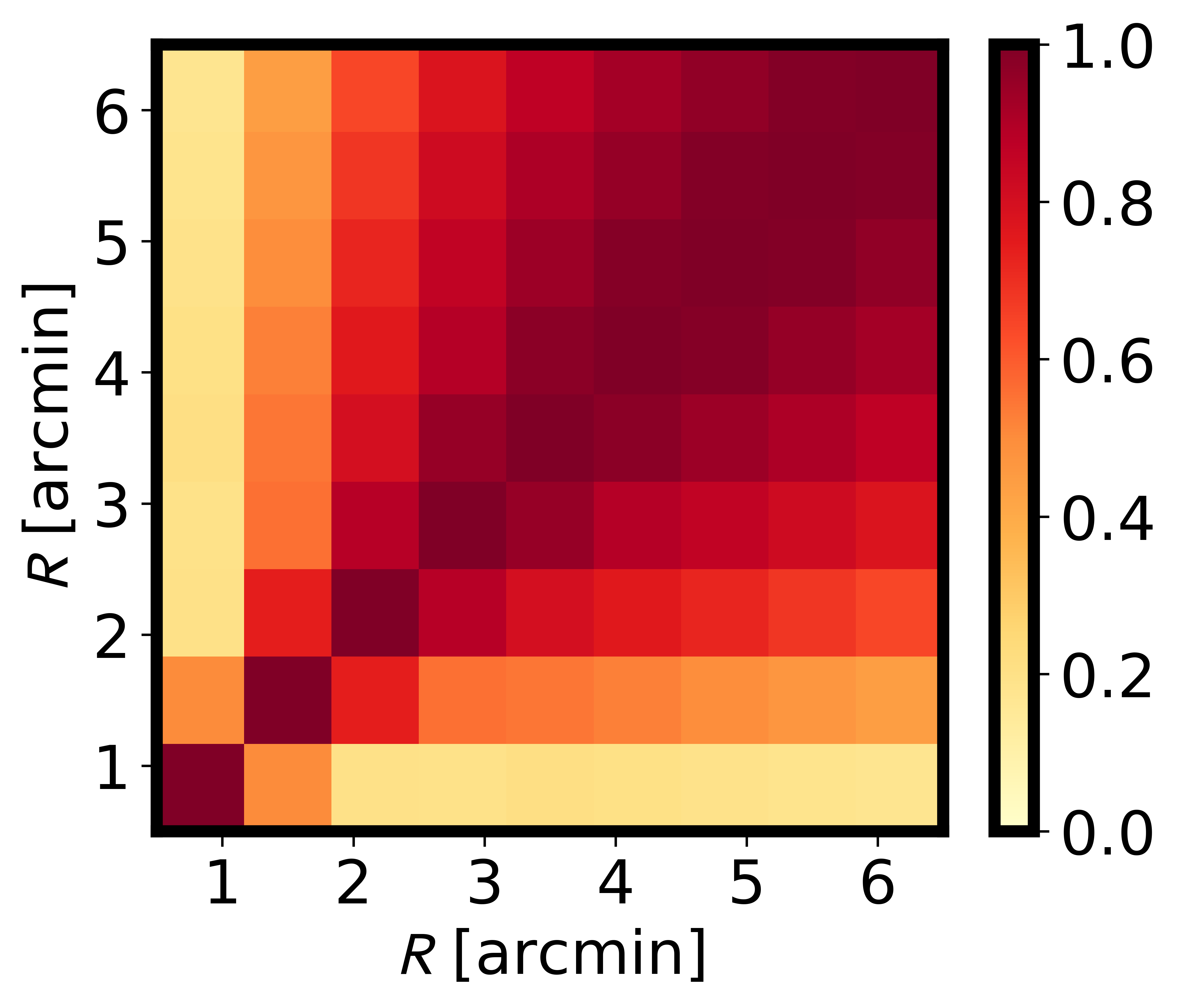} \\ 
    \caption{Correlation coefficient matrix between kSZ measurements at different CAP filter radii, estimated from 10,000 block bootstrap realizations. 
    The measurements show strong correlations at large apertures ($R \gtrsim 3'$) due to residual CMB fluctuations, while smaller apertures exhibit weaker correlations.}
    \label{fig.corr_coeff}
\end{figure}

We estimate the configuration space covariance matrix using a block bootstrap procedure that accounts for the spatial correlations inherent in galaxy clustering. We divide the survey footprint into non-overlapping $2^\circ \times 2^\circ$ patches, yielding 1837 independent spatial blocks. For each bootstrap realization, we randomly resample these blocks with replacement to construct a resampled galaxy catalog containing the same total number of objects as the original dataset. From each resampled catalog, we measure the stacked kSZ CAP profile following the procedure described in Section~\ref{harmonic_CAP}. We repeat this process 10,000 times to generate an ensemble of profile measurements from which we construct the covariance matrix (Fig.~\ref{fig.corr_coeff}). By sampling on patches of sky instead of on individual galaxies, we produce an unbiased estimate of the covariance matrix that takes into account the noise correlation between the galaxies. This is particularly important for high-density galaxy samples, where bootstrapping individual galaxies underestimates the covariance by $\sim 10\%$ on the scales of interest.



\section{Harmonic space measurements and results}\label{sec.harmonic}

\begin{figure*}[t!]
    \centering
    \includegraphics[width=\linewidth]{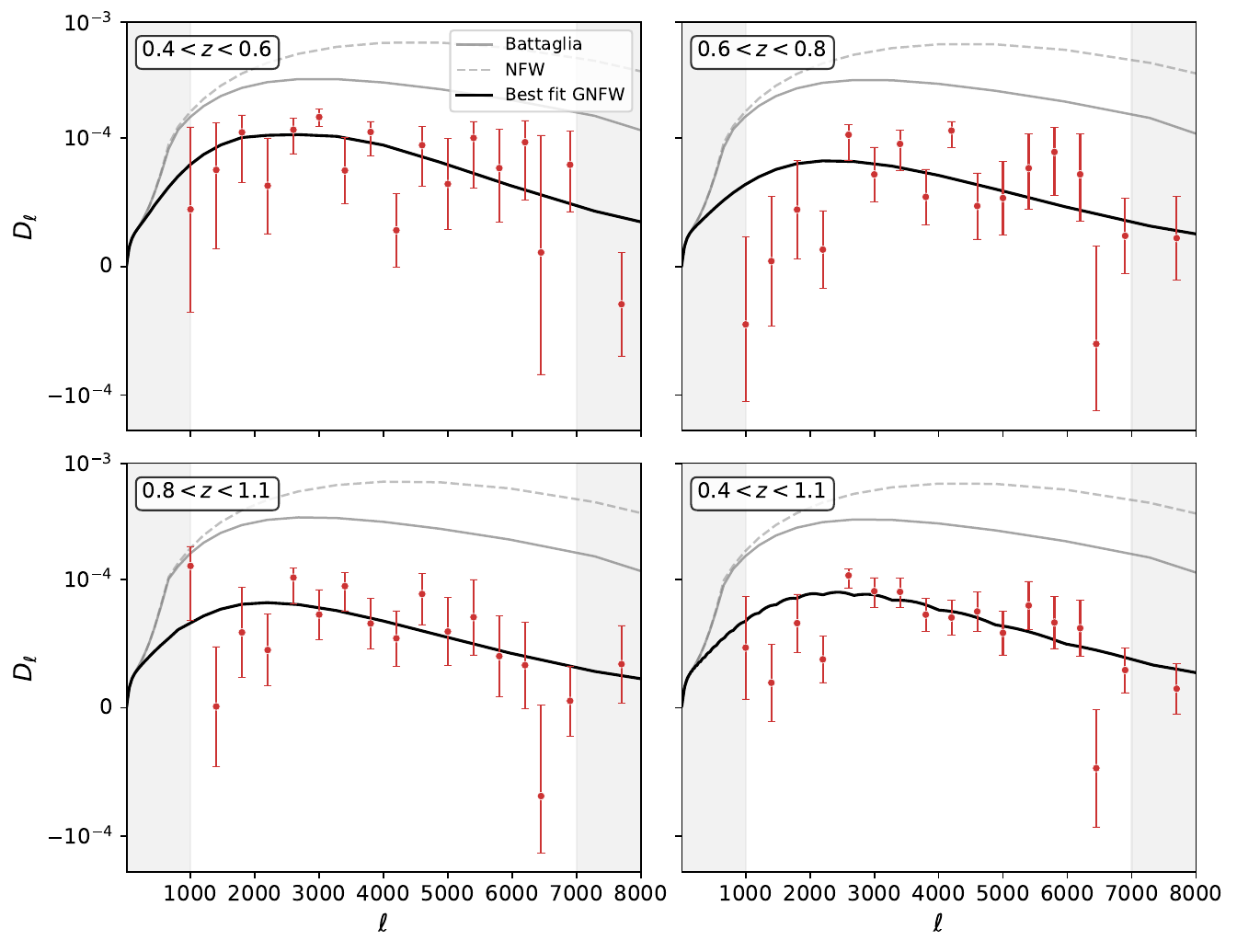}
    \caption{Measured kSZ-galaxy power spectrum from cross-correlating DESI galaxies with ACT CMB temperature maps across three redshift bins and the combined sample. Red points show the measurements with error bars; the shaded grey regions indicate multipoles excluded from the fit ($\ell < 1000$ and $\ell > 7000$). The black solid line shows the best-fit generalized Navarro-Frenk-White (GNFW) gas profile model, which provides acceptable fits across all redshift bins (PTE $= 0.38$, $0.06$, $0.38$, and $0.07$ for $z_1$, $z_2$, $z_{3}$, and combined, respectively). The solid grey line shows predictions using the \citet{Battaglia_2016} gas profile, which is strongly disfavored across all redshift bins (PTE $< 0.001$ for all). The dashed grey line shows predictions using an NFW dark matter-only profile, which significantly overestimates the signal as expected due to the absence of baryonic feedback.}

    \label{fig.measurement_fits}
\end{figure*}

We present harmonic-space measurements of the cross-power spectrum between the kSZ template and CMB maps. This approach complements traditional real-space measurements and provides a direct probe of the gas profile across different angular scales. Figure~\ref{fig.measurement_fits} shows the measured cross-power spectra $D_\ell$ for each of the three individual redshift bins ($0.4<z<0.6$, $0.6<z<0.8$, and $0.8<z<1.1$) as well as for the combined $0.4<z<1.1$ sample. The measurements span multipoles $\ell \sim 1000$--$10000$, corresponding to angular scales sensitive to the gas distribution from the halo cores to the outskirts. We restrict our analysis to the multipole range $1000 < \ell < 7000$ (unshaded region in Figure~\ref{fig.measurement_fits}). This range is chosen based on two criteria: first, it captures the scales where the bulk of the signal-to-noise is concentrated, as extending to either larger or smaller scales yields negligible improvement in total SNR; second, our null tests pass across this range, keeping systematic effects under control.

We compare our measurements to predictions from the \citet{Battaglia_2016} model (solid grey) and an NFW profile (dashed grey). The \citep{Battaglia_2016} model, calibrated from hydrodynamical simulations including feedback processes, provides a useful benchmark. However, it is strongly disfavored across all redshift bins, with PTE $< 0.001$ in each case. This suggests that the gas profiles in DESI LRG halos differ significantly from this simulation-calibrated prediction. As expected, the measurements show clear deviation from the NFW profile, demonstrating that the gas distribution traced by the kSZ effect does not simply follow the dark matter distribution.

Fitting a single amplitude rescaling to the Battaglia profile against our combined-sample bandpowers over $1000 < \ell < 7000$, we find $A_{\rm Batt} = 0.249 \pm 0.015$ ($\chi^2 = 17.3$ for $15$ degrees of freedom, $\text{PTE} = 0.30$), indicating that the Battaglia AGN feedback model overpredicts the kSZ signal by a factor of $\approx 4$. This result is qualitatively consistent with kSZ velocity reconstruction analyses that adopt the Battaglia AGN profile as their fiducial $C_\ell^{\tau g}$.
Ref.~\citep{lai2025kszvelocityreconstructionact} measure $A = 0.39 \pm 0.04$ relative to the Battaglia model using a quadratic maximum likelihood cross-power spectrum estimator applied to ACT DR6 and DESI Legacy Survey DR9 photometric LRGs, while \citep{hotinli2025velocityreconstructionkszmeasuring} find $b_v = 0.45^{+0.06}_{-0.05}$ with an independent 3D Cartesian pipeline on the same data combination. In their framework, the quadratic velocity estimator is normalized using the Battaglia AGN profile as the fiducial $C_\ell^{\tau g}$, so $b_v$ (or equivalently $A$) parametrizes the ratio of the true to predicted electron-galaxy cross-power, averaged over the small scales ($\ell \sim 1000$--$10000$) entering the estimator. Since our measured bandpowers also scale linearly with $C_\ell^{\tau g}$ (Eq.~\ref{eq.theory}), and the Battaglia profile shape is an acceptable fit to the data when the amplitude is freed, $A_{\rm Batt}$ and $b_v$ can be meaningfully compared as measurements of the same underlying suppression factor. More quantitative agreement between these estimates is not expected, since the $b_v$ parameter also depends on the CMB noise level and angular resolution as well as the galaxy survey shot noise, and is thus expected to vary between experiments.

\begin{figure}[t!]
    \centering
    \includegraphics[width=\columnwidth]{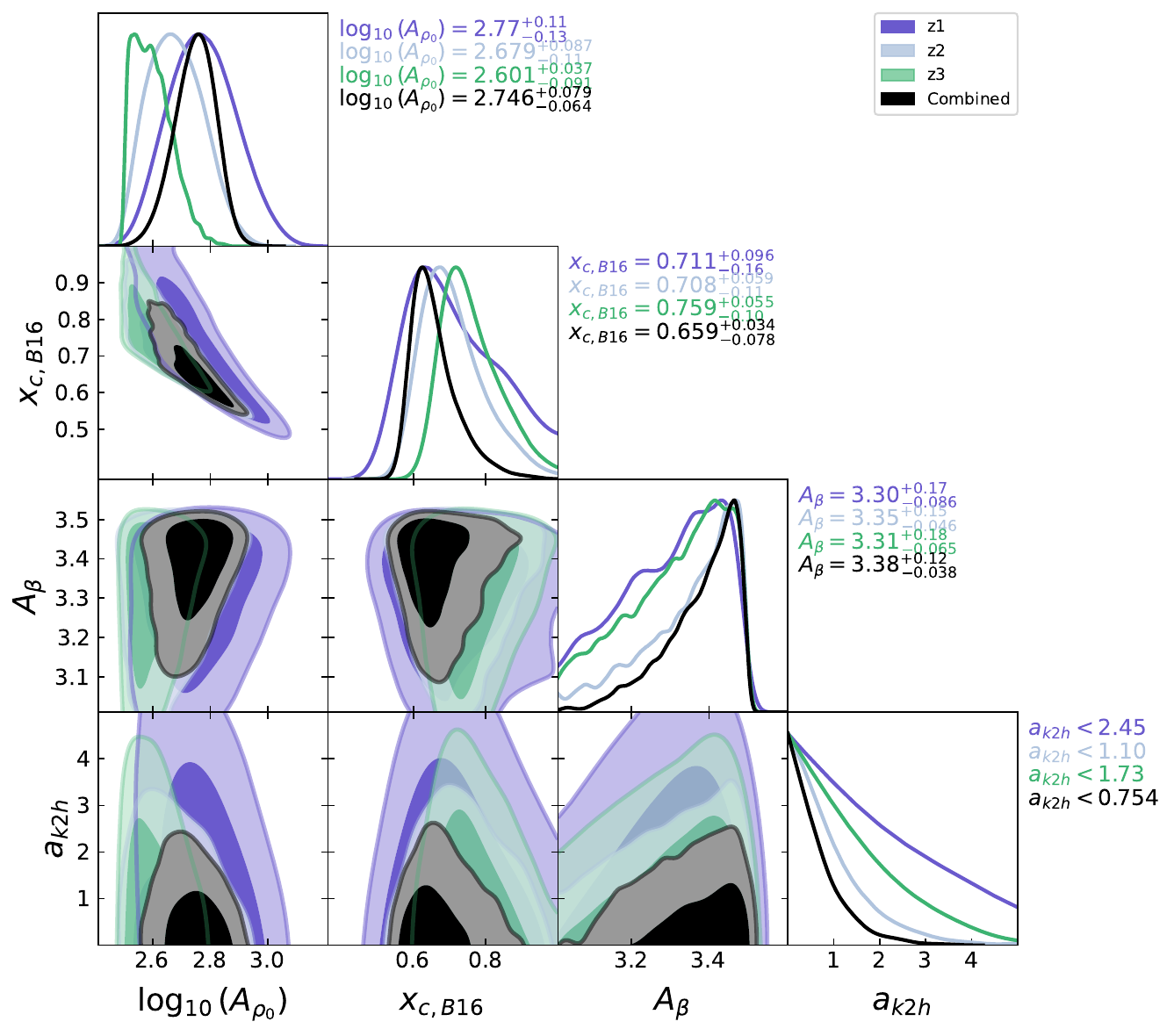}
    \caption{Marginalized posteriors for the four GNFW gas profile parameters (numerical constraints in Table~\ref{table.prior}). Diagonal panels show one-dimensional marginalized posteriors; off-diagonal panels show two-dimensional contours at 68\% and 95\% confidence levels. The four parameters varied are the normalization $\log_{10}(A_{p_0})$, the concentration parameter $x_{\rm c, B16}$, the outer slope $A_\beta$, and the two-halo amplitude $a_{\rm k2h}$. Purple, light blue, and green contours correspond to the three individual redshift bins $0.4 < z < 0.6$, $0.6 < z < 0.8$, and $0.8 < z < 1.1$, respectively, while black contours show the combined $0.4 < z < 1.1$ sample. The combined analysis yields tight constraints on the gas profile normalization and outer slope, demonstrating the power of kSZ measurements to constrain the gas distribution in LRG host halos.}
    \label{fig:Corner}
\end{figure}

To characterize the gas distribution, we constrain the GNFW density profile defined in Eq.~\ref{eq.1h} by fitting four parameters: the density amplitude $\rho_0$, the core radius $x_{c,k}$, the power-law index $\beta_k$ describing the asymptotic profile falloff, and the two-halo term amplitude $A_{k2h}$. We fix two parameters that primarily affect the small-radius behavior, where our measurements have limited sensitivity: $\gamma_k=-0.2$ and $\alpha_k=1$, following \citep{PhysRevD.103.063514,Battaglia_2016}. Table~\ref{table.prior} presents the marginalized constraints with $1\sigma$ uncertainties for each redshift bin individually and for the joint analysis. Figure~\ref{fig:Corner} shows the one-dimensional marginalized posteriors and two-dimensional contours for the four fitted parameters. The gas profile normalization $\log_{10}(\rho_0)$ and outer slope $\beta_k$ are particularly well constrained. The concentration parameter $x_{c,k}$ shows broader degeneracies with other parameters, as expected given our sensitivity range.

The best-fit GNFW profile provides a good description of the measurements
across all redshift bins within our analysis range. For the individual bins,
we obtain $\chi^2 = 12.9$, $20.5$, and $12.8$ for $z_1$, $z_2$, and $z_3$,
respectively, all for 12 degrees of freedom, corresponding to PTE values of
$0.38$, $0.06$, and $0.38$. The joint constraint yields
$\chi^2_{\rm bf} = 19.7$ for 12 degrees of freedom (PTE $= 0.07$). The
acceptable PTE values across all redshift bins indicate no evidence for
unmodeled systematics or redshift-dependent biases within the fitting range.

\begin{table}[htbp]
    \begin{center}
    \caption{Marginalized constraints on GNFW gas profile parameters from fitting the kSZ cross-power spectrum. We report the median and 68\% credible intervals for each parameter; one-sided limits indicate the 68\% upper or lower bound when the posterior is unconstrained on one side. The HOD parameters are fixed to the best-fit values from the DESI DR2 clustering analysis in each redshift bin. For the combined sample, no single HOD fit is used; instead, the theory prediction is constructed as a weighted sum of the three individual-bin predictions, $\bar{C}_\ell^{ge,\mathrm{comb}} = \sum_i w_i , C_\ell^{ge,i}$, where the weights $w_i$ are proportional to the galaxy number in each bin. The best-fit $\chi^2$ is the minimum value found across the converged chain samples, with $\nu = 12$ degrees of freedom (16 data points after $\ell$-range cuts minus 4 gas parameters). The last row shows the uniform priors adopted for each parameter.
    The full parameter correlation structure is shown in Fig.~\ref{fig:Corner}.
}
    \label{table.prior}
    \renewcommand{\arraystretch}{1.2}
    \small
    \begin{tabular}{l|c c c c|c}
        \hline
        Bin & $\log_{10}\rho_{0}$ & $x_{c,k}$ & $\beta_k$ & $A_{\rm k2h}$ & $\chi^2$ \\
        \hline
        $z_1$ & $2.77^{+0.11}_{-0.13}$ & $0.711^{+0.096}_{-0.16}$ & $3.30^{+0.17}_{-0.086}$ & $<2.45$ & 12.92 \\
        $z_2$ & $2.679^{+0.087}_{-0.11}$ & $0.708^{+0.059}_{-0.11}$ & $3.35^{+0.15}_{-0.046}$ & $<1.10$ & 20.50 \\
        $z_3$ & $2.601^{+0.037}_{-0.091}$ & $0.759^{+0.055}_{-0.10}$ & $3.31^{+0.18}_{-0.065}$ & $<1.73$ & 12.83 \\
        Comb. & $2.746^{+0.079}_{-0.064}$ & $0.659^{+0.034}_{-0.078}$ & $3.38^{+0.12}_{-0.038}$ & $<0.754$ & 19.74 \\
        \hline
        Prior & $\mathcal{U}$[1, 5] & $\mathcal{U}$[0.1, 1] & $\mathcal{U}$[3.01, 6.0] & $\mathcal{U}$[0, 5] & -- \\
        \hline
    \end{tabular}
    \end{center}
\end{table}




\subsection{Null and systematic tests}\label{sec.null_tests}

We perform a velocity shuffling null test to verify that our kSZ detection is driven by the coherent line-of-sight velocity field rather than by random scatter or systematic effects. Under velocity randomization, the positive and negative weights average to zero, so any signal that is not correlated with the true galaxy velocities (including residual tSZ or CIB contamination) is expected to vanish in expectation. A nonzero measurement from a shuffled-velocity template would therefore indicate either a velocity-independent systematic or a failure of the shuffling procedure.

\begin{figure}
    \centering
    \includegraphics[width=\linewidth]{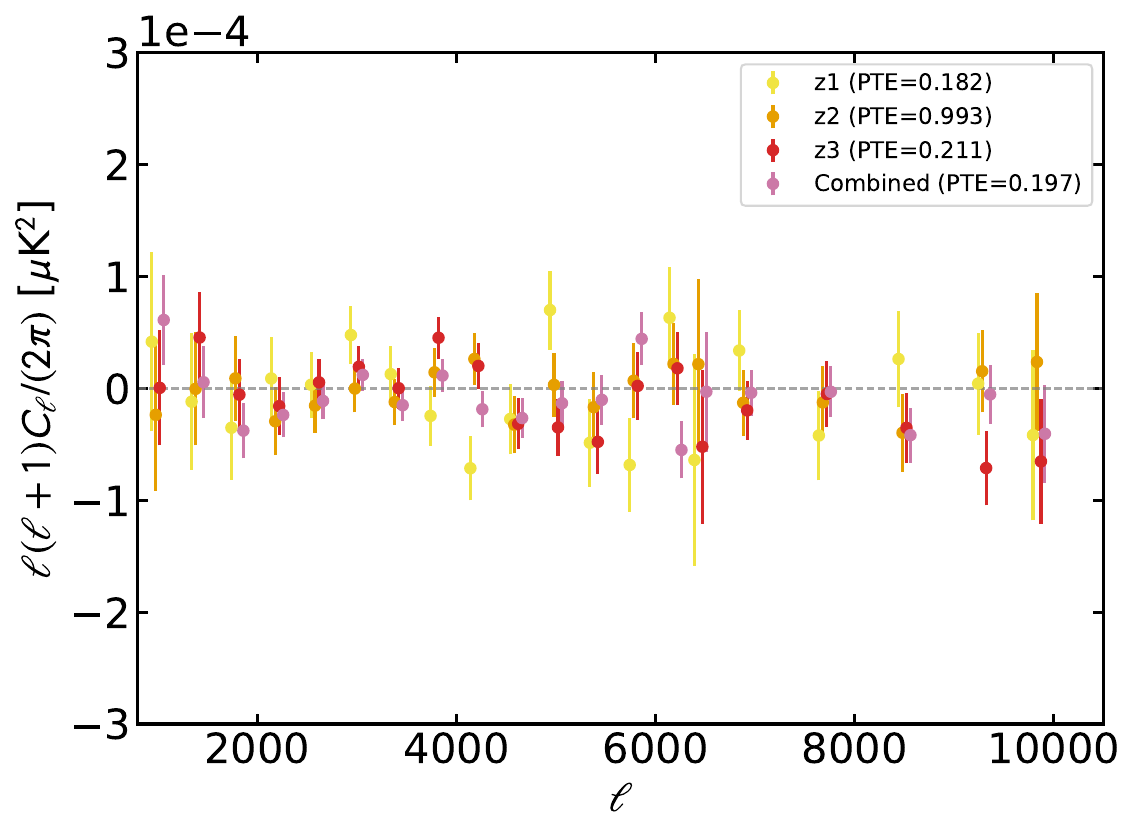} \\ 
\caption{Null test using random velocity templates in harmonic space. We construct kSZ templates by assigning random line-of-sight velocities to the true DESI galaxy positions at three different redshift slices ($z_1$, $z_2$, $z_3$; yellow, orange, and red, respectively) as well as for the combined sample (pink), then cross-correlate each template with the ACT DR6 CMB temperature map. All random templates yield measurements consistent with zero across all multipoles (PTE $= 0.182$, $0.993$, $0.211$, and $0.197$ for $z_1$, $z_2$, $z_3$, and combined, respectively).}
\label{fig:random_template}
\end{figure}

To test for such contamination, we construct null templates by assigning random line-of-sight velocities to the true DESI galaxy positions at three different redshift slices ($z_1$, $z_2$, $z_3$) as well as for the combined sample, then cross-correlate these templates with the ACT DR6 CMB maps following our fiducial analysis pipeline. As shown in Fig.~\ref{fig:random_template}, all randomized templates yield measurements consistent with zero across all multipoles, with PTE values of 0.182, 0.993, 0.211, and 0.197 for $z_1$, $z_2$, $z_3$, and the combined sample, respectively. This confirms that the detected signal originates from the coherent peculiar velocity field of DESI galaxies rather than from a velocity-independent foreground or systematic. The consistency of this null test across different redshift slices and for the full combined sample further validates the robustness of our kSZ measurement.

\subsection{Impact of the HOD parameters on theory prediction}

\begin{table}[htbp]
    \begin{center}
    \caption{Same as Table~\ref{table.prior}, but marginalizing over the HOD uncertainty from the DESI DR2 clustering analysis; one-sided entries are the 68\% limit, and the $\log_{10}\rho_0$ prior is $\mathcal{U}[2.1,4.5]$ (the marginalized runs adopt the tighter amplitude prior). The constraints on the amplitude parameter $\rho_0$ are degraded by a factor of $\sim$1.4--3.5$\times$ compared to the fixed HOD case, with the largest degradation in $z_2$ where the HOD--gas degeneracy is strongest. The shape parameter $\beta_k$ is an upper limit in all bins, including the combined sample. The best-fit $\chi^2$ values are reported for $\nu = 12$ degrees of freedom.}
    \label{table.marginalize_hod}
    \renewcommand{\arraystretch}{1.2}
    \small
    \begin{tabular}{l|c c c c|c}
        \hline
        Bin & $\log_{10}\rho_{0}$ & $x_{c,k}$ & $\beta_k$ & $A_{\rm k2h}$ & $\chi^2$ \\
        \hline
        $z_1$ & $2.88^{+0.11}_{-0.24}$ & $0.62\pm0.21$ & $<3.43$ & $<2.38$ & 12.80 \\
        $z_2$ & $3.10^{+0.29}_{-0.40}$ & $0.391^{+0.065}_{-0.28}$ & $<3.21$ & $<1.55$ & 16.52 \\
        $z_3$ & $2.72^{+0.10}_{-0.24}$ & $0.63^{+0.26}_{-0.19}$ & $<3.35$ & $<2.21$ & 12.32 \\
        Comb. & $3.01^{+0.17}_{-0.25}$ & $0.428^{+0.085}_{-0.21}$ & $<3.27$ & $<1.32$ & 17.85 \\
        \hline
    \end{tabular}
    \end{center}
\end{table}

We adopt the halo occupation distribution (HOD) framework to model the galaxy-halo connection for the DESI LRG sample. The HOD specifies the probability distribution for the number of galaxies occupying a dark matter halo of mass $M$, distinguishing between central and satellite galaxies. We use the baseline HOD model from the AbacusHF-v2 mock catalogs \citep{Zhang_2026}, which follows the functional form of \citep{Zheng_2007} with additional velocity bias parameters. The mean occupation of central galaxies is given by
\begin{equation}
\langle N_{\rm cen} \rangle = \frac{f_{\rm cen}}{2} \left[ 1 + {\rm erf}\left( \frac{\log M - \log M_{\rm min}}{\sigma_{\log M}} \right) \right],
\end{equation}
where $M_{\rm min}$ is the characteristic minimum mass for hosting a central galaxy, $\sigma_{\log M}$ controls the transition width, and $f_{\rm cen}$  allows for incomplete central occupation. The satellite occupation follows
\begin{equation}
\langle N_{\rm sat} \rangle = \langle N_{\rm cen} \rangle \left( \frac{M - \kappa M_{\rm min}}{M_1} \right)^{\alpha_s} \Theta(M - \kappa M_{\rm min}),
\end{equation}
where $M_1$ sets the satellite mass scale, $\alpha_s$ is the power-law slope, and $\kappa$ determines the mass threshold below which halos host no satellites. Since $\langle N_{\rm sat} \rangle \propto \langle N_{\rm cen} \rangle$, the completeness factor $f_{\rm cen}$ rescales the central and satellite occupations uniformly and cancels in the $\bar{n}_g$-normalized cross-spectrum; the kSZ prediction is therefore independent of its value. Satellite positions follow an NFW profile truncated at the virial radius. The HOD parameters for each redshift bin are obtained by fitting the projected two-point correlation function of DESI LRGs, with best-fit values provided by the AbacusHF-v2 fitting pipeline. All halo masses are defined as $M_{200c}$, converted from CompaSO virial masses using the \citet{Duffy_2008} concentration-mass relation.\footnote{This conversion is not accurate at the percent level for CompaSO halo masses. However, since the HOD and gNFW profile parameters are significantly degenerate, any residual mass uncertainties are largely absorbed by the HOD fit and do not bias the inferred gas profile.}

For our harmonic-space analysis, we adopt three redshift bins ($0.4 < z < 0.6$, $0.6 < z < 0.8$, and $0.8 < z < 1.1$) that are consistent with the redshift bins for which HOD parameters have been calibrated from galaxy clustering. This differs from our configuration-space analysis presented in Sec ~\ref{sec.real_space}, where we use four finer redshift bins ($0.4 < z < 0.6$, $0.6 < z < 0.8$, $0.8 < z < 0.95$, and $0.95 < z < 1.1$). The three-bin scheme in harmonic space enables direct marginalization over HOD uncertainties using the calibrated posteriors, while the four-bin scheme in configuration space maximizes our sensitivity to potential redshift evolution of the kSZ signal.

The kSZ cross-correlation signal depends on both the gas profile parameters we aim to constrain and the HOD parameters that describe the galaxy-halo connection.\footnote{We fix the background cosmology to \textit{Planck} 2018 TTTEEE best-fit values throughout this analysis. We have verified that this choice has negligible impact on our results: drawing 50 correlated samples from the \textit{Planck} 2018 TTTEEE posterior, spanning $\sim 4\sigma$ in $\Omega_m$, $H_0$, and $\sigma_8$, the resulting spread in $D_\ell^{\mathrm{kSZ}}$ is 2--3\% ($\sigma/\mu$) across our fitting range ($\ell = 1000$--$5000$), over an order of magnitude smaller than the $\sim$30\% spread from gas profile parameter variations (see Fig.~\ref{fig:cosmo_spread}).} Previous analyses have typically fixed HOD parameters at their best-fit values from galaxy clustering measurements. However, this approach neglects HOD parameter uncertainties, which can introduce systematic biases in the derived gas profile constraints. To properly account for these uncertainties, we implement a Gaussian Process (GP) emulator that enables efficient marginalization over the HOD parameter space. The emulator is trained on 1000 \texttt{class-sz} evaluations spanning the
joint 10-dimensional space of HOD and gas parameters, achieving a
median relative accuracy of 0.16--0.19\% on independent test samples, and
0.8\% within the MCMC prior volume and fitting range (see
Appendix~\ref{emulator} for training details). This allows us to
marginalize over 100 HOD realizations at each MCMC step while maintaining
computational efficiency comparable to a single \texttt{class-sz}
evaluation.

\subsubsection{Likelihood Implementation}

We then modify our likelihood to include HOD marginalization via Monte Carlo
integration. For each MCMC step with gas parameters
$\boldsymbol{\theta}_{\rm gas}$, we draw $N_{\rm HOD} = 100$ HOD parameter
vectors uniformly within the 5th--95th percentile bounds of the weighted
posterior from small-scale galaxy clustering fits. For each HOD sample
$\boldsymbol{\theta}_{\rm HOD}^{(j)}$, we predict
$C_\ell^{ge}(\boldsymbol{\theta}_{\rm HOD}^{(j)},
\boldsymbol{\theta}_{\rm gas})$ using the GP emulator and compute the
marginalized prediction as

\begin{equation}
\bar{C}_\ell^{ge} = \frac{1}{N_{\rm HOD}} \sum_{j=1}^{N_{\rm HOD}}
C_\ell^{ge}(\boldsymbol{\theta}_{\rm HOD}^{(j)},
\boldsymbol{\theta}_{\rm gas}).
\end{equation}

We then apply the beam convolution and window function to this marginalized
spectrum and evaluate the likelihood as

\begin{equation}
\ln \mathcal{L} = -\frac{1}{2} \sum_{b} \frac{(\hat{C}_b -
\bar{C}_b)^2}{\sigma_b^2},
\end{equation}
where $\hat{C}_b$ are the measured bandpowers, $\bar{C}_b$ are the theory
predictions after windowing, and $\sigma_b$ are the uncertainties. The HOD
samples are pre-generated at likelihood initialization with a fixed random
seed to ensure reproducibility. The uniform distribution over the
clustering-derived bounds is a conservative choice: since the galaxy
clustering analysis yields well-constrained HOD parameters (typical
uncertainties $\approx 0.05$\,dex in log-masses) and the kSZ signal is only
weakly sensitive to HOD parameters relative to gas parameters, assigning
equal weight across the credible interval captures the dominant systematic
uncertainty from HOD modeling without biasing toward the best-fit values.

Table~\ref{table.marginalize_hod} shows the marginalized posteriors on GNFW
gas parameters after integrating over HOD parameter uncertainties. The
central values are consistent with the fixed-HOD results in
Table~\ref{table.prior} within $1\sigma$, but the uncertainties on
$\log_{10}A_{\rho_0}$ increase by a factor of
$\sim$1.4--3.5$\times$ depending on the redshift bin, and the slope
parameter $\beta_k$ becomes an upper limit in every bin, including the
combined sample. Appendix~\ref{app:hod_marg} compares the fixed-HOD and
marginalized best-fit spectra bin by bin (Fig.~\ref{fig:fit_marg}) and shows
the marginalized parameter contours (Fig.~\ref{fig:marg_corner}); the two fits
agree within the statistical errors, and the marginalized contours broaden
relative to the fixed-HOD case without shifting the medians beyond $1\sigma$.

\section{Configuration space measurement and results} \label{sec.real_space}

In this section, we present our configuration space measurement of the kSZ CAP profiles from the DESI LRG DR2 sample that overlaps with ACT DR6. We perform a series of consistency checks, foreground and null tests that we describe in more detail in Appendix \ref{appendix_A}.

Figure~\ref{fig.profile_real} shows the stacked CMB temperature map around galaxy positions, weighted by reconstructed velocities without the CAP filter applied. The inner and outer white circles indicate the beam FWHM (1.6 arcmin) and the halo virial radius ($0.56$~Mpc$/h$ at $z=0.8$, corresponding to 1.2 arcmin), respectively. The kSZ signal extends well beyond both scales, with gas distributed out to radii of $\sim3.5$~Mpc$/h$ (corresponding to $\sim7$ arcmin at the median redshift of $z=0.8$), spanning much larger scales than expected from virialized halo gas alone.

\begin{figure}[!htbp]
    \centering
    \includegraphics[width=\linewidth]{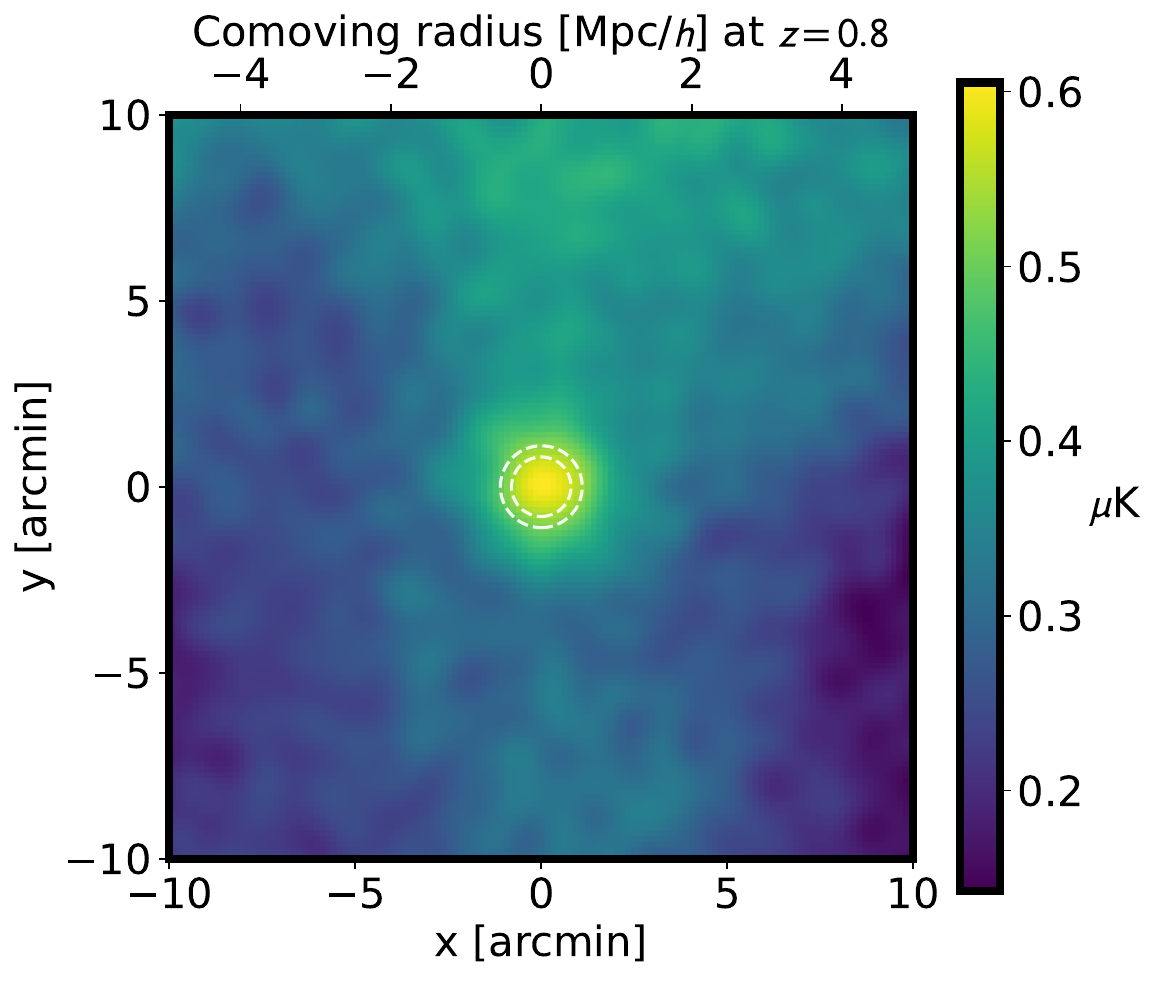} \\ 
\caption{Stacked CMB temperature map around galaxy positions, weighted by reconstructed velocities without application of the CAP filter. The central bright region shows the kSZ signal. The large-scale variations across the map and the overall mean offset of $\sim 0.4\,\mu$K reflect residual primary CMB fluctuations that do not perfectly cancel in the stack, consistent with expectations given the finite sky area; this residual averages down with the number of independent degree-scale patches. The inner dashed white circle indicates the ACT beam FWHM of 1.6 arcmin, while the outer circle marks the typical halo virial radius of $0.64$~Mpc$/h$ at $z=0.8$ (corresponding to 1.1 arcmin). The kSZ signal clearly extends beyond both scales, tracing gas out to several virial radii.}

    \label{fig.profile_real}
\end{figure}

\begin{figure}[!htbp]
    \centering
    \includegraphics[width=\linewidth]{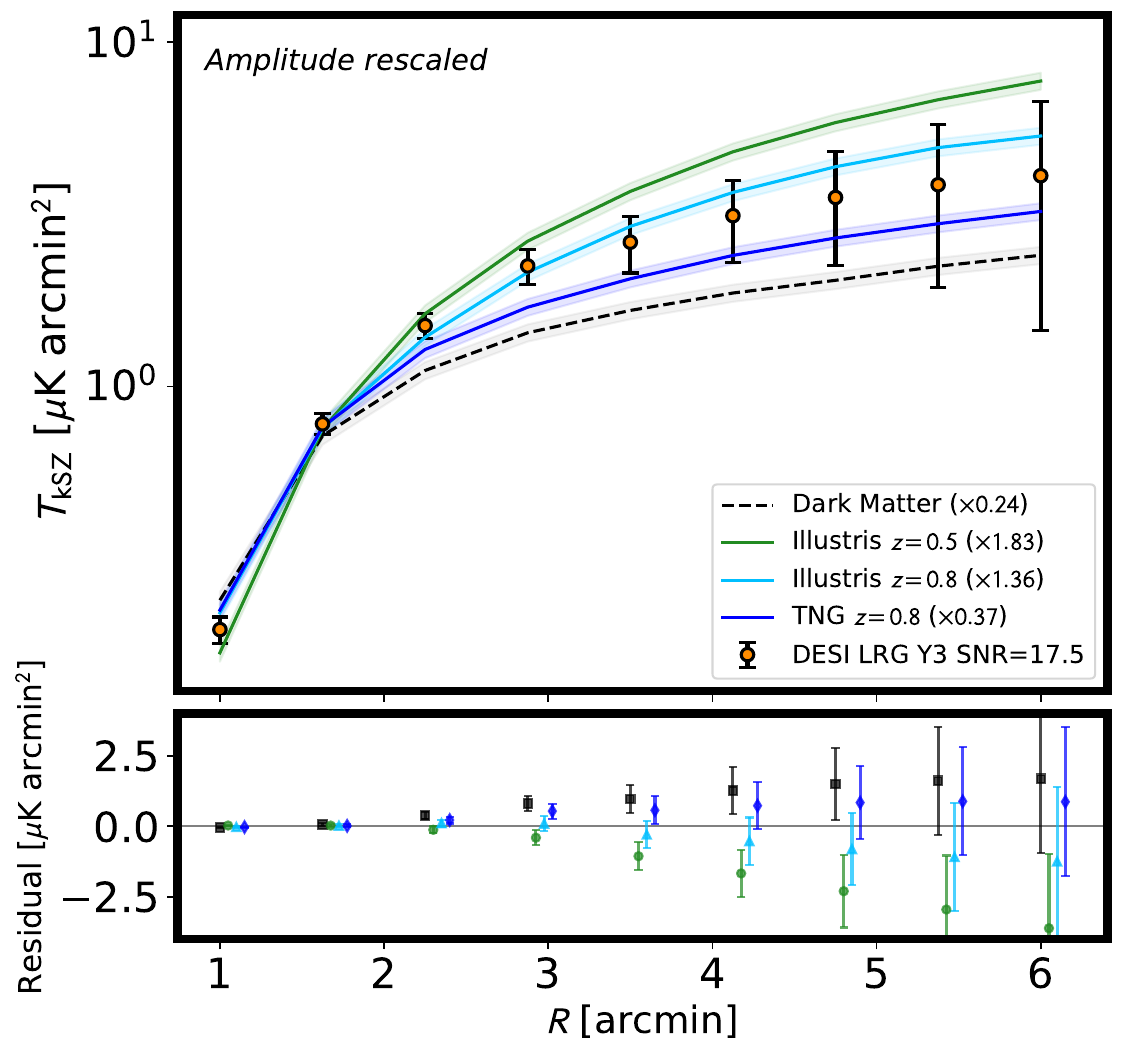}
    \caption{ The mean stacked kSZ signal in $\mu$K\,arcmin$^2$ as a function 
of CAP filter radius $R$, measured at SNR $= 18$ using DESI LRG DR2 galaxies  (orange points with error bars). \textbf{Top panel:} Simulation profiles rescaled by a free amplitude factor (Table~\ref{tab:rescaling}): dark matter (dashed black), Illustris at $z = 0.5$ (green) and $z = 0.8$ (cyan), and 
IllustrisTNG at $z = 0.8$ (blue). Shaded bands show $1\sigma$ amplitude uncertainties. \textbf{Bottom panel:} Residuals between data and models. In the absence of amplitude information, the profile shape alone is not sufficient to distinguish between different feedback scenarios. For example, 
IllustrisTNG requires a rescaling factor of $0.37$, implying halo masses $\sim 3\times$ lower than expected from HOD and lensing constraints \citep{Hadzhiyska:2025mvt}. With the improved DR2 precision, none of the profiles provide statistically acceptable fits to the shape alone. These comparisons are shown for illustration; a detailed confrontation with hydrodynamical simulations is deferred to future work.}
    \label{fig.measurements}
\end{figure}

\begin{table}[h]
\centering
\begin{tabular}{|l|c|c|c|}
\hline
\multicolumn{4}{|c|}{\textbf{Freeing the simulated profile amplitude}} \\
\hline
Simulation & Rescaling factor & $\chi^2_{\text{best-fit}}$ & $\sigma_\mathrm{PTE}$ \\
\hline
Illustris ($z = 0.5$) & $\mathbf{1.828\pm0.105}$& 33.2 &3.66 \\
\hline
Illustris ($z = 0.8$) &$\mathbf{1.355 \pm 0.077}$ & 28.2& 3.13\\
\hline
IllustrisTNG ($z = 0.8$) &$\mathbf{0.367 \pm0.021}$& 22.9 &2.48 \\
\hline
DM TNG ($z = 0.8$) & $\mathbf{0.235\pm0.013}$ & 34.1& 3.76 \\
\hline
\end{tabular}
\caption{Best-fit amplitude rescaling factors, $\chi^2$ goodness-of-fit 
statistics, and PTE-equivalent significance for different simulation 
profiles fitted to the measured kSZ signal. Each profile is rescaled 
in amplitude to isolate the shape comparison, with $\chi^2_\mathrm{best-fit}$ 
quantifying the shape match quality. However, when the rescaling factor 
is large, amplitude and shape are not fully independent: rescaling by a 
large factor effectively compares to halos of different mass, since gas 
profiles are mass-dependent. For example, the DM and IllustrisTNG rescaling 
factors of $0.24$ and $0.37$ would imply halo masses $3$--$4\times$ lower 
than expected for DESI LRGs from HOD modeling and weak lensing 
\citep{Hadzhiyska:2025mvt}, and are thus physically implausible.}

\label{tab:rescaling}
\end{table}

In Fig.~\ref{fig.measurements}, we present the mean stacked kSZ signal as a function of CAP filter radius, measured in $\mu\mathrm{K},\mathrm{arcmin}^2$. We detect the kSZ signal at $\mathrm{SNR} = 18$, calculated using $\mathrm{SNR}=\sqrt{\chi^2_\mathrm{null}-\chi^2_\mathrm{bf}}$, where $\chi^2_\mathrm{bf}$ is obtained from fitting the amplitude of our measurements to the Illustris-1 gas profile at $z=0.5$. We compare these measurements to ACT beam-convolved CAP profiles extracted from the Illustris-1 simulation \citep{Nelson_2015}, spanning a comoving volume of approximately $(106,\text{cMpc})^3$, and from IllustrisTNG300-1 \citep{nelson2021illustristngsimulationspublicdata}, with a substantially larger box of $(302.6,\text{cMpc})^3$. For both simulations, we select the $z=0.8$ snapshot as the closest available output to the median redshift of our DESI LRG DR2 sample ($\mathrm{Med}(z) = 0.76$). We additionally include the Illustris-1 profile at $z=0.5$, which \citep{guachalla2025backlightingextendedgashalos} showed to have the best match in radial shape to the DR1 measurements. Beyond the gas profiles, we also extract the dark matter profile from TNG300-1 at $z=0.8$.

One caveat when comparing our measurements to simulations arises from the finite size of the simulation boxes: the relatively modest box sizes lead to an undersampling of the most massive halos that contribute to our LRG-selected sample. Since the kSZ signal amplitude scales with optical depth and hence with halo mass ($\delta T_{\rm kSZ} \propto \tau \propto M_{\rm halo}$), this volume effect primarily impacts the overall normalization rather than the radial profile shape. Accordingly, we adopt a conservative approach by marginalizing over the absolute amplitude, fitting each simulated profile to the data with a free multiplicative factor (rescaling factor in Table~\ref{tab:rescaling}) and focusing the comparison on the radial shape. The best-fit rescaling factors, $\chi^2$ values, probability-to-exceed (PTE) statistics, and corresponding Gaussian-equivalent significance ($\sigma$) are summarized in Table~\ref{tab:rescaling}.

The comparison disfavors gas distributions that follow the dark matter profile,
with a PTE of $9 \times 10^{-5}$ ($\chi^2_\mathrm{best-fit} = 34.1$), providing
evidence for baryon feedback effects that redistribute gas relative to dark matter.
The dark matter profile is too concentrated at small radii to provide a good fit,
in agreement with \citep{PhysRevD.103.063513, hadzhiyska2025evidencelargebaryonicfeedback,
guachalla2025backlightingextendedgashalos}.

We emphasize that this comparison focuses solely on the radial profile \emph{shape},
with each simulation rescaled by a free amplitude factor (Table~\ref{tab:rescaling}).
This approach is agnostic to the absolute normalization but cannot distinguish
between simulations that achieve good shape agreement through physically implausible
rescaling. In particular, the IllustrisTNG profile requires a rescaling factor of 
$0.37$, implying halo masses $\sim 3\times$ lower than those inferred 
from HOD modeling and CMB lensing for DESI LRGs 
\citep{yuan2023desionepercentsurveyexploring, Hadzhiyska:2025mvt}. Such a large 
discrepancy is unlikely, and instead suggests that the simulation with weaker 
feedback produces profiles that are too concentrated. This illustrates the 
importance of combining both shape and amplitude information. While previous 
works have attempted to incorporate amplitude constraints from HOD modeling 
or lensing \citep{Hadzhiyska:2025mvt, PhysRevD.103.063514, Bigwood:2025kur, 
Siegel:2025ivd, Sunseri:2025hhj, DES:2024iny}, 
a definitive interpretation requires characterizing the redshift and mass dependence of the velocity correlation coefficient $r$, which directly sets the measurement normalization (Section~\ref{sec.r}); we defer this calibration to future work. For reference, the unrescaled simulation 
profiles are shown in Appendix~\ref{app:no_rescaling}.

With this caveat in mind, we find that our improved statistical precision now reveals
that \emph{none} of the hydrodynamical simulation profiles provide 
an excellent statistical
fit to the data in terms of profile shape alone. The Illustris $z = 0.8$
and $z = 0.5$ profiles, which provided reasonable agreement in previous studies with
lower signal-to-noise \citep{hadzhiyska2025evidencelargebaryonicfeedback,
guachalla2025backlightingextendedgashalos}, are now disfavored at the $\sim 3\sigma$
level. The IllustrisTNG $z = 0.8$ profile yields the lowest $\chi^2$,
but only achieves the marginally acceptable fit of $2.5\sigma$ through an extreme
amplitude rescaling that is inconsistent with independent mass constraints.
The ranking in Table~\ref{tab:rescaling} may appear counterintuitive given 
that Illustris $z = 0.8$ visually tracks the data more closely at small radii; 
we explain the origin of the $\chi^2$ differences in Appendix~\ref{appendix_chi2}.

\subsection{Redshift dependence}

\begin{figure}[!htbp]
    \centering
    \includegraphics[width=\linewidth]{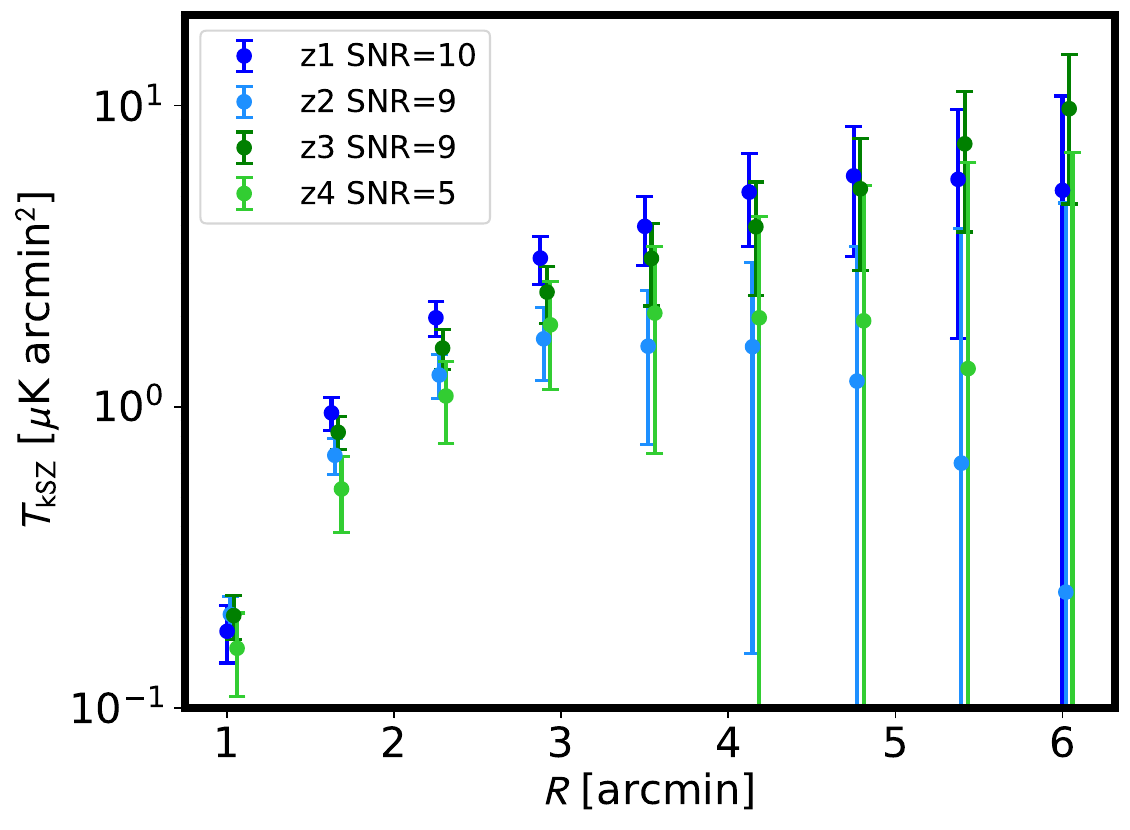}
    \caption{Mean kSZ temperature profiles as a function of projected radius for DESI LRGs in four redshift bins: $z_1$ ($0.4 < z < 0.6$, dark blue), $z_2$ ($0.6 < z < 0.8$, light blue), $z_3$ ($0.8 < z < 0.95$, dark green), and $z_4$ ($0.95 < z < 1.1$, light green). The detection significance corresponds to SNR $\approx$ 10, 9, 9, and 5 for bins $z_1$ through $z_4$, respectively. This is an improvement of $\sim 2\times$ compared to \citep{guachalla2025backlightingextendedgashalos}.}
    \label{fig.redshift_profiles}
\end{figure}

We divide the DESI LRG DR2 sample into four redshift bins to examine potential redshift evolution of the kSZ signal: $z_1$ ($0.4 < z < 0.6$), $z_2$ ($0.6 < z < 0.8$), $z_3$ ($0.8 < z < 0.95$), and $z_4$ ($0.95 < z < 1.1$). The stacked kSZ profiles for each bin are shown in Fig.~\ref{fig.redshift_profiles}, achieving detection significances of SNR $\approx$ 10, 9, 9, and 5 for $z_1$ through $z_4$, respectively.

\begin{figure}[!htbp]
    \centering
    \includegraphics[width=\linewidth]{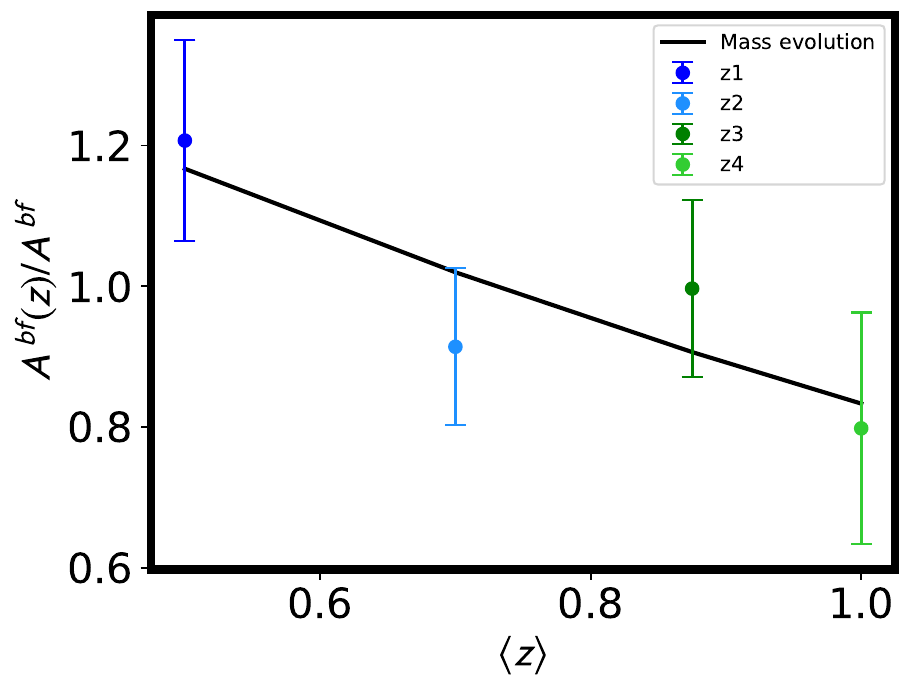}
    \caption{Fractional kSZ amplitude $A^{\mathrm{bf}}(z)/A^{\mathrm{bf}}$ as a function of mean redshift. The solid line shows the expected evolution from the decrease in mean halo mass at fixed comoving number density, computed from the halo occupation distribution integrated over the halo mass function. The data are consistent with this mass-driven trend ($\chi^2/\mathrm{dof} = 1.5/4$, PTE $= 0.82$). This trend is qualitatively expected: maintaining constant comoving number density requires selecting halos further down the mass function at higher redshift, where fewer massive halos have formed, resulting in typical halo masses that decrease with redshift. However, the current data cannot cleanly separate this mass-driven evolution from possible redshift dependence of the velocity reconstruction correlation coefficient $r$, which also directly scales the measured amplitude (Section~\ref{sec.r}). Disentangling these effects will require improved calibration of $r(z)$ from simulations.}
    \label{fig.z_evolution}
\end{figure}

Figure~\ref{fig.z_evolution} shows the fractional amplitude $A^{\mathrm{bf}}(z)/A^{\mathrm{bf}}$ for each redshift bin, defined as the ratio of the best-fit amplitude in each bin to that of the combined baseline analysis. The data reveal a decreasing kSZ signal with redshift, as expected 
from the decrease in mean halo mass of the LRG sample toward 
higher redshift 
\citep{yuan2023desionepercentsurveyexploring, Hadzhiyska:2025mvt}. 
The data reveal a decreasing kSZ signal with redshift, as expected 
from the decrease in mean halo mass of the LRG sample toward 
higher redshift 
\citep{yuan2023desionepercentsurveyexploring, Hadzhiyska:2025mvt}. 
The DESI LRG target selection maintains an approximately constant 
comoving number density out to $z \sim 0.8$ 
\citep{2023AJ....165...58Z}; at higher redshift, the halo mass 
function is less developed, so matching the same number density 
requires reaching to lower-mass halos, naturally producing a 
weaker kSZ signal.\footnote{At $z \gtrsim 0.8$, the LRG number 
density declines and sample completeness drops substantially 
\citep{yuan2023desionepercentsurveyexploring}. Both clustering-based 
HOD fits and independent CMB lensing measurements 
\citep{Hadzhiyska:2025mvt} confirm that the mean halo mass 
continues to decrease toward higher redshift, and the kSZ 
amplitude follows the same downward trend.} We quantify this by computing the mean halo mass in each bin by integrating the HOD over the halo mass function, finding that $\langle M \rangle$ decreases by approximately 29\% from $z_1$ to $z_4$ (solid line). The measured kSZ amplitudes follow this trend ($\chi^2/\mathrm{dof} = 1.5/4$, PTE $= 0.82$), consistent with mass-driven evolution and no additional redshift dependence in gas properties.

We note that this interpretation assumes the CAP aperture is sufficiently large to capture the integrated gas signal, such that the measured kSZ amplitude scales with total halo mass. At smaller apertures where only the inner profile is probed, the mass dependence may differ due to variations in profile shape. Additionally, the velocity reconstruction correlation coefficient $r$ may vary with redshift. Since the kSZ amplitude scales directly with $r$, variations in reconstruction performance would manifest as a multiplicative factor on the measured profiles that is partially degenerate with mass evolution. Disentangling these effects will require calibration of $r(z)$ from simulations across the full LRG redshift range.

\subsection{Mass dependence}

The impact of baryonic feedback on gas distributions is expected to vary with halo mass, as processes such as AGN activity and supernovae couple differently to halos of different depths and baryon content. Physically, since the kSZ signal is proportional to the projected electron number density, the stacked amplitude scales with the total gas mass enclosed within the aperture. Because stellar mass correlates with halo mass (albeit not one-to-one), we expect more massive galaxies to yield stronger kSZ signals. This mass dependence has been observed in previous analyses \citep{guachalla2025backlightingextendedgashalos, hadzhiyska2025evidencelargebaryonicfeedback}, and our improved sample size now allows us to confirm this trend at substantially higher significance.

\begin{figure}[!htbp]
    \centering
    \includegraphics[width=\linewidth]{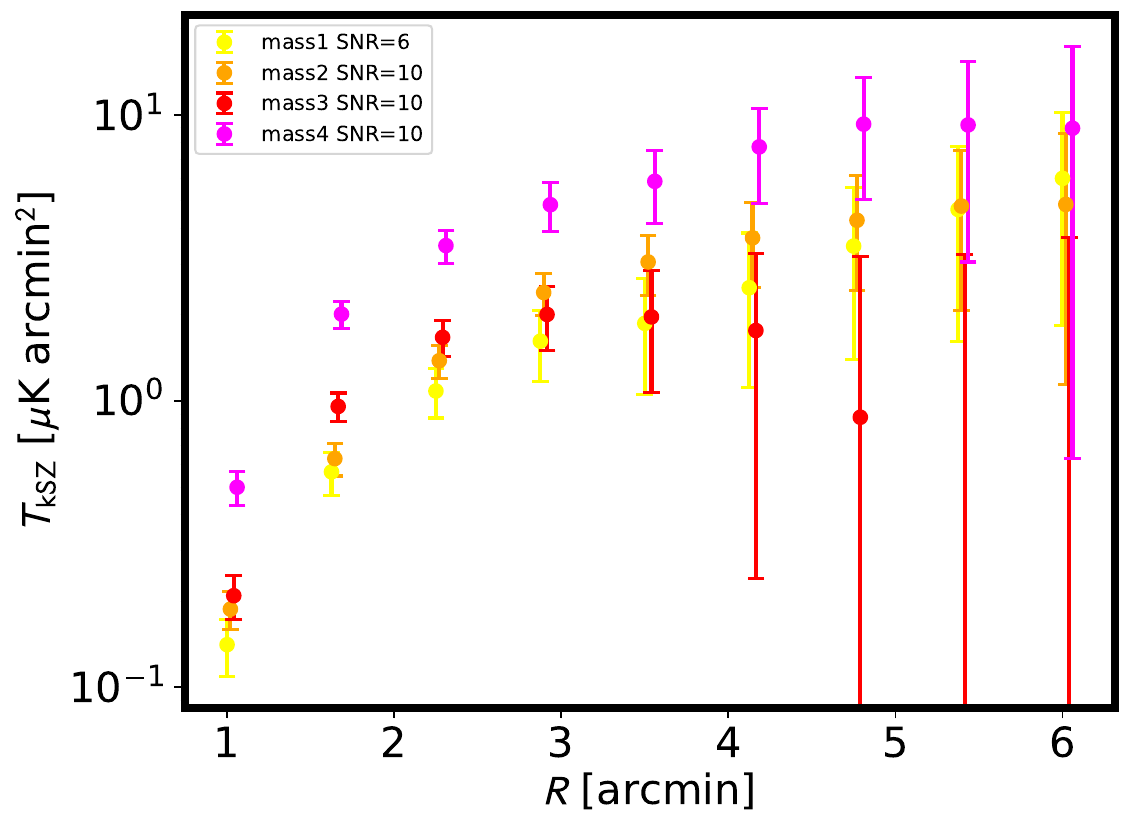} \\ 
    \vspace{-5pt} 
    \includegraphics[width=\linewidth]{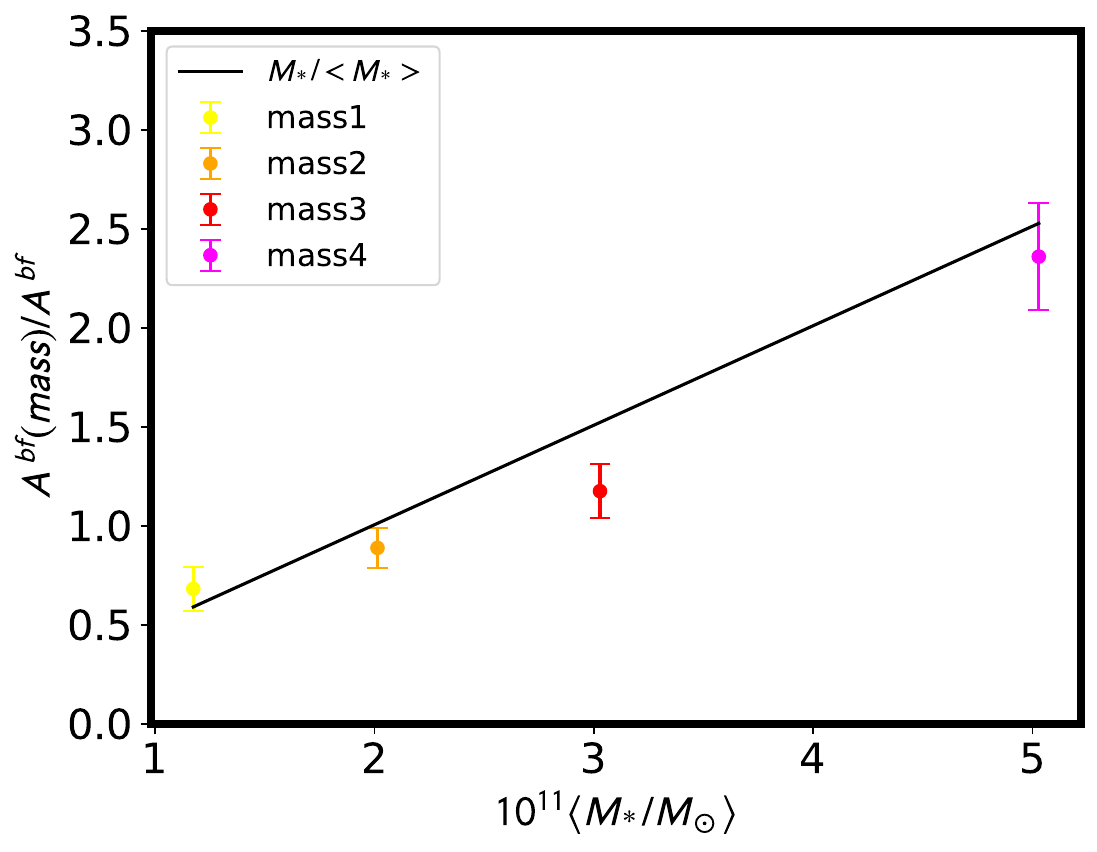}
    \caption{\textbf{Top:} kSZ stacked profile for the different stellar mass bins, denoted by mass. \textbf{Bottom:} Mass dependence of the measured kSZ amplitude. The galaxy sample is divided into four stellar mass bins using DESI-calibrated photometric stellar masses \citep{Zhou_2023}. Each bin is detected with high significance (SNR $\simeq 6$, 10, 10, and 10). The points show the ratio of the best-fit kSZ amplitude in each mass bin to that of the combined baseline analysis, $A^{\mathrm{bf}}(M_\star)/A^{\mathrm{bf}}$, while the black line indicates the expected linear scaling $M_\star/\langle M_\star \rangle$, where $\langle M_\star \rangle$ is the mean stellar mass of the DESI DR2 sample. The data broadly follow the linear trend (PTE $= 0.06$), confirming that higher-mass galaxies exhibit a stronger kSZ signal, with the most massive bin showing a factor of $\sim 3.5$ enhancement relative to the least massive bin.}
    \label{fig.mass_evol}
\end{figure}

As outlined in Section~\ref{sec.desi_data}, we divide our galaxy sample into four stellar mass bins using the DESI-calibrated photometric masses from \citet{Zhou_2023}. The high quality of our DR2 measurement yields detections in each bin with SNR $\simeq 6$, 10, 10, and 10, enabling robust differential comparisons across the full mass range. In Fig.~\ref{fig.mass_evol} we find that galaxies in higher stellar mass bins exhibit larger kSZ amplitudes, with the highest-mass bin showing a signal $3.5\times$ larger than the lowest bin. This trend is physically expected: the kSZ signal scales with total gas mass, which broadly increases with halo mass, and halo mass correlates with stellar mass. Over a relatively narrow mass range one can approximate the relation between stellar and halo mass as linear, suggesting a simple model of the form
\begin{equation}
A^{\mathrm{bf}}(M_\star)/A^{\mathrm{bf}} = \frac{M_\star}{\langle M_\star \rangle},
\end{equation}
where $\langle M_\star \rangle$ is the mean stellar mass of the DESI DR2 sample. Comparing this zero-free-parameter prediction to the four stellar mass bins yields $\chi^2 = 9.2$ for 4 degrees of freedom, corresponding to a probability-to-exceed (PTE) of 0.06, so the data are consistent with this scaling at the $2\sigma$ level.
As with the redshift-dependent analysis, however, the velocity correlation coefficient $r$ may also vary with halo mass (Section~\ref{sec.r}), introducing a degeneracy between intrinsic mass scaling of the gas content and mass-dependent reconstruction performance.

\subsection{Correspondence between configuration and harmonic estimators}

To verify consistency between our harmonic-space and configuration-space analyses, we perform a Fourier transform of the best-fit harmonic-space power spectrum to predict the real-space stacked profile.

To convert from harmonic space to real space, we compute the Fourier transform of the binned bandpowers. The harmonic-space window function for the CAP filter is
\begin{equation}
W_\ell(\theta_d) = A_{\rm disk} \frac{4 J_1(\ell \theta_d) - 2\sqrt{2} J_1(\sqrt{2} \ell \theta_d)}{\ell \theta_d},
\end{equation}
where $J_1$ is the Bessel function of the first kind.

The real-space profile is obtained via
\begin{equation}\label{eq.ft_real}
\Delta T(\theta_d) = \frac{1}{r \, \sigma_{\rm rec}} \sum_b \Delta\ell_b \, \frac{2\ell_b + 1}{4\pi} \, W_{\ell_b}(\theta_d) \, \tilde{C}_b^{\hat{\pi} T},
\end{equation}
where $\Delta\ell_b$ is the width of bin $b$, $A_{\rm disk} = \pi \theta_d^2$ is the disk area, and the factor $1/(r \, \sigma_{\rm rec})$ converts from the harmonic convention to the stacking convention (which measures $\bar{\tau} \, \sigma_{\rm true} / c$).

The bin width factor $\Delta\ell_b$ arises because the NaMaster bandpower windows are normalized such that each row sums to unity (averaging over $\ell$ within the bin) rather than summing the contributions. This must be accounted for when performing the discrete Fourier sum.

We fit a single amplitude parameter $A$ to the model prediction, minimizing
\begin{equation}
\chi^2 = (\mathbf{d} - A \mathbf{m})^T \mathbf{C}^{-1} (\mathbf{d} - A \mathbf{m}),
\end{equation}
where $\mathbf{d}$ is the measured real-space profile, $\mathbf{m}$ is the Fourier-transformed model, and $\mathbf{C}$ is the bootstrap covariance matrix. A best-fit value of $A = 1$ indicates consistency between the harmonic and configuration-space measurements.

\begin{figure}
    \centering
    \includegraphics[width=\columnwidth]{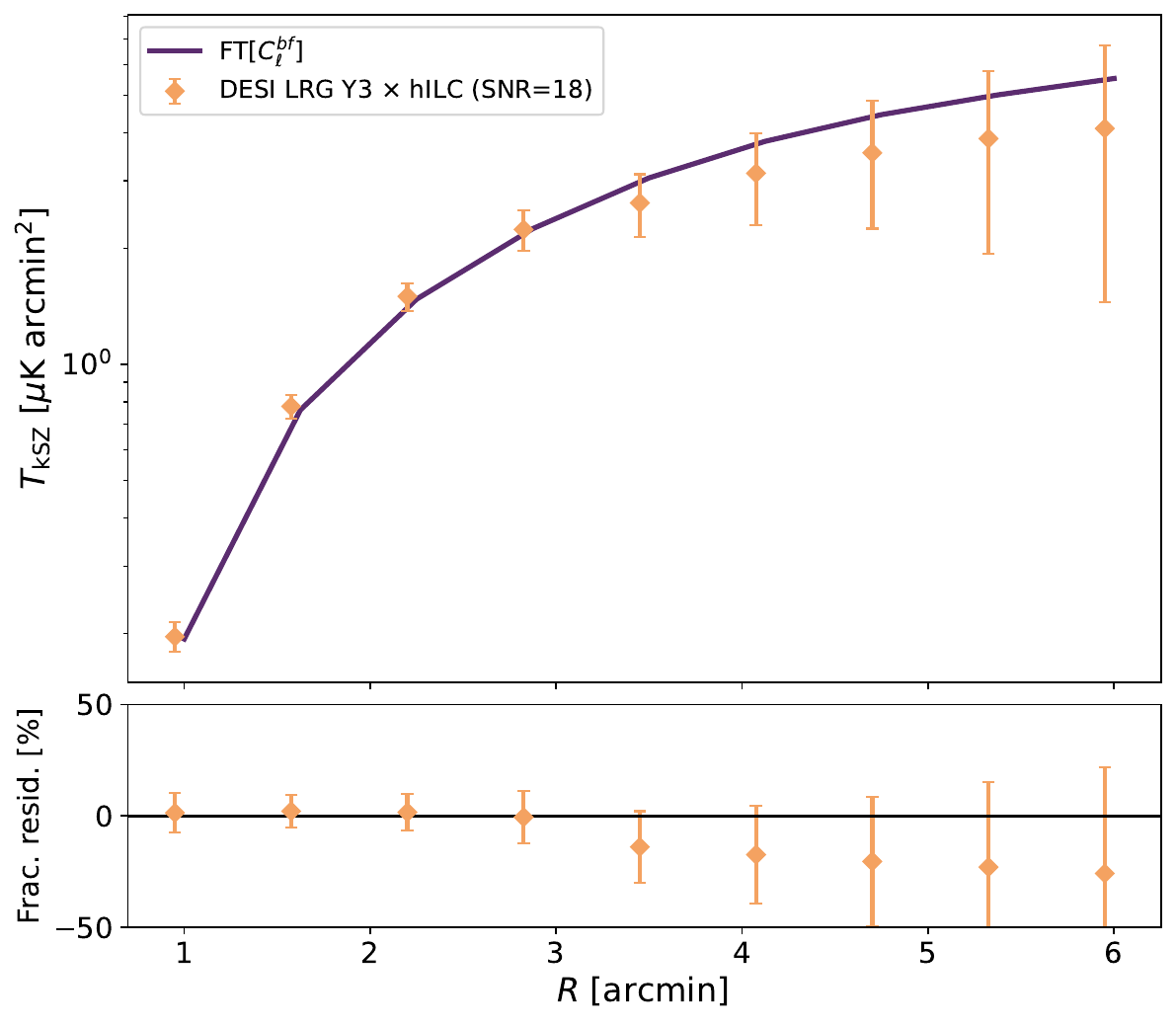}
    \caption{Comparison between the real-space CAP measurement and the Fourier transform of the best-fit harmonic power spectrum. \textit{Top panel:} The velocity-weighted stacked kSZ signal as a function of aperture radius. Orange points show the DESI LRG DR2 $\times$ ACT hILC measurement with bootstrap uncertainties. The purple curve shows the Fourier transform of the best-fit $C_\ell^{\hat{\pi}T}$ from the harmonic analysis (the HOD-marginalized combined fit of Table~\ref{table.marginalize_hod}), converted to real space using Equation~(\ref{eq.ft_real}). \textit{Bottom panel:} Fractional residuals between the data and the Fourier-transformed model. Perfect agreement is not expected at large scales due to the profile truncation scheme employed in the harmonic analysis.}
    \label{fig:harmonic_realspace_comparison}
\end{figure}

For the combined redshift sample we obtain $A = 0.92 \pm 0.05$, consistent with unity at the $1.6\sigma$ level. Here we adopt the HOD-marginalized combined best-fit spectrum (Table~\ref{table.marginalize_hod}), rather than the fixed-HOD best fit used for the black curve of Fig.~\ref{fig.measurement_fits}. The Fourier transform and comparison with the CAP measurement is shown in Fig.~\ref{fig:harmonic_realspace_comparison}. We observe some deviation at large aperture radii, likely due to the profile truncation scheme in the harmonic analysis which enforces that the enclosed gas mass equals $f_b M_{200c}$ at the halo boundary, a constraint not imposed in the CAP measurement.

We also use this Fourier-transformed profile as a template to compute the signal-to-noise ratio of the real-space measurement, obtaining results consistent with both the Illustris-based template and the harmonic-space detection significance of $18\sigma$.

\subsection{Comparison with previous work}

\subsubsection{Redshift dependence comparison with DR1 spec-z}

To validate the robustness of our kSZ measurements, we perform a consistency check by comparing our DESI DR2 results with a reanalysis of the DESI DR1 spectroscopic redshift sample. For this comparison, we reprocess the DR1 spec-z catalog using the same DR1 mask\footnote{This has the same footprint as the DR2 mask except it also masks all of the point sources from the ACT DR6 catalogue, which is not done in the DR2 analysis.} that removes all identified sources and clusters.

\begin{figure}[!htbp]
    \centering
    \includegraphics[width=\linewidth]{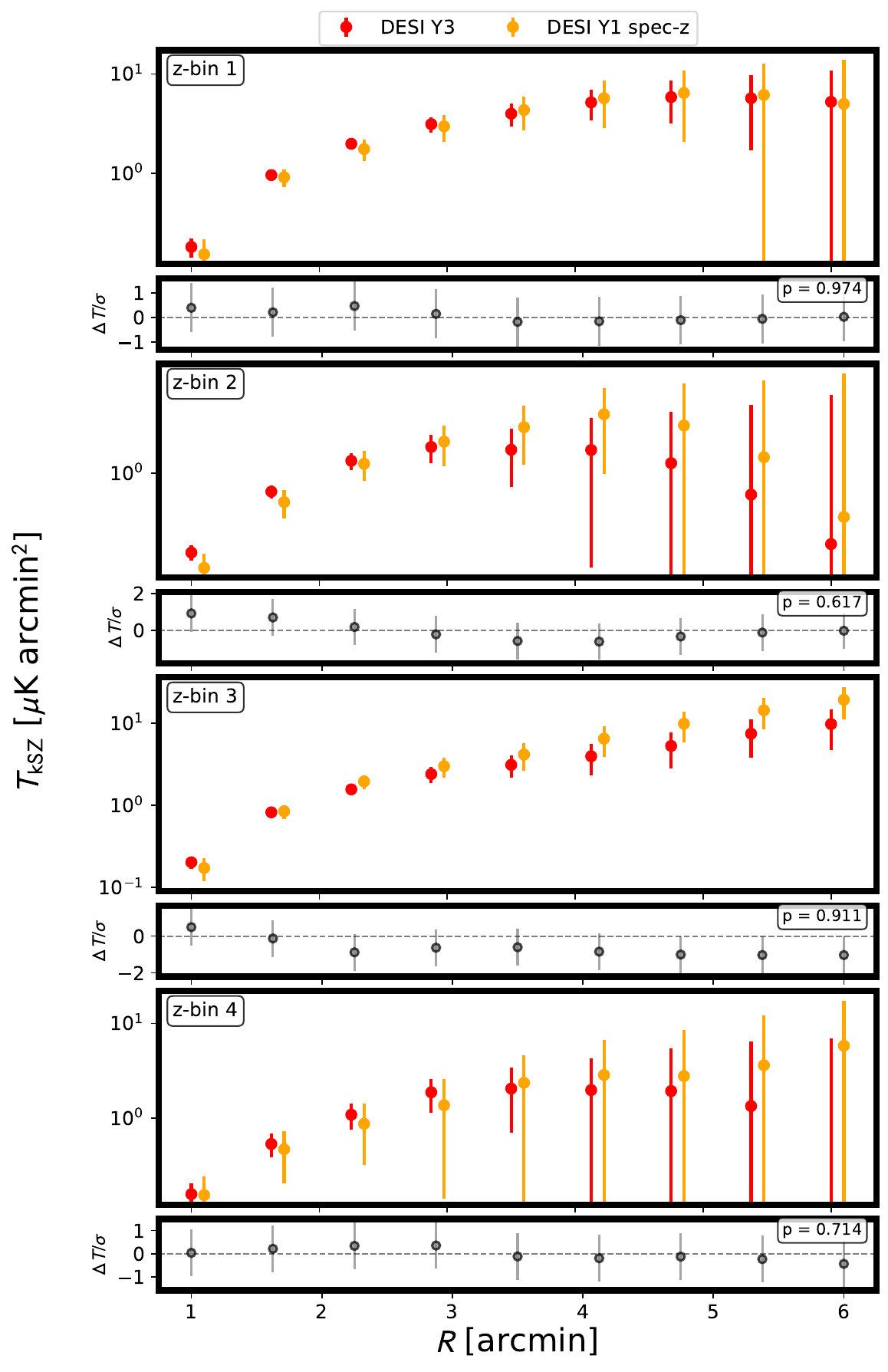}
    \caption{Stacked kSZ signal versus angular scale in four redshift bins. Top panels: kSZ temperature measurements for DESI DR2 (red) and reanalyzed DESI DR1 spec-z (orange). The DR1 data points have been offset by $0.1$ arcmin along the horizontal axis for visual clarity. Bottom panels: difference between DR2 and DR1 measurements normalized by the difference uncertainty with null test p-values. The PTE values account for the correlations arising from using the same patch of CMB sky and the substantial overlap in galaxies between the two samples (DR1 galaxies are a subset of DR2) by drawing through the same realizations for both analyses during the bootstrapping. The excellent agreement across all redshift bins demonstrates robust consistency between data releases despite these correlations.}
    \label{fig.y1y3comparison}
\end{figure}

Fig.~\ref{fig.y1y3comparison} presents the stacked kSZ temperature profiles as a function of angular scale for four redshift bins, with each bin displayed as a pair of panels. The top panels show the measured kSZ signals for both DESI DR2 (red) and the reanalyzed DR1 spec-z sample (orange), while the bottom panels display the difference between these measurements along with null test p-values. The PTE values are computed using a covariance matrix that properly accounts for the correlations between the DR2 and DR1 measurements arising from two sources: the use of the same CMB map and the substantial overlap in galaxy samples (DR1 galaxies form a subset of DR2). We account for these correlations by drawing through the same bootstrap realizations for both the DR2 and DR1 analyses, ensuring that the covariance matrix captures the correlated noise and sample variance between the two measurements. Across all four redshift bins, we observe excellent agreement between the DR2 and DR1 measurements, with differences consistent with zero within the measurement uncertainties.

\begin{figure}[!htbp]
    \centering
    \includegraphics[width=\linewidth]{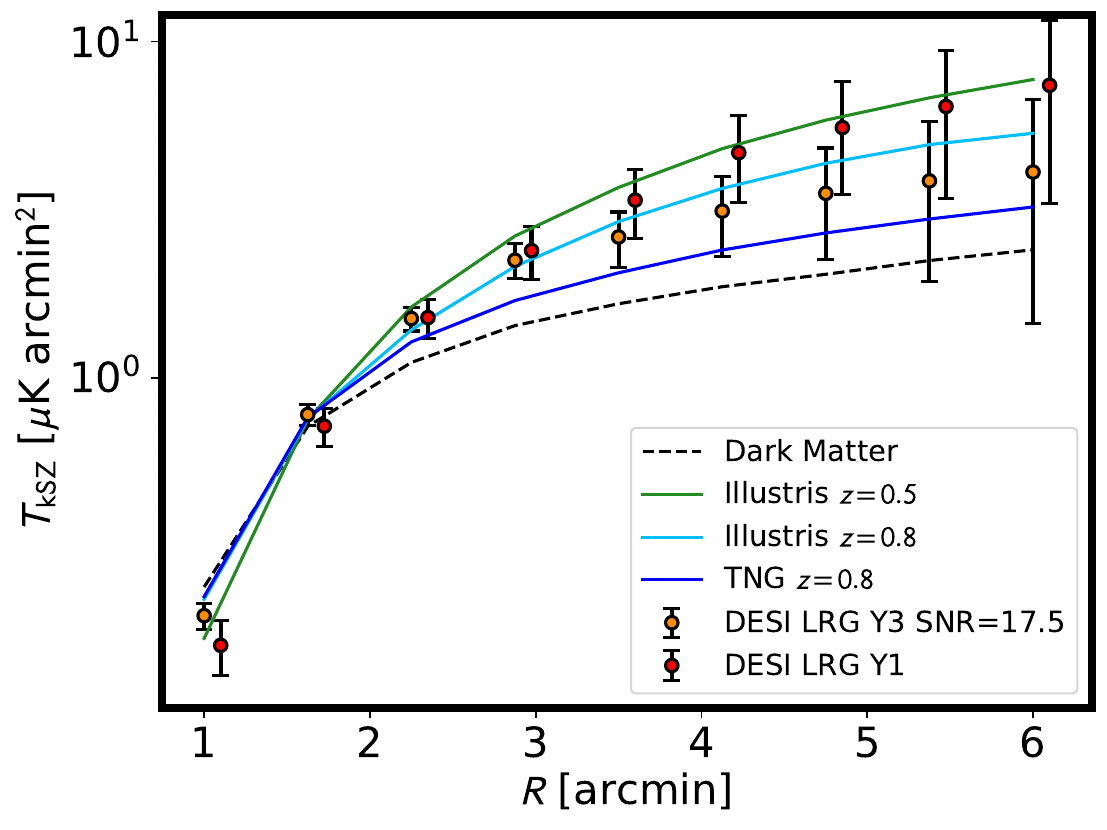}
    \caption{Stacked kSZ signal versus angular scale comparison between DR1 and DR2.}
    \label{fig.y1y3overall}
\end{figure}

\begin{figure}
    \centering
    \includegraphics[width=\linewidth]{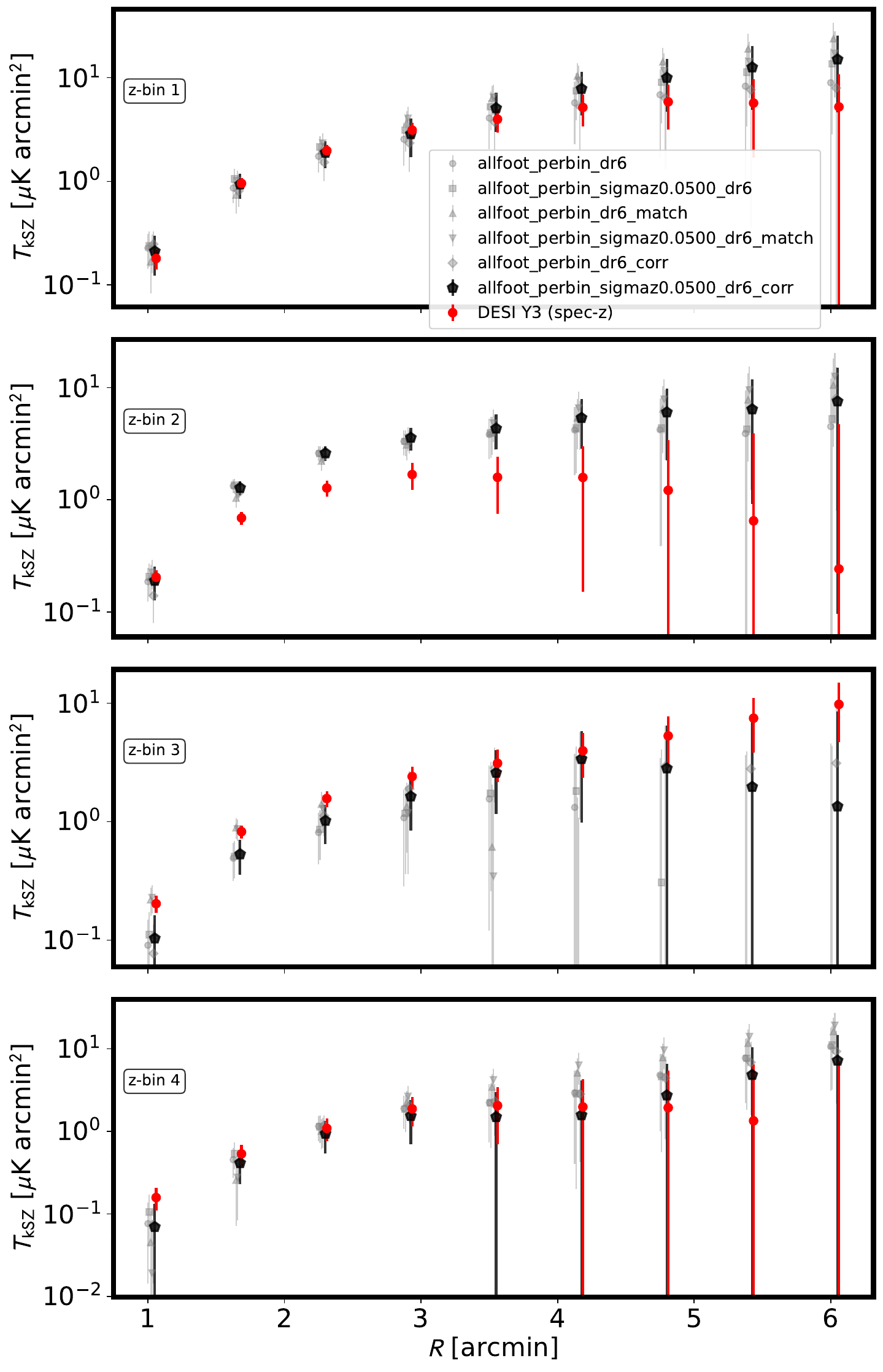}
    \caption{Comparison of kSZ measurements using DESI DR2 spectroscopic redshifts (red points) with various photo-z implementations from the DESI DR9 legacy imagiing survey\footnote{The legend labels encode the analysis configuration of the photometric sample from \citep{hadzhiyska2025evidencelargebaryonicfeedback}: ``allfoot'' denotes the full DESI imaging footprint (combining the DES, North, and South sub-regions); ``perbin'' indicates that the velocity reconstruction is performed independently in each redshift bin; ``dr6'' refers to the use of ACT DR6 hILC temperature maps; ``sigmaz0.0500'' indicates that a photometric redshift quality cut of $\sigma_z/(1+z) < 0.05$ has been applied; ``match'' denotes area matching between the DR9 and DR10 footprints; and ``corr'' indicates that the reconstructed velocity outlier correction (3$\sigma$ clipping) has been applied. The fiducial analysis of \citep{hadzhiyska2025evidencelargebaryonicfeedback} corresponds to ``allfoot\_perbin\_sigmaz0.0500\_dr6\_corr.''}. Grey points represent different analysis choices explored in \citet{hadzhiyska2025evidencelargebaryonicfeedback}, including variations in photometric redshift error modeling, sky coverage, and calibration corrections. The black points correspond to the configuration most similar to our spectroscopic analysis: the DR9-only photo-$z$ sample with a $\sigma_z < 0.05$ quality cut applied. Note that many configurations in \citet{hadzhiyska2025evidencelargebaryonicfeedback} combine both DR9 and DR10 imaging data, and are not directly comparable to the spectroscopic sample presented here. The key methodological differences between the black photo-z measurement and our spec-z analysis are additional corrections applied only to the photo-z sample to account for photometric redshift uncertainties, including outlier removal and velocity symmetrization (see Section~\ref{compare_z} for details).}
    \label{fig.ksz_spec_vs_photo}
\end{figure}

\subsubsection{Comparison with photo-z DR9}\label{compare_z}

To understand the differences between spectroscopic and photometric kSZ measurements, we perform a comparison with photo-z-based measurements using the DESI DR9 legacy imaging survey \citep{hadzhiyska2025evidencelargebaryonicfeedback}. Figure~\ref{fig.ksz_spec_vs_photo} presents our DESI DR2 spectroscopic results (red points) alongside various photo-z
implementations (grey points) explored in \citet{hadzhiyska2025evidencelargebaryonicfeedback}. These include variations in photometric redshift error modeling ($\sigma_z$), sky coverage (footprint definitions), and calibration corrections. The black points correspond to the configuration most closely matching our spectroscopic analysis: the DR9-only photo-z sample with a redshift quality cut of $\sigma_z < 0.05$ and the DR6 calibration correction applied. The primary methodological differences between the photo-z measurements and our spec-z analysis stem from corrections necessary to account for photometric redshift uncertainties. Specifically, the photo-z analysis implements: (1) removing photometric $z$ outliers when the estimated noise is above $\sigma_z/(1+z) > 0.5$ and velocity outliers that are above $3\sigma$; (2) velocity symmetrization through random down-sampling to ensure equal numbers of galaxies in each velocity bin around the mean, which guarantees zero mean signal in the absence of kSZ and avoids biases from massive clusters. As demonstrated in \citet{hadzhiyska2025evidencelargebaryonicfeedback}, these corrections have mostly negligible effects on the final measurements, with uncorrected measurements yielding marginally higher SNR due to larger sample sizes, except in redshift bin 3 where the slightly asymmetric line-of-sight velocity distribution
benefits from outlier removal. Furthermore, many results in \citet{hadzhiyska2025evidencelargebaryonicfeedback} combine both DR9 and DR10 imaging data (thanks to the improved photo-$z$ quality when using DR10 data), whereas our comparison focuses
specifically on DR9 to maintain consistency with the spectroscopic sample selection. We note that only a qualitative comparison is possible here, as the two analyses share the same CMB map and have substantial galaxy overlap, and we do not have access to the cross-covariance needed to assess the significance of any differences. Moreover, perfect agreement is not expected: our redshift bins are defined by hard cuts in spectroscopic redshift, whereas the photo-z analysis applies hard cuts in photometric redshift. The mean spectroscopic redshifts in our four bins are $\bar{z} = 0.51,\,
0.71,\, 0.87,\, 1.01$, while the corresponding DR9 photo-z
samples have mean redshifts $\bar{z} = 0.47,\, 0.63,\, 0.79,\,
0.92$, reflecting the systematic offset introduced by photo-z scatter leaking lower-redshift galaxies into higher bins. Due to photo-z scatter, the true redshift distribution of the photo-z selected sample has tails extending beyond the nominal bin edges. Furthermore, the spectroscopic sample is a subset of the photometric one and is undersampled in overdense regions due to fiber collisions. With these caveats, the spectroscopic and photometric measurements show broadly consistent amplitudes and radial dependence across all four redshift bins, with differences typically within $1\sigma$ of the individual error bars. The second redshift bin shows the largest discrepancy, which is not surprising given the combination of differing sample selections and analysis corrections described above.

\section{Conclusion}\label{sec.conclusion}

We present the most precise measurements of the kSZ effect to date for luminous red galaxies, utilizing $\sim 2.4$ million spectroscopically confirmed LRGs from DESI DR2  cross-correlated with ACT DR6 CMB temperature maps. We measure the kSZ signal at $18\sigma$ significance in both harmonic and configuration space, representing nearly a factor of three improvement in sample size compared to previous DESI Year 1 measurements.

We introduce a novel harmonic-space analysis methodology that constructs an explicit momentum-weighted kSZ template by projecting galaxy line-of-sight velocities, reconstructed via the continuity equation, onto the sky. Cross-correlating this template with CMB maps yields bandpower measurements with reduced correlations compared to traditional real-space stacking. This formalism also enables straightforward combination with other angular power spectrum probes, facilitating future multi-probe constraints on baryonic feedback. Our measurements span angular scales sensitive to gas distributions from halo cores to the outskirts, providing a probe of baryonic structure around massive galaxies at redshifts $0.4 < z < 1.1$.

A key methodological contribution of this work is the conversion of measured galaxy-centered gas profiles into halo-centered profiles through explicit incorporation of the HOD. By forward-modeling our measurements within the HOD framework and propagating uncertainties from galaxy clustering constraints, we obtain gas thermodynamic parameters that characterize the baryon distribution around dark matter halos rather than around galaxies alone. This framework facilitates interpretation in terms of halo mass and enables more direct constraints on feedback prescriptions in theoretical models. We parameterize the gas distribution using GNFW profiles, finding that the measured kSZ signal deviates significantly from dark matter-only predictions. The GNFW fits presented here provide a compact description of the gas profiles that can be used to benchmark and inform hydrodynamical simulations, and enable direct forward modeling in future galaxy-galaxy lensing and kSZ analyses involving DESI LRG samples.

When split by stellar mass, we find that the kSZ amplitude scales approximately linearly with stellar mass, confirming that more massive galaxies host more extended gas halos. Splitting by redshift reveals a decreasing kSZ amplitude toward higher redshift that is consistent with the expected evolution in mean halo mass ($\chi^2/\mathrm{dof} = 1.5/4$, PTE $= 0.82$). Maintaining constant comoving number density requires selecting lower-mass halos at higher redshift, and this mass-driven trend fully accounts for the observed amplitude evolution with no evidence for additional redshift dependence in the gas-to-halo-mass relation.

Comparison with hydrodynamical simulations reveals that none of the current state-of-the-art models provide 
an excellent statistical fit
to our radial profile shape, with Illustris ruled out at $\sim 3\sigma$ and IllustrisTNG disfavored at $2.5\sigma$. The large amplitude rescaling required for IllustrisTNG ($\sim 0.37$) further suggests that weaker feedback models produce overly concentrated profiles. These comparisons remain preliminary, as the simulated profiles do not incorporate our LRG selection function or span the full redshift range of the sample. Applying consistent galaxy selection via abundance matching or other techniques in future simulation comparisons will be essential for drawing robust conclusions about feedback physics.
An important systematic in the physical interpretation of our measurements is the velocity correlation coefficient $r$ between reconstructed and true velocities. Since $r$ enters as an overall amplitude scaling of the kSZ signal, it does not affect our detection significance or the shape of the inferred gas profile, but it does rescale the inferred gas density normalization $\rho_0$ (and derived quantities such as enclosed gas mass) proportionally. For our fiducial combined analysis, we adopt $r = 0.65$ from the light cone analysis of \citep{PhysRevD.109.103534}, which averages over the full DESI LRG sample and provides the most robust single estimate available. When subdividing by redshift or stellar mass, $r$ may deviate from this fiducial value due to redshift-dependent nonlinear growth, evolving satellite fractions, and changes in galaxy bias (Section~\ref{sec.r}). While \citep{PhysRevD.109.103534} does not provide bin-by-bin estimates, these physical drivers evolve continuously with redshift, suggesting that departures from the fiducial value are gradual rather than abrupt across our redshift range. We caution, however, that amplitude comparisons across subsamples carry an additional systematic uncertainty beyond the statistical errors reported, and that relative trends (e.g., the direction of redshift evolution) are more robust than absolute amplitudes in individual bins. Calibrating $r$ in narrow redshift and mass bins using mock catalogs that incorporate evolving HOD models and match the DESI survey selection function is an important direction for future work.

Together with measurements of the BGS and ELG samples presented in a companion paper \citep{HadzhiyskaBGSELG}, this work provides kSZ constraints spanning nearly three orders of magnitude in stellar mass and the redshift range $0.1 \lesssim z \lesssim 1.6$. This lever arm, covering the transition from stellar to AGN feedback-dominated halos, offers a powerful program for testing how feedback efficiency depends on halo mass and cosmic epoch, and can be used in conjunction with thermal SZ (tSZ) measurements of the same sample \cite{Liu:2025zqo} to constrain the thermodynamic properties of the host halos.

The next generation of CMB experiments, combined with completed DESI observations and complementary data from Rubin Observatory, Euclid, and SPHEREx, will further increase the statistical power of kSZ measurements. The statistical precision achieved here is no longer the limiting factor for interpreting kSZ measurements. Further progress now necessitates improved handling of modeling systematics, particularly the calibration of velocity reconstruction across redshift and mass bins, and the application of consistent galaxy selection when comparing with simulations. Addressing these challenges is the focus of ongoing work. With systematic uncertainties under control, kSZ observations will provide stringent tests of galaxy formation models and tight constraints on baryonic feedback processes.

\section*{Acknowledgements}

\input{acknowledgements}

\section{Data availability}
The data products associated with this work are publicly available on Zenodo at \url{https://zenodo.org/records/21289019} \citep{data_zenodo}.

%

\vspace{5mm}




\appendix




\section{Configuration space null tests}\label{appendix_A}

\subsection{Foreground tests}\label{sec:freq_null}

\label{sec:freq_null}

The kinematic Sunyaev-Zel'dovich effect produces a spectral distortion that is indistinguishable from the primary CMB blackbody spectrum in thermodynamic temperature units. In contrast, astrophysical foregrounds exhibit distinct frequency dependence: the thermal SZ effect scales as $f(\nu) = x\coth(x/2) - 4$ where $x = h\nu/k_B T_{\rm CMB}$, producing a decrement at 90\,GHz and an increment at frequencies above $\sim$220\,GHz, while thermal dust emission rises steeply with frequency following a modified blackbody spectrum. Radio point sources and the cosmic infrared background also contribute frequency-dependent contamination.

To test for foreground contamination in our kSZ measurement, we repeat our stacking analysis using single-frequency maps at 90\,GHz and 150\,GHz.

\begin{figure}
    \centering
    \includegraphics[width=0.5\textwidth]{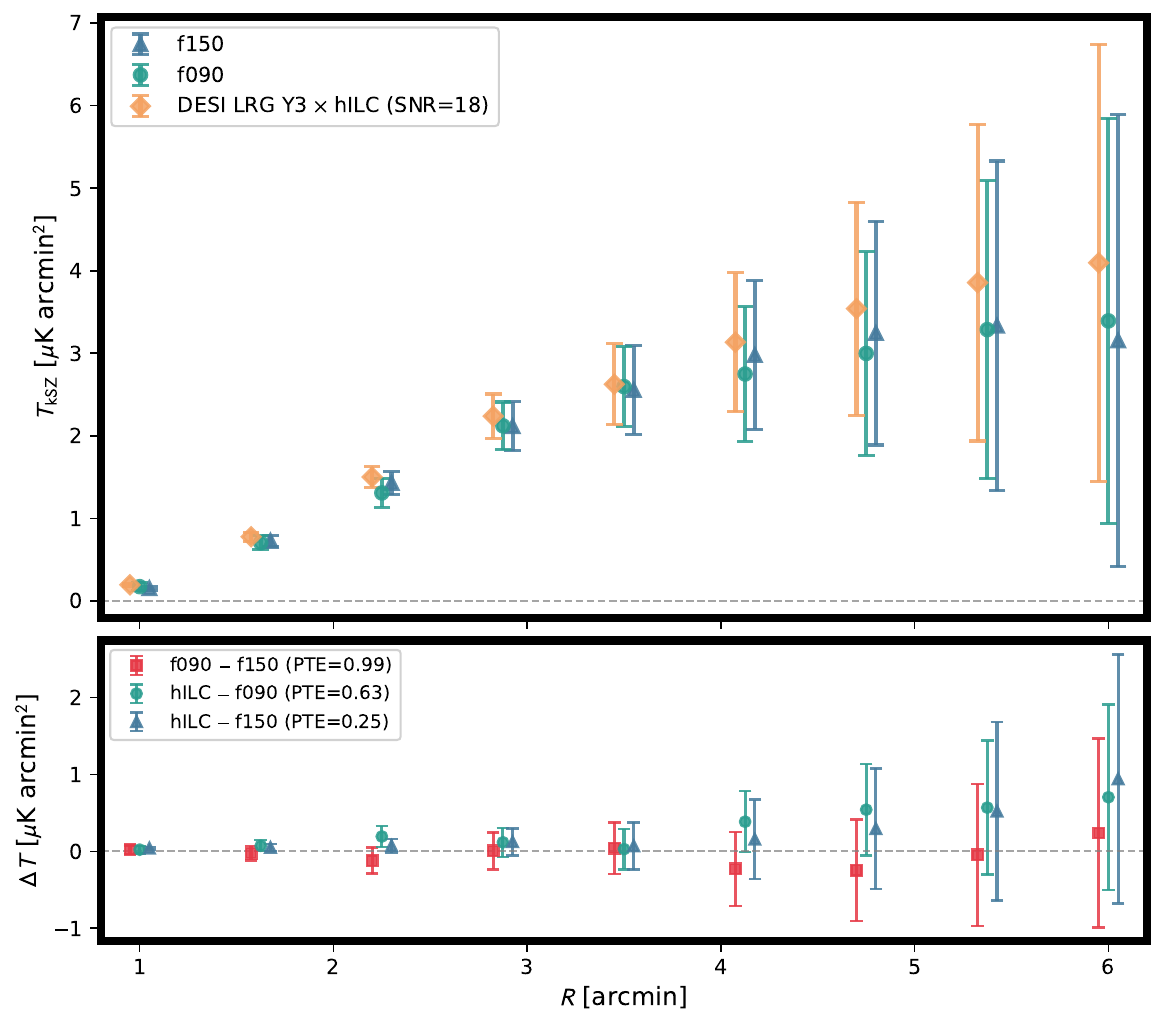}
    \caption{Foreground test for the kSZ signal. \textit{Upper panel:} Velocity-weighted stacked temperature profiles measured using the baseline harmonic ILC map (orange diamonds, SNR\,$=$\,18), single-frequency 150\,GHz (blue triangles), and 90\,GHz (teal circles) maps. All three measurements are consistent within uncertainties. \textit{Lower panel:} Pairwise differences between the measurements. The 90$-$150\,GHz difference (red squares) tests for frequency-dependent foreground contamination from the tSZ effect and dust, which would produce a non-zero signal. The differences between the hILC and single-frequency maps (teal circles and blue triangles) verify the consistency of the ILC combination. All differences are consistent with zero, with probability-to-exceed values of 0.99, 0.63, and 0.25 respectively.}
    \label{fig:freq_null}
\end{figure}

Figure~\ref{fig:freq_null} presents the results of this comparison. The upper panel shows the velocity-weighted stacked profiles for all three maps: our baseline hILC measurement (orange diamonds), the 150\,GHz map (blue triangles), and the 90\,GHz map (teal circles). All three measurements are in excellent agreement across the full range of aperture radii.

The lower panel displays the pairwise differences between these measurements. The 90$-$150\,GHz difference (red squares) is consistent with zero across all scales with a probability-to-exceed (PTE) of 0.99, indicating no significant frequency-dependent signal. Similarly, the differences between the hILC and single-frequency maps yield PTE values of 0.63 (hILC$-$90\,GHz) and 0.25 (hILC$-$150\,GHz), both consistent with null.

These results rule out significant contamination from frequency-dependent foregrounds such as the tSZ effect or dust emission at the precision of our measurement.

\subsubsection{NGC vs SGC}

We perform a null test to verify the isotropy of the kSZ signal by dividing our galaxy sample into two disjoint regions: the North Galactic Cap (NGC) and South Galactic Cap (SGC). These regions have similar signal-to-noise ratios but probe different areas of the ACT DR6 footprint. Since the kSZ effect arises from the Doppler shift of CMB photons by large-scale velocity flows, which are statistically isotropic in the standard cosmological model, we expect the measured kSZ amplitude to be consistent between these two independent sky regions.

\begin{figure*}[!htbp]
    \centering
    \includegraphics[width=\linewidth]{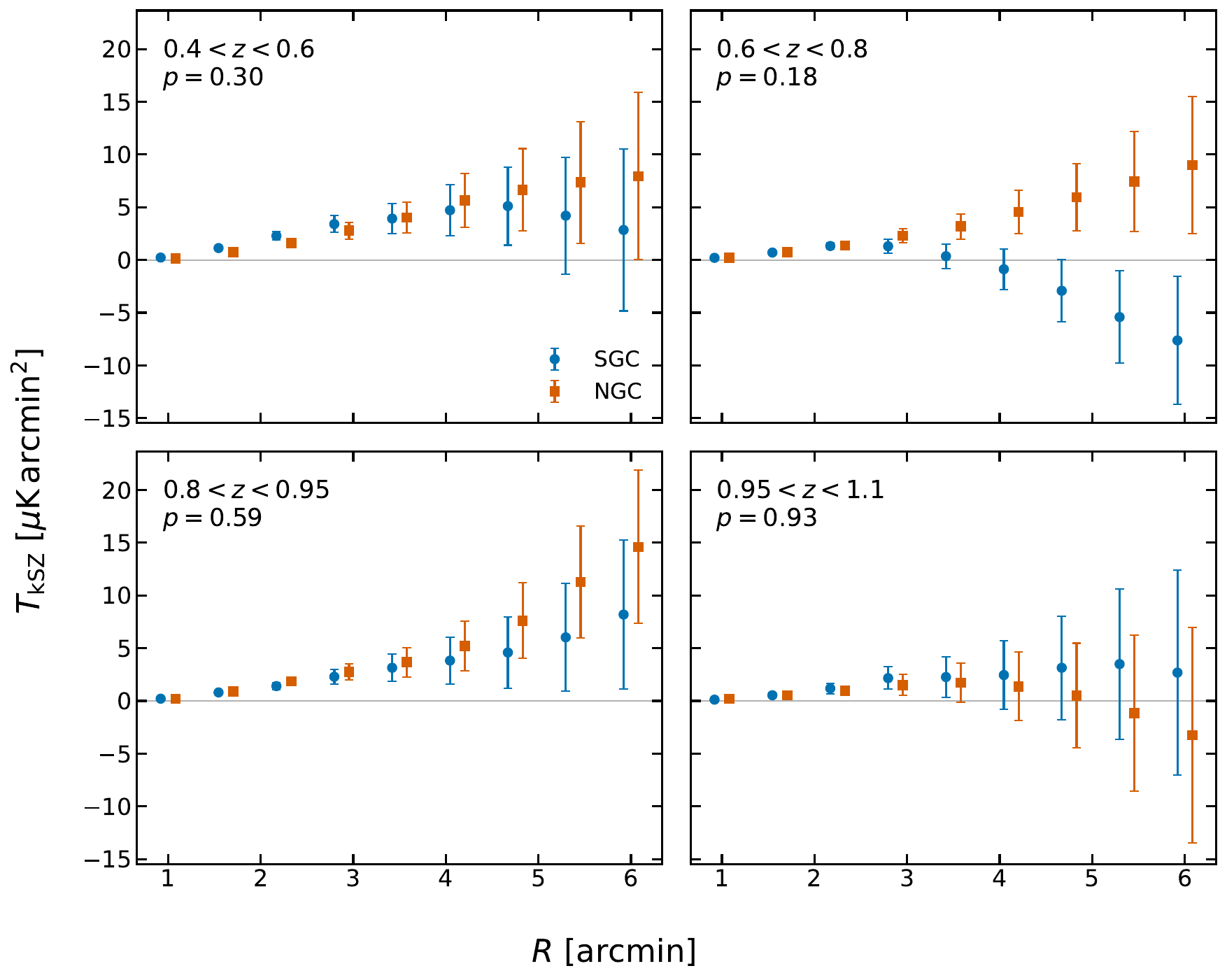}
    \caption{North$-$South isotropy test in redshift bins. Stacked kSZ CAP profiles measured separately on the SGC (blue circles) and NGC (orange squares) footprints in the four configuration-space redshift bins; the annotated $p$-values are the PTEs of the NGC$-$SGC difference in each bin, computed from the paired bootstrap difference covariance over all nine apertures. The apparent large-aperture divergence in the $0.6<z<0.8$ bin reflects the strong correlation between neighbouring CAP apertures at large $R$ and is consistent with the null hypothesis (PTE $= 0.18$).}
    \label{fig:ngc_sgc_zbins}
\end{figure*}

\begin{figure}[!htbp]
    \centering
    \includegraphics[width=0.5\textwidth]{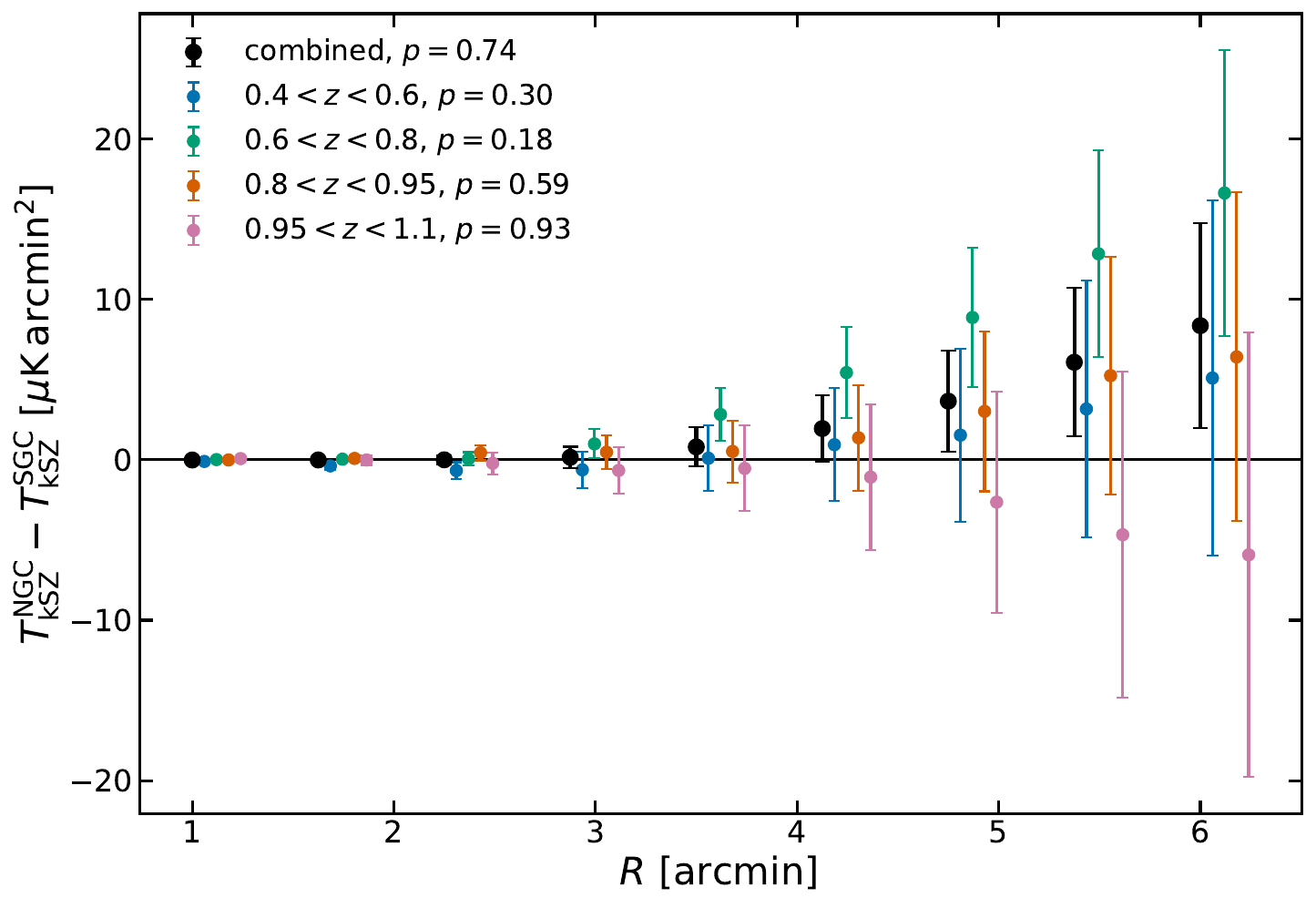}
    \caption{North$-$South isotropy null test. Difference between the kSZ CAP profiles measured in the NGC and SGC, for the combined sample (black) and for each redshift bin (colors, offset horizontally for clarity). All differences are consistent with zero: PTE $= 0.74$ for the combined sample and $0.30$, $0.18$, $0.59$, $0.93$ for $z_1$ through $z_4$.}
    \label{fig:ngc_sgc_comparison}
\end{figure}

Figure~\ref{fig:ngc_sgc_zbins} compares the CAP profiles measured separately on the two caps in each of the four configuration-space redshift bins, and Figure~\ref{fig:ngc_sgc_comparison} shows the NGC$-$SGC difference for the combined sample and for each bin. For every difference we estimate the covariance from the paired bootstrap realizations of the two caps, apply the \citet{Hartlap_2006} correction to its inverse, and quote the PTE of the null hypothesis over the nine aperture bins. The two caps are consistent in every case: PTE $= 0.30$, $0.18$, $0.59$, and $0.93$ for $z_1$ through $z_4$, and PTE $= 0.74$ for the combined sample. In the $0.6 < z < 0.8$ bin the SGC profile falls below the NGC profile at large apertures ($R \gtrsim 4'$); this apparent divergence carries little statistical weight because successive CAP filters at large $R$ share most of their sky area, so neighbouring aperture bins are strongly correlated and a visually coherent offset corresponds to far fewer independent degrees of freedom than the number of points suggests. The full-covariance null test for this bin yields PTE $= 0.18$, consistent with statistical isotropy.

\subsubsection{Random template tests}

We construct a velocity shuffling null test to validate that the measured signal originates from the kSZ effect rather than correlated contaminants such as the cosmic infrared background (CIB) or thermal SZ (tSZ). In this test, we randomly shuffle the sample of reconstructed velocities 10,000 times and perform the CAP stacking of Eq.~\ref{eq.stacking}, but assign each galaxy a shuffled velocity instead of its true velocity. Since the kSZ effect is proportional to the line-of-sight velocity, shuffling the velocities destroys the correlation between the CMB temperature and the actual galaxy motions, and should therefore yield a null result. Any residual non-zero signal would indicate contamination from sources that are spatially correlated with galaxy positions but independent of their velocities, such as residual tSZ or CIB emission in the ACT maps.

\begin{figure}[!htbp]
    \centering
    \includegraphics[width=0.5\textwidth]{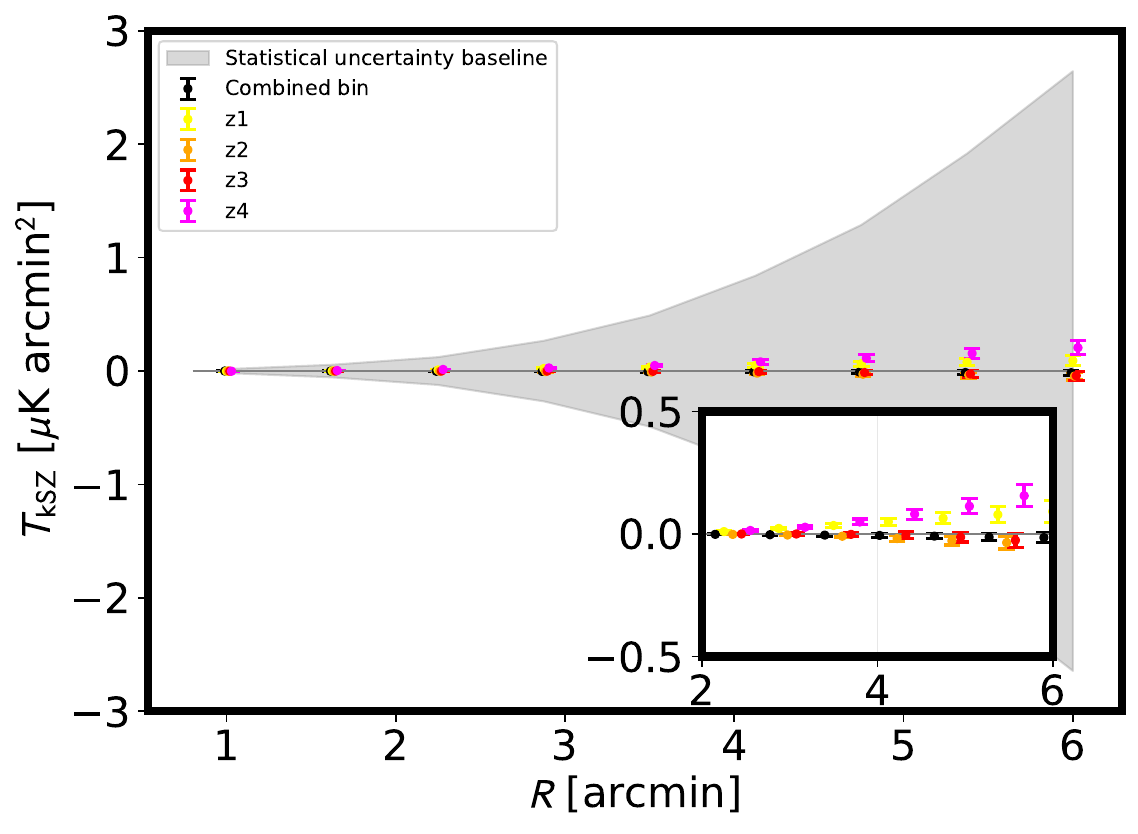} \\ 
    \caption{Null test for systematic errors in the kSZ measurement. CMB temperature cutouts are weighted with randomly shuffled galaxy velocities before stacking. The kSZ temperature as a function of angular separation shows results consistent with zero for both the combined sample (black points) and individual redshift bins z1--z4 (colored points), as expected when the velocity field is uncorrelated with the galaxy positions. The gray band represents the statistical uncertainty of the baseline measurement with unshuffled velocities. The inset shows a zoomed view of the outer radial bins. This demonstrates that the pipeline does not spuriously generate kSZ-like signals from noise or systematic effects.}
    \label{fig.Random_velocity}
\end{figure}

Figure~\ref{fig.Random_velocity} shows the results of this null test. The mean shuffled profiles for both the combined sample and individual redshift bins are consistent with zero across all scales, with the largest deviations well within the statistical uncertainties of the measurement.

\subsubsection{Robustness to the CMB map used}
\begin{figure}[!htbp]
    \centering
    \includegraphics[width=0.5\textwidth]{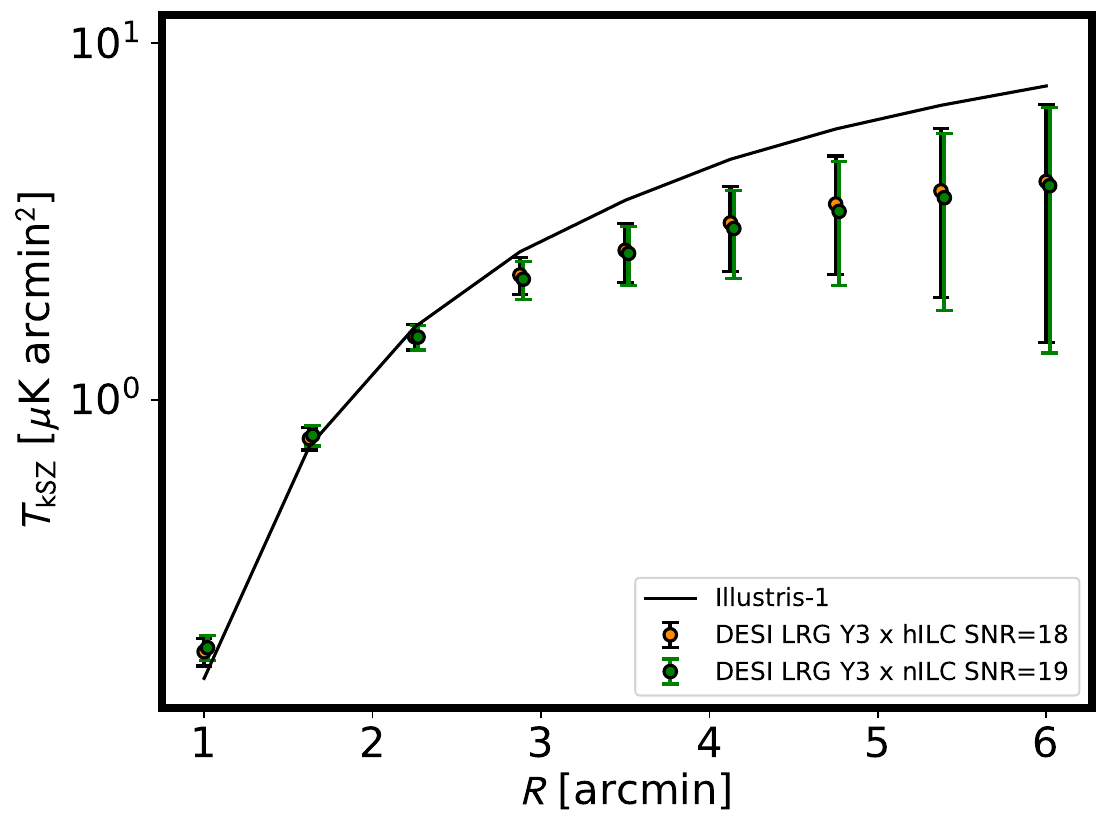} \\ 
    \caption{Comparison of the kSZ radial temperature profile measured using the harmonic ILC (hILC, orange) and needlet ILC (nILC, green) CMB temperature maps, both constructed from ACT data. The stacking is performed on DESI LRG DR2 galaxies with the velocity reconstruction method described in Sec.~\ref{sec.methodology}. The solid black curve shows the prediction from the Illustris-1 simulation. Both analyses yield consistent profiles with comparable signal-to-noise ratios (SNR$=$18 for hILC and SNR$=$19 for nILC), demonstrating that our kSZ detection is robust to the choice of component separation methodology. The agreement validates that the signal is not dominated by residual foregrounds or systematic artifacts associated with either map-making approach.}

    \label{fig.hilcnilc}
\end{figure}

To assess the robustness of our kSZ measurement to the choice of CMB map processing, we perform the stacking analysis using two CMB temperature maps prepared with different component separation methods: our baseline harmonic ILC (hILC) map and an alternative needlet ILC (nILC) map. The nILC map is produced following the methodology of Ref.~\cite{PhysRevD.109.063530}, which combines DR6 data using a wavelet-based internal linear combination pipeline. This approach provides joint localization in real and harmonic space, enabling efficient suppression of spatially varying Galactic contamination while accounting for the inhomogeneous ACT noise properties. As shown in Fig.~\ref{fig.hilcnilc}, the kSZ profiles obtained from both maps are in excellent agreement across all radial bins, with signal-to-noise ratios of 18 and 19 for the hILC and nILC analyses, respectively. The consistency between these two component separation approaches, which differ in their treatment of foreground removal and noise weighting, provides a strong validation that our kSZ signal is not driven by residual foreground contamination or artifacts specific to either method.

\subsubsection{Configuration pipeline verification: Comparison with DR1 LRG}

\begin{figure}[!htbp]
    \centering
    \includegraphics[width=0.5\textwidth]{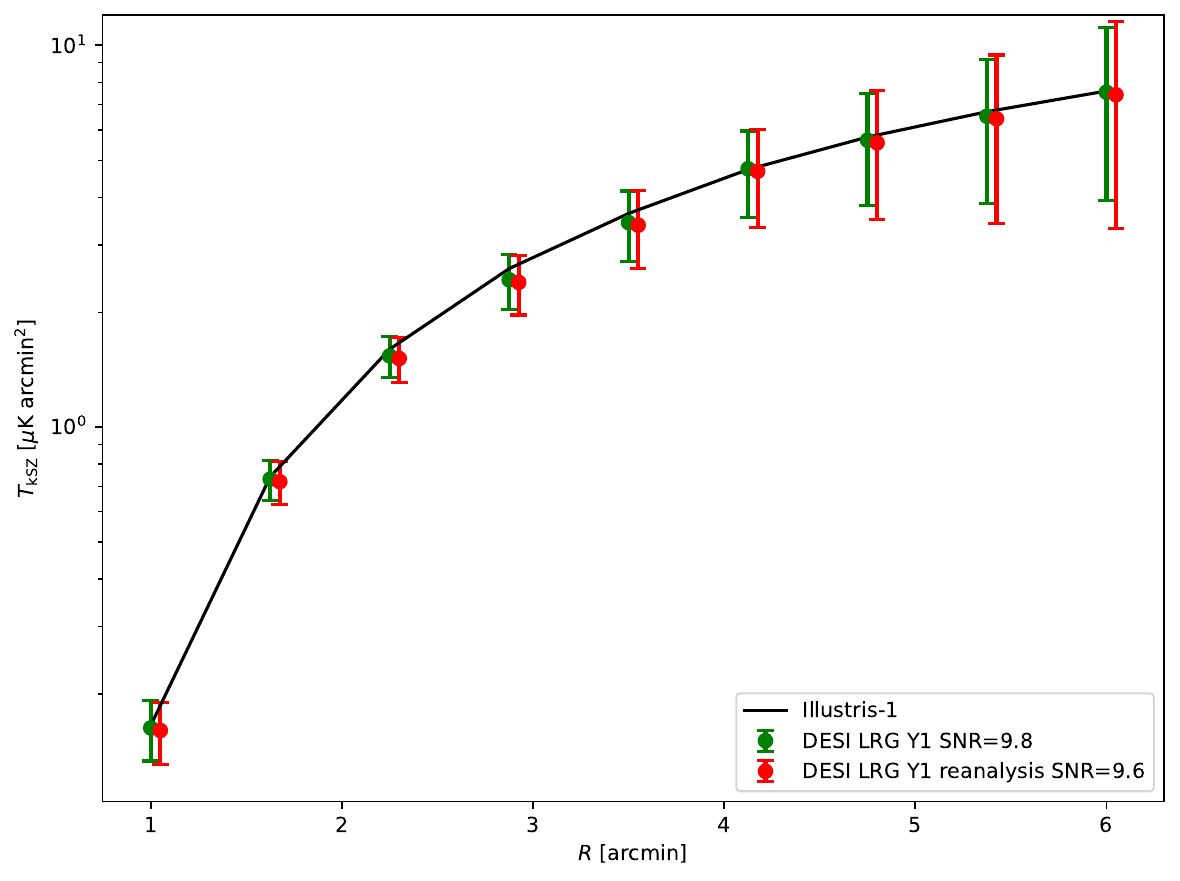} \\ 
    \caption{Comparison of our DR1 reanalysis (red, SNR = 9.6) with the published DR1 results from \citep{guachalla2025backlightingextendedgashalos} (green, SNR = 9.8). The solid black curve shows the Illustris-1 prediction. The two measurements agree within uncertainties at all angular scales, validating our pipeline. The marginally lower SNR in our reanalysis reflects the use of correlated block bootstrapping for uncertainty estimation.}
    \label{fig.y1y3}
\end{figure}

 As a sanity check of our analysis pipeline, we reprocessed the DESI DR1 spectroscopic LRG catalog using identical methods to those applied to the DR2 data. Figure~\ref{fig.y1y3} compares our reanalysis of the DR1 sample (red points, SNR = 9.6) with the published results from \citet{guachalla2025backlightingextendedgashalos} (green points, SNR = 9.8). The two measurements agree within uncertainties across all angular scales, with differences consistent with statistical fluctuations. The close agreement validates our pipeline implementation and confirms that our DR2 analysis employs consistent methodology with previous work. The marginally lower SNR in our reanalysis (9.6 vs 9.8) reflects the use of correlated bootstrapping which improves the uncertainty estimation compared to the DR1 analysis.

\subsection{Cumulative $\chi^2$ analysis} \label{appendix_chi2}

\begin{figure}[!htbp]
    \centering
    \includegraphics[width=\linewidth]{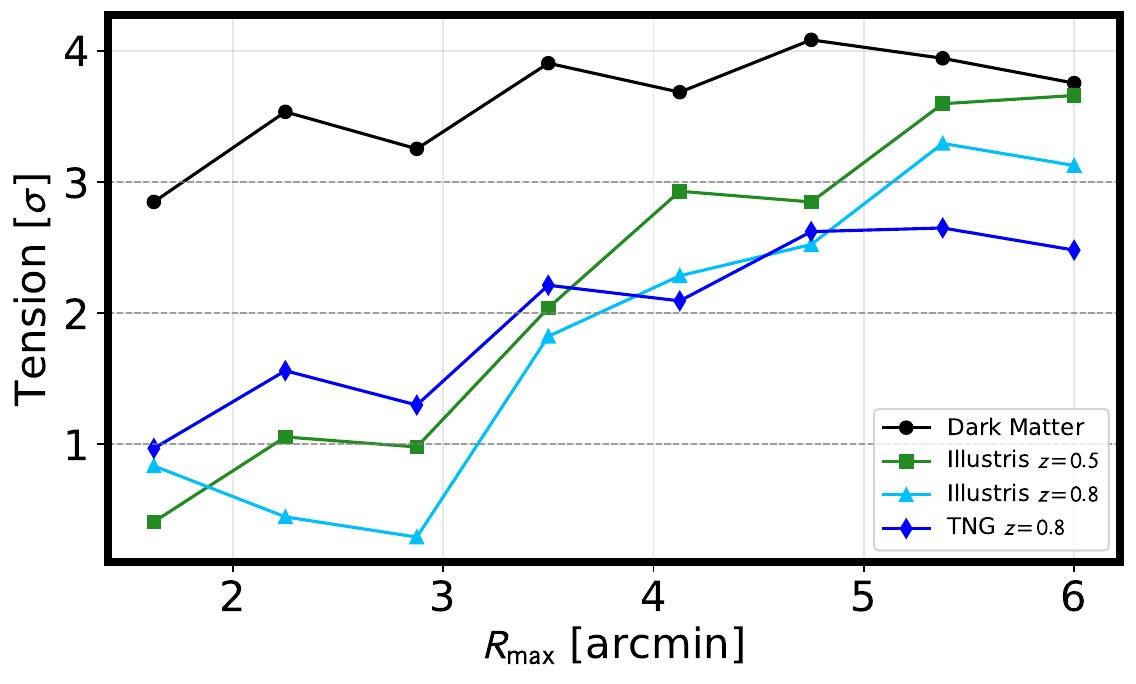}
    \caption{Cumulative significance for simulation gas profiles as a function of maximum aperture radius. At each $R_{\rm max}$, we fit a free amplitude to each model using all data points with $R \leq R_{\rm max}$ and compute the $\chi^2$ of the residuals, which is then converted to Gaussian-equivalent significance via the probability-to-exceed. This tests the radial shape of each profile independent of the overall amplitude.}
    \label{fig.chi_sigma}
\end{figure}

Figure~\ref{fig.chi_sigma} shows the cumulative tension as a function of 
maximum radius $R_{\rm max}$ included in the fit. At small radii 
($R \lesssim 3$ arcmin), both Illustris profiles show lower tension than 
IllustrisTNG, consistent with visual inspection of Figure~\ref{fig.measurements}. 
However, at larger radii ($R \gtrsim 4$ arcmin), the Illustris profiles 
develop increasingly large residuals that accumulate into the total $\chi^2$, 
while IllustrisTNG maintains more moderate tension across the full radial 
range. This explains the ranking in Table~\ref{tab:rescaling} despite the 
visual impression. We note that the largest aperture radii are particularly 
susceptible to noise from primary CMB fluctuations, requiring care when 
interpreting the relative discrimination power between simulation profiles.

\section{Simulation Profile Comparison Without Amplitude Rescaling}
\label{app:no_rescaling}

\begin{figure}[h]
    \centering
    \includegraphics[width=\columnwidth]{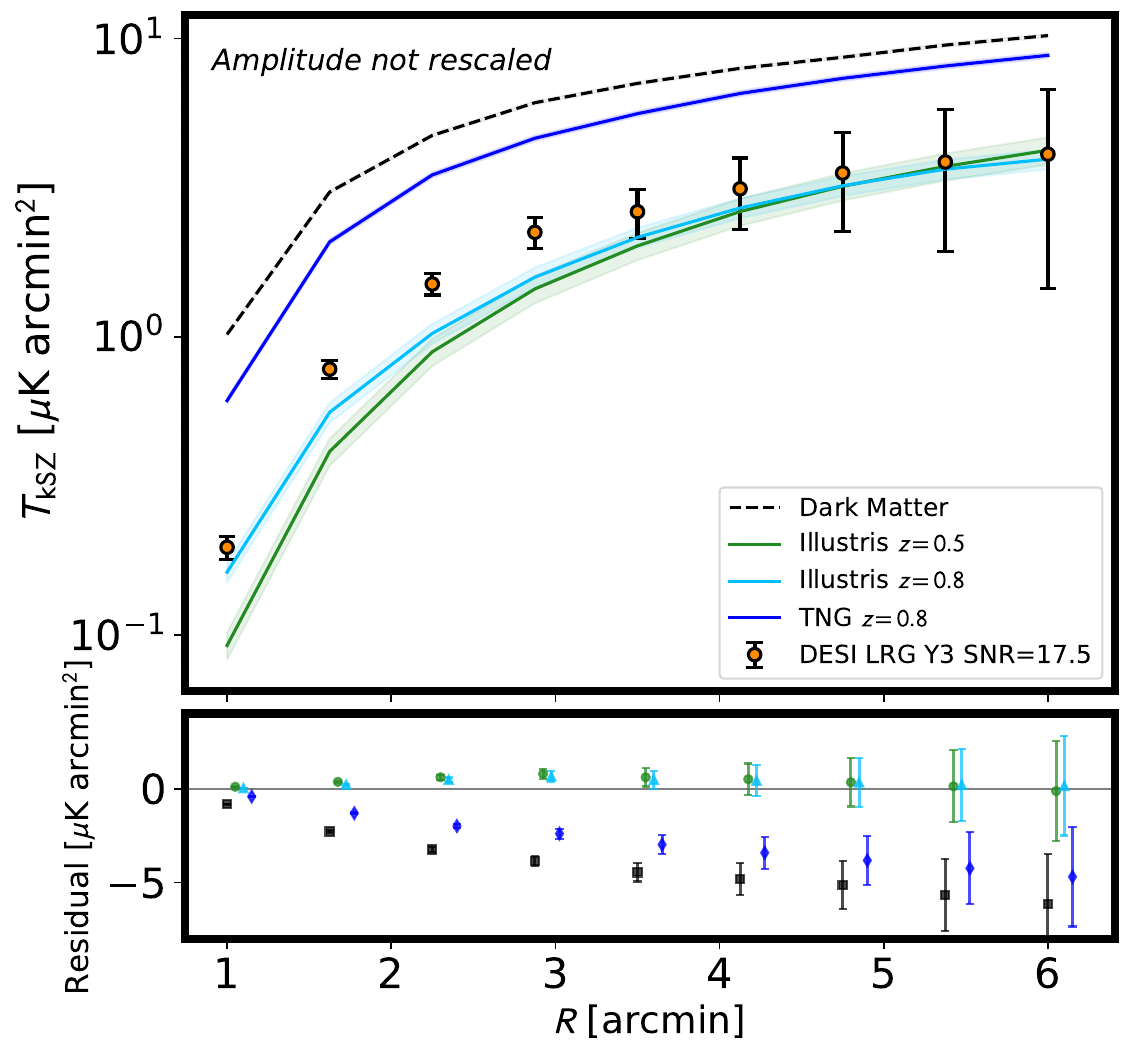}
    \caption{ Same as Fig.~\ref{fig.measurements}, but without rescaling the 
    simulation profile amplitudes. \textbf{Top panel:} The Illustris-1 profiles 
    at $z=0.5$ (green) and $z=0.8$ (cyan) are closest to the data, while the 
    dark matter profile (dashed black) and IllustrisTNG at $z=0.8$ (blue) 
    overpredict the signal by factors of $\sim 4$ and $\sim 3$, respectively. 
    \textbf{Bottom panel:} Residuals between data and unrescaled models.}
    \label{fig.profile_no_rescaling}
\end{figure}

In the main text (Fig.~\ref{fig.measurements}), we compare our kSZ measurements 
to simulation profiles after fitting a free overall amplitude to each curve. 
Here we show the same comparison without any amplitude rescaling 
(Fig.~\ref{fig.profile_no_rescaling}). The Illustris-1 profiles at $z=0.5$ and 
$z=0.8$ lie closest to the data, while the dark matter and IllustrisTNG profiles 
significantly overpredict the signal. This reflects the well-known sensitivity of 
the kSZ amplitude to the gas fraction within the halo: the dark matter profile 
assumes all baryons trace the dark matter, while IllustrisTNG retains a larger 
fraction of gas within $r_{200c}$ compared to Illustris-1, due to differences in 
their feedback implementations \citep{Nelson_2015, 
nelson2021illustristngsimulationspublicdata}. The rescaling factors required to 
match the data are listed in Table~\ref{tab:rescaling}.

\section{Normalization of the template}

Relating the 2D density $\delta^{2D}_g$ contrast to the 3D density contrast $\delta^{3D}_g$. We define the 2D density contrast as:

\begin{equation}
    \delta^{2D}_g(\boldsymbol{\theta})=\frac{n^{2D}_g(\boldsymbol{\theta})}{\bar{n}^{2D}_g}-1
\end{equation}

This quantity is unit-less since $n^{2D}_g(\boldsymbol{\theta})$ and ${\bar{n}^{2D}_g}\equiv{N_g}/\Omega_{survey}$ are in units of $sr^{-1}$.

Starting with $n^{2D}_g(\boldsymbol{\theta})$, the angular number density of galaxies at position $\boldsymbol{\theta}$ on the sky, one can obtain the number of galaxies per pixel once we know the solid angle of the pixel element.

\begin{equation}
    N_g^{\mathrm{pix}} = n_g^{2D}(\boldsymbol{\theta}) \, \Omega_{\mathrm{pix}} = \sum_{b} n_g^{3D}({\chi_b}, \boldsymbol{\theta}) \times \Delta V({\chi_b})
\end{equation}

where:
\begin{itemize}
    \item $N_g^{\mathrm{pix}}$ is the total number of galaxies in a pixel centered at angular position $\boldsymbol{\theta}$
    \item $n_g^{2D}(\boldsymbol{\theta})$ is the 2D angular number density [galaxies/steradian] at position $\boldsymbol{\theta}$
    \item $\Omega_{\mathrm{pix}}$ is the solid angle subtended by the pixel [steradians]
    \item $n_g^{3D}({\chi_b}, \boldsymbol{\theta})$ is the comoving 3D number density [galaxies/Mpc$^3$] at comoving distance $\chi_b$ along direction $\boldsymbol{\theta}$
    \item The sum is over discrete radial bins indexed by $b$
\end{itemize}

The comoving volume element for bin $b$ spanning from $z_b - \Delta z_b/2$ to $z_b + \Delta z_b/2$ is:
\begin{equation}
    \Delta V(z_b) = \Omega_{\mathrm{pix}} \chi^2(z_b) \frac{d\chi}{dz}\bigg|_{z_b} \Delta z_b = \Omega_{\mathrm{pix}} \chi^2(z_b) \frac{c}{H(z_b)} \Delta z_b
\end{equation}

where $\chi(z)$ is the comoving distance to redshift $z$, and $H(z)$ is the Hubble parameter.

Substituting back:
\begin{equation}
    n_g^{2D}(\boldsymbol{\theta}) \, \Omega_{\mathrm{pix}} = \sum_{b} n_g^{3D}({\chi_b}, \boldsymbol{\theta}) \, \Omega_{\mathrm{pix}} \chi^2(z_b) \frac{c}{H(z_b)} \Delta z_b
\end{equation}

Dividing both sides by $\Omega_{\mathrm{pix}}$:
\begin{equation}
    n_g^{2D}(\boldsymbol{\theta}) = \sum_{b} n_g^{3D}({\chi_b}, \boldsymbol{\theta}) \chi^2(z_b) \frac{c}{H(z_b)} \Delta z_b
\end{equation}

Taking the limit as $\Delta z_b \rightarrow 0$, the sum becomes an integral:
\begin{equation}
    n_g^{2D}(\boldsymbol{\theta}) = \int n_g^{3D}(\chi, \boldsymbol{\theta}) \chi^2 \, d\chi
\end{equation}

We therefore also know the relation between the mean 2D and 3D mean number densities

\begin{equation}
    \bar{n}_g^{2D} = \int{d\chi}\chi^2 \bar{n}_g^{3D}(\chi)
\end{equation}

Now we can derive the relation between the 2D density $\delta^{2D}_g$ contrast to the 3D density contrast $\delta^{3D}_g$.

\begin{widetext}
\begin{equation}
    \delta^{2D}_g(\boldsymbol{\theta})
    =\frac{\int d\chi \,\chi^2 n_g^{3D}(\chi,\boldsymbol{\theta})}{\int d\chi \,\chi^2 \bar{n}_g^{3D}(\chi)}-1
    =\frac{\int d\chi \,\chi^2 \bar{n}_g^{3D}(\chi)\,\delta^{3D}_g({\chi,\boldsymbol\theta})}{\int d\chi \,\chi^2 \bar{n}_g^{3D}(\chi)}
    =\int d\chi \Bigg[\frac{\chi^2\bar{n}_g^{3D}(\chi)}{\int d\chi^\prime \chi^{\prime2} \bar{n}_g^{3D}(\chi^\prime)}\Bigg] \delta^{3D}_g({\chi,\boldsymbol\theta})
    =\int d\chi \,\frac{dp}{d\chi}\,\delta^{3D}_g(\chi,\boldsymbol\theta)
\end{equation}
\end{widetext}

\begin{equation}
\frac{dp}{d\chi} = \frac{\chi^2 \bar{n}_g^{3D}(\chi)}{\int \chi'^2 \bar{n}_g^{3D}(\chi') \, d\chi'}
\end{equation}

is the weight function that has units of $[Mpc^{-1}]$ that ensures the RHS is kept unit-less.

Note that this makes sense since

\begin{equation}
N_{\mathrm{total}} = {\Omega}_{\mathrm{survey}} \int_0^{\infty} \bar{n}_g^{3D}(\chi) \chi^2 \, d\chi
\end{equation}

and the above guarantees

\begin{equation}
\int_0^{\infty} \frac{dp}{d\chi} d\chi = 1
\end{equation}

\subsection{Building the kSZ momentum template}

\begin{equation}
\hat{\pi}(\boldsymbol{\theta}) = \sum_{i \in \mathrm{pixel}(\boldsymbol{\theta})} \frac{v_{r,i}}{c}
\end{equation}

\begin{equation}
\hat{\pi}(\boldsymbol{\theta}) = \hat{\pi}^{\,2D}(\boldsymbol{\theta}) \times \Omega_{\mathrm{pix}}
\end{equation}

here, note that $\hat{\pi}$ is an extensive quantity and depends on the size of the pixel.

\begin{align}
\hat{\pi}(\boldsymbol{\theta})
&= \int_{\mathrm{l.o.s.}} n_g^{3D}(\chi, \boldsymbol{\theta}) \,
    \frac{v_r(\chi, \boldsymbol{\theta})}{c} \, dV \\[6pt]
&= \Omega_{\mathrm{pix}} \int n_g^{3D}(z, \boldsymbol{\theta}) \,
    \frac{v_r(z, \boldsymbol{\theta})}{c} \,
    \chi^2(z) \, \frac{c}{H(z)} \, dz \\[6pt]
&= \Omega_{\mathrm{pix}} \int n_g^{3D}(\chi, \boldsymbol{\theta}) \,
    \frac{v_r(\chi, \boldsymbol{\theta})}{c} \,
    \chi^2 \, d\chi \\[6pt]
&= \Omega_{\mathrm{pix}} \int \bar{n}_g^{3D}(\chi)\,
    [1 + \delta_g(\chi, \boldsymbol{\theta})] \,
    \frac{v_r(\chi, \boldsymbol{\theta})}{c} \,
    \chi^2 \, d\chi \\[6pt]
&= \Omega_{\mathrm{pix}} \, \bar{n}_g^{2D} 
    \int_0^{\infty} \frac{dp}{d\chi} \,
    [1 + \delta_g(\chi, \boldsymbol{\theta})] \,
    \frac{v_r(\chi, \boldsymbol{\theta})}{c} \, d\chi .
\end{align}

Thus we can define an improved intensive template as:

\begin{equation}
\hat{\pi}(\boldsymbol{\theta}) = \frac{1}{\Omega_{pix}\bar{n}^{2D}_g}\sum_{i \in \mathrm{pixel}(\boldsymbol{\theta})} \frac{v_{r,i}}{c}
\end{equation}

where the normalization is the inverse of the mean number of galaxies per pixel.

\subsection{Robustness in building the template}

\begin{figure}[!htbp]
    \centering
    \includegraphics[width=0.5\textwidth]{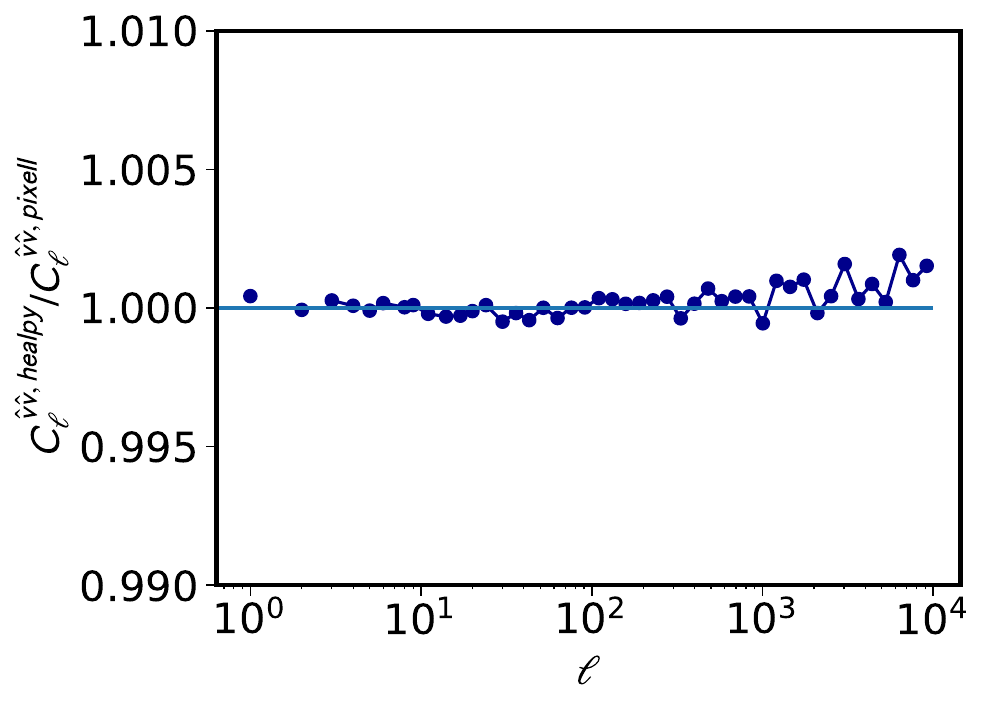} \\ 
\caption{Robustness to pixelization. Ratio of the auto-power spectra of the kSZ velocity template constructed on an equal–area \texttt{HEALPix} grid and on a CAR grid using \texttt{pixell}. The two constructions use the same galaxy sample, sky footprint (including the $\delta>-20^\circ$ cut), and normalization. The ratio is consistent with unity at the sub-percent level; the small residual fluctuations are attributable to differing pixel-window functions and edge assignments. This demonstrates that the template traces the same projected velocity field independent of the chosen pixelization.}
    \label{fig:hp_vs_car}
\end{figure}

We verified that our kSZ template is insensitive to the choice of sky pixelization by constructing it in two independent ways: (i) on the sphere using an equal–area \texttt{HEALPix} grid (after projecting the CAR maps and mask to \texttt{HEALPix}), and (ii) directly on a CAR grid using \texttt{pixell}, enforcing the same declination cut ($\delta>-20^\circ$), sky footprint, and overall normalization. In both cases the template represents the same projected line–of–sight velocity field per unit solid angle; hence, modulo the pixel-window function, it is a geometric scalar and should be invariant under a change of pixelization. As shown in Fig.~\ref{fig:hp_vs_car}, the ratio of the auto-spectra obtained from the two constructions is consistent with unity at the sub-percent level with the small residual fluctuations attributable to finite-resolution effects (pixel-window differences and boundary assignment near mask edges).

\section{Galaxy mask used}

We construct a binary survey mask from the DESI random catalogs, which trace the survey selection function. 
Random objects from both the North Galactic Cap (NGC) and South Galactic Cap (SGC) are combined, and a 
declination cut of $\delta > -20^\circ$ is applied. Each random object position is converted to a HEALPix 
pixel index at resolution $N_{\mathrm{side}} = 8192$. The mask is then defined as
\begin{equation}
    M_p = \Theta\left(\sum_{i \in p} 1\right),
\end{equation}
where $\Theta(x)$ is the Heaviside step function, and the sum runs over all random objects falling within 
pixel $p$. This binary mask defines the survey footprint but, unlike a density-weighted mask of the form 
$\bar{n}_p \propto \sum_{i \in p} w_i^r / \Omega_p$, does not capture completeness variations within the 
observed region.

\section{Impact of cosmological parameter uncertainty}
\label{sec:cosmo_spread}

Throughout this work, we fix the background cosmology to the \textit{Planck} 2018 TTTEEE best-fit values \citep{Planck:2018vyg}. Here we verify that this assumption has negligible impact on our gas profile constraints as the uncertainties are currently dominated by the degeneracies between the gas parameters and the HOD as opposed to cosmology.

\begin{figure}
    \centering
    \includegraphics[width=\columnwidth]{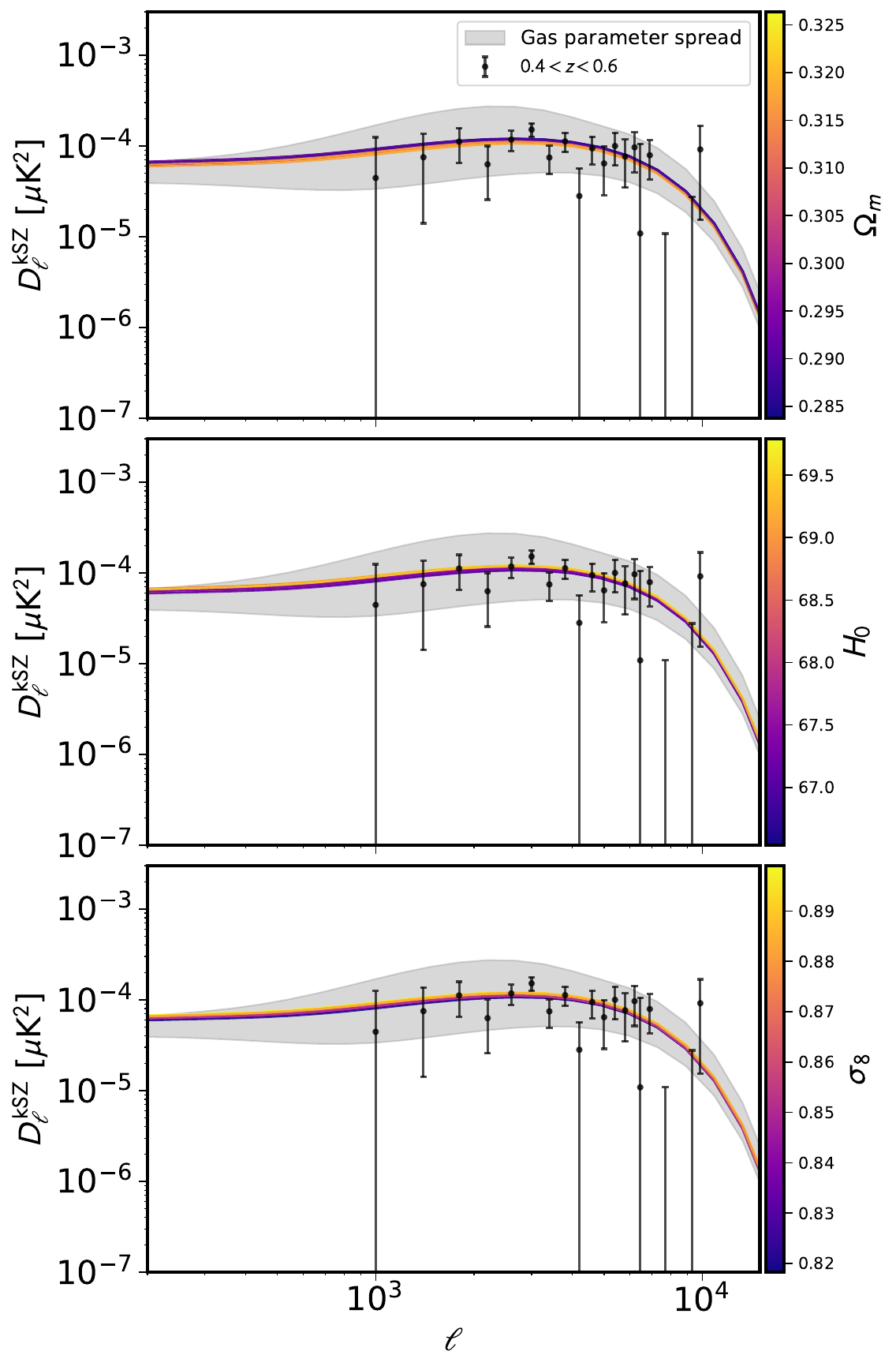}
    \caption{Impact of cosmological parameter uncertainty on the predicted kSZ-galaxy power spectrum. The gray band shows the spread from varying gas profile parameters ($\rho_0$, $x_c$, $\beta$) across their prior range at fixed \textit{Planck} 2018 cosmology. Colored curves show 50 correlated draws from the \textit{Planck} 2018 TTTEEE posterior at fixed best-fit gas parameters, color-coded by $\Omega_m$ (top), $H_0$ (middle), and $\sigma_8$ (bottom). Black data points show the $0.4 < z < 0.6$ measurement. The cosmological uncertainty is subdominant by over an order of magnitude.}
    \label{fig:cosmo_spread}
\end{figure}

We assess the sensitivity of the predicted $D_\ell^{\mathrm{kSZ}}$ to cosmological parameter uncertainty by drawing 50 correlated samples from the \textit{Planck} 2018 TTTEEE posterior, spanning approximately $4\sigma$ in $\Omega_m$, $H_0$, and $\sigma_8$. For each cosmological realization, we compute the full halo model prediction using \texttt{class-sz} with the gas profile parameters fixed at their best-fit values. The resulting spread in $D_\ell^{\mathrm{kSZ}}$ is shown as the colored curves in Fig.~\ref{fig:cosmo_spread}, color-coded by $\Omega_m$ (top), $H_0$ (middle), and $\sigma_8$ (bottom). For comparison, the gray band shows the spread obtained by varying the gas profile parameters ($\rho_0$, $x_c$, $\beta$) across their prior range at fixed cosmology.

 The fractional variation due to cosmology is 2--3\%, compared to $\sim$30\% from gas parameter variations. The cosmological uncertainty is therefore subdominant by over an order of magnitude and does not affect our conclusions.

\section{Profile truncation and physically allowed GNFW parameters}\label{app:zbrent}

The GNFW gas profile (Eq.~\ref{eq.1h}) is normalized by requiring that
the enclosed gas mass equals the cosmic baryon fraction times the halo
mass,
\begin{equation}\label{eq:mass_conserve}
M_{\rm gas}(<r_{\rm cut}) = f_b\,M_{200c},
\end{equation}
where $r_{\rm cut}$ is the truncation radius determined by this
condition. In practice, \texttt{class-sz} solves
Eq.~(\ref{eq:mass_conserve}) numerically using Brent's root-finding
method. For certain parameter combinations no solution exists, and
the root finder fails. Here we derive the analytic condition
under which a valid truncation radius is guaranteed.

Substituting $u = r'/(x_c\,r_{200c})$, the enclosed gas mass within
radius $r$ is
\begin{align}\label{eq:mgas_enclosed}
M_{\rm gas}(<r) &= 4\pi\,\rho_0\,\rho_{\rm cr}\,f_b\,
(x_c\,r_{200c})^3 \nonumber\\
&\quad\times \int_0^{r/(x_c r_{200c})}
\frac{u^{2+\gamma}\,du}
{\left[1 + u^\alpha\right]^{(\beta+\gamma)/\alpha}}.
\end{align}
Integrating to infinity and defining the shape integral
\begin{equation}\label{eq:J_def}
\mathcal{J}(\alpha,\beta,\gamma)
\equiv \int_0^{\infty}
\frac{u^{2+\gamma}\,du}
{\left[1 + u^\alpha\right]^{(\beta+\gamma)/\alpha}},
\end{equation}
the total gas mass of the untruncated profile is
\begin{equation}\label{eq:mgas_total}
M_{\rm gas}(\infty) = 4\pi\,\rho_0\,\rho_{\rm cr}\,f_b\,
(x_c\,r_{200c})^3\,\mathcal{J}(\alpha,\beta,\gamma).
\end{equation}
The substitution $t = u^\alpha$ transforms
Eq.~(\ref{eq:J_def}) into a standard Beta-function integral,
yielding the closed form
\begin{equation}\label{eq:J_closed}
\mathcal{J}(\alpha,\beta,\gamma)
= \frac{1}{\alpha}\,
\frac{\Gamma\!\left(\frac{3+\gamma}{\alpha}\right)\,
      \Gamma\!\left(\frac{\beta-3}{\alpha}\right)}
     {\Gamma\!\left(\frac{\beta+\gamma}{\alpha}\right)}.
\end{equation}
This integral converges if and only if $\beta > 3$, since the
convergence of the Beta function requires
$(\beta - 3)/\alpha > 0$. For $\beta \leq 3$, the outer profile
slope is too shallow: the total mass integral diverges and
\texttt{class-sz} encounters a numerical overflow. We therefore
restrict our analysis to $\beta > 3$ throughout.

A truncation radius satisfying Eq.~(\ref{eq:mass_conserve}) exists
if and only if $M_{\rm gas}(\infty) \geq f_b\,M_{200c}$.
Substituting $M_{200c} = (4\pi/3) \times 200\,\rho_{\rm cr}\,
r_{200c}^3$ and canceling common factors yields the necessary
condition
\begin{equation}\label{eq:stability}
\rho_0\,x_c^3\,\mathcal{J}(\alpha,\beta,\gamma) \geq
\frac{200}{3}.
\end{equation}
When this condition is violated, the profile is too dilute or too
steeply falling for the enclosed mass to reach $f_b\,M_{200c}$ at
any finite radius. During sampling, we exclude parameter
combinations that violate Eq.~(\ref{eq:stability}).

\section{Emulator training}\label{emulator}

Direct marginalization over HOD parameters would require approximately $10^6$ 
evaluations of \texttt{class-sz} \citep{Bolliet:2023eob} for a typical MCMC analysis, 
which is computationally prohibitive. Our GP emulator reduces this computational 
cost by several orders of magnitude while maintaining high accuracy.

We train a joint emulator in the combined 10-dimensional parameter space consisting 
of 6 HOD parameters and 4 gas profile parameters. The HOD parameters are 
$\log_{10}M_{\rm min}$, $\log_{10}M_1$, $\sigma_{\log M}$, $\kappa$, $\alpha_c$, 
and $\alpha_s$. We convert the HOD constraints from our galaxy clustering analysis, originally expressed in virial masses, to $M_{200c}$ using the \texttt{Colossus} package 
\citep{Diemer_2018} with the \citep{Duffy_2008} concentration-mass relation evaluated 
at the effective redshift of each sample.

The gas profile parameters varied in the emulator are $\log_{10}A_{\rho_0}$, 
$x_c$, $A_\beta$, and $a_{\rm k2h}$. We fix $A_\alpha = 0.88$ 
as in \citet{Battaglia_2016}. The HOD parameter ranges are determined from the 
5th to 95th percentile of our galaxy clustering posteriors obtained via PocoMC 
\citep{karamanis2022accelerating,karamanis2022pocomc}, while gas parameter ranges 
are set by the priors in our MCMC analysis.

We generate training samples using Latin Hypercube Sampling \citep{McKay1979} 
in the 10-dimensional parameter space. Before training, we filter out parameter 
combinations that produce numerically unreliable spectra due to root-finding 
failures in the internal density profile solver of \texttt{class-sz} 
(Appendix~\ref{app:zbrent}). Approximately 77\% of the na\"ive samples fall in 
this unstable region. We oversample by a factor of 5 to compensate, yielding 
1000 stability-filtered training points and 100 independent test points. For 
each sample, we evaluate \texttt{class-sz} to compute the galaxy-electron 
cross-power spectrum $C_\ell^{ge}$ over the multipole range 
$\ell \in [2, 20000]$ with logarithmic spacing $\Delta\log\ell = 0.2$. We then 
restrict to the range $\ell \in [300, 10000]$, which encompasses our fitting 
range with margin, retaining 17 multipole bins.

We train independent GP emulators for each of the 17 multipole bins. For each 
multipole $\ell_i$, we model
\begin{equation}
C_\ell^{ge}(\boldsymbol{\theta}_{\rm HOD}, \boldsymbol{\theta}_{\rm gas}) = 
\mathcal{GP}(\boldsymbol{\theta}; \mu, k),
\end{equation}
where $\boldsymbol{\theta} = [\boldsymbol{\theta}_{\rm HOD}, \boldsymbol{\theta}_{\rm gas}]$ 
is the combined 10-dimensional parameter vector, $\mu$ is the mean function (set 
to the training sample mean), and $k$ is the kernel function. We use a 
Mat\'ern-3/2 kernel \citep{tazi2023intuitionframeworkapplyinggps} with a white noise term:
\begin{equation}
k(\boldsymbol{\theta}, \boldsymbol{\theta}') = \sigma_f^2 
\left(1 + \sqrt{3}\,r\right)\exp\left(-\sqrt{3}\,r\right) 
+ \sigma_n^2\,\delta_{\boldsymbol{\theta}\boldsymbol{\theta}'},
\end{equation}
where $r^2 = \sum_{i=1}^{10}(\theta_i - \theta_i')^2/\ell_i^2$, $\sigma_f^2$ 
is the signal variance, $\ell_i$ are the per-dimension length scales, and 
$\sigma_n^2$ is the noise level. We choose the Mat\'ern-3/2 kernel over the 
squared exponential (RBF) kernel because the spectrum depends non-smoothly on 
parameters near the stability boundary, and the Mat\'ern-3/2 kernel accommodates 
functions that are once-differentiable rather than infinitely smooth. 
Hyperparameters are optimized via maximum likelihood with 5 random restarts. 
Length-scale bounds are set to $[10^{-2}, 10^{3}]$ in the standardized parameter 
space; the optimized length scales confirm that the spectrum depends primarily 
on the gas parameters ($\ell_i \sim 8$--$30$) while the HOD parameters have 
minimal influence ($\ell_i \sim 300$--$1000$). All input parameters are 
standardized to zero mean and unit variance before training. We implement the 
emulators using \texttt{scikit-learn} \citep{scikit-learn}, with training 
requiring approximately 5 minutes for all 17 multipoles on a single CPU.

We validate the emulator accuracy on the 100 independent test samples. Across 
all three redshift bins, the median relative error is 0.16--0.19\%, with a 
95th percentile error of 1.9--2.2\%. The small number of outliers with larger 
errors occur exclusively near the stability boundary at low $A_\beta$ and low 
$A_{\rho_0}$; these regions are excluded by our MCMC prior 
(App~\ref{app:zbrent}). Within the MCMC prior volume and fitting multipole 
range $\ell \in [999, 7000]$, the median emulator uncertainty is 0.8\%, well 
below our statistical measurement uncertainties of 5--10\% per bandpower. We 
verify that emulator errors show no systematic trends with either HOD or gas 
parameters across the prior volume. The GP also provides prediction uncertainties 
$\sigma_{\rm GP}(\boldsymbol{\theta})$, which we verify are well calibrated by 
confirming that actual prediction errors on test samples are consistent with the 
GP uncertainties. Since $\sigma_{\rm GP}$ is much smaller than our measurement 
uncertainties, we do not propagate emulator uncertainties into our final 
constraints.

For our combined redshift analysis, we train separate emulators for each of the 
three redshift bins: $z \in [0.4, 0.6]$, $[0.6, 0.8]$, and $[0.8, 1.1]$. 
Each bin has its own HOD parameter posterior from galaxy clustering, but shares 
the same gas profile parameters across redshift. The final prediction for the 
combined sample is
\begin{equation}
C_\ell^{ge,\rm comb}(\boldsymbol{\theta}_{\rm gas}) = \sum_{i=1}^3 w_i 
\langle C_\ell^{ge,i}(\boldsymbol{\theta}_{\rm HOD}^{(i)}, \boldsymbol{\theta}_{\rm gas}) 
\rangle_{\boldsymbol{\theta}_{\rm HOD}^{(i)}},
\end{equation}
where $w_i$ are weights proportional to the galaxy number density in each bin
($w = [0.235, 0.361, 0.405]$ for bins 1, 2, and 3 respectively, computed from the $N_{\rm gal} = 557{,}590$, $857{,}243$, and $962{,}037$ galaxies entering the corresponding per-bin templates), and the angle
brackets denote marginalization over HOD parameters within each bin.

\section{HOD marginalization: best-fit spectra and parameter contours}\label{app:hod_marg}
\begin{figure*}[htbp]
    \centering
    \includegraphics[width=\linewidth]{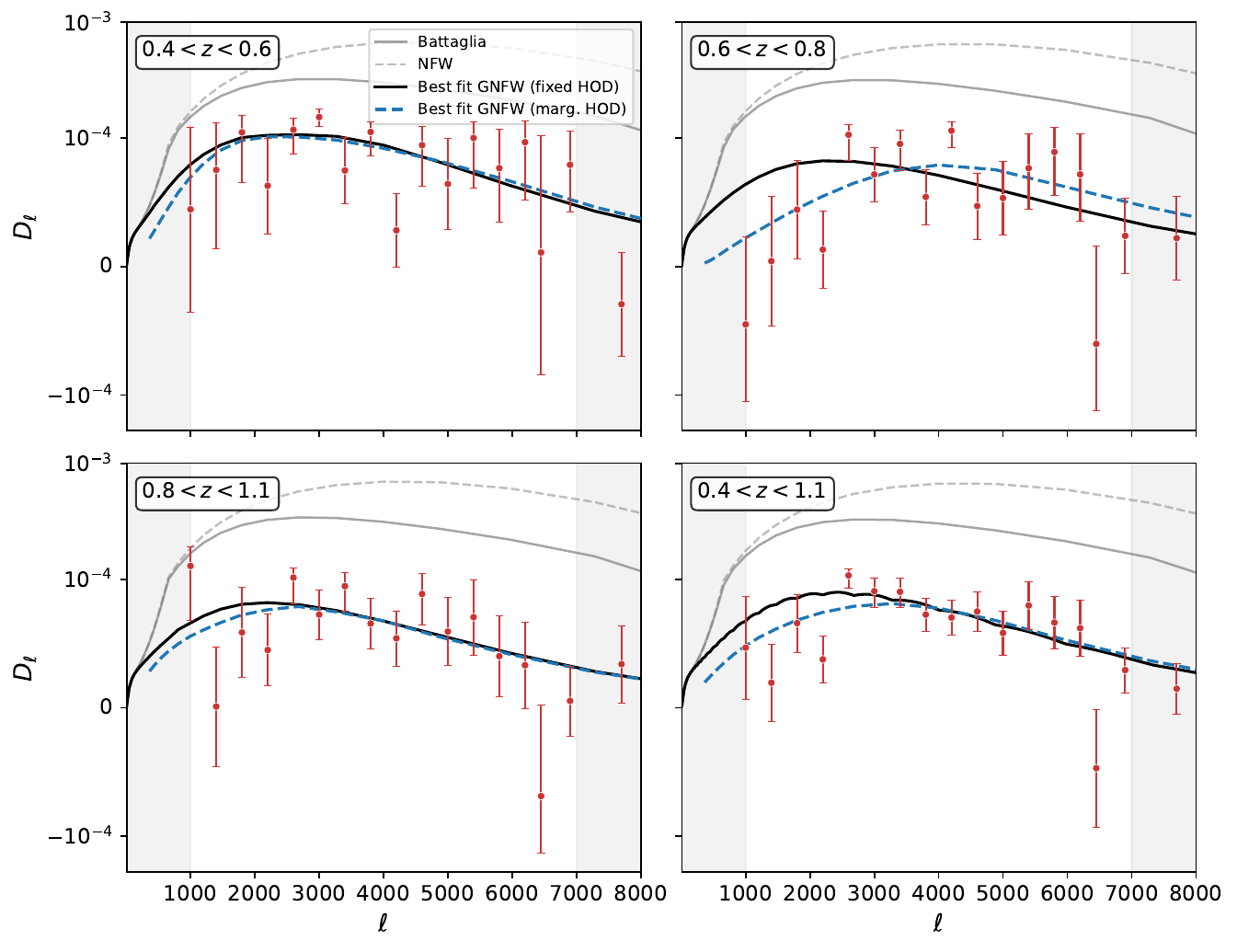}
    \caption{Best-fit kSZ cross-power spectra for the fixed-HOD (black solid) and HOD-marginalized (blue dashed) analyses, overlaid on the measured bandpowers (red points) in the three redshift bins and the combined sample. Grey curves show the Battaglia (solid) and NFW (dashed) reference profiles as in Fig.~\ref{fig.measurement_fits}; shaded bands mark the multipoles excluded from the fit ($\ell < 1000$ and $\ell > 7000$). The marginalized curve is drawn over the emulator training range ($\ell \gtrsim 360$). The two best fits agree within the statistical errors in all panels.}
    \label{fig:fit_marg}
\end{figure*}

We compare here the fixed-HOD and HOD-marginalized fits summarized in Table~\ref{table.prior} and Table~\ref{table.marginalize_hod}. Figure~\ref{fig:fit_marg} overlays the two best-fit GNFW spectra on the measured bandpowers in each redshift bin and for the combined sample; the fixed-HOD curve (black) and the marginalized-HOD curve (blue dashed) agree to within the statistical errors across the fit range, with the largest difference in the $0.6 < z < 0.8$ bin where the HOD-gas degeneracy is strongest. The marginalized best fit is evaluated on the emulator multipole grid ($\ell \gtrsim 360$) at the maximum-posterior point of the corrected chains, so it is drawn only over the range where the emulator is trained. The best-fit goodness of fit is comparable in the two cases: $\chi^2 = 12.9, 20.5, 12.8$ and $19.7$ (fixed HOD) versus $12.8, 16.5, 12.3$ and $17.9$ (marginalized), for $z_1$, $z_2$, $z_3$, and the combined sample respectively, all for $\nu = 12$; marginalization improves the $z_2$ and combined fits slightly by relaxing the profile shape.

\begin{figure}[htbp]
    \centering
    \includegraphics[width=\linewidth]{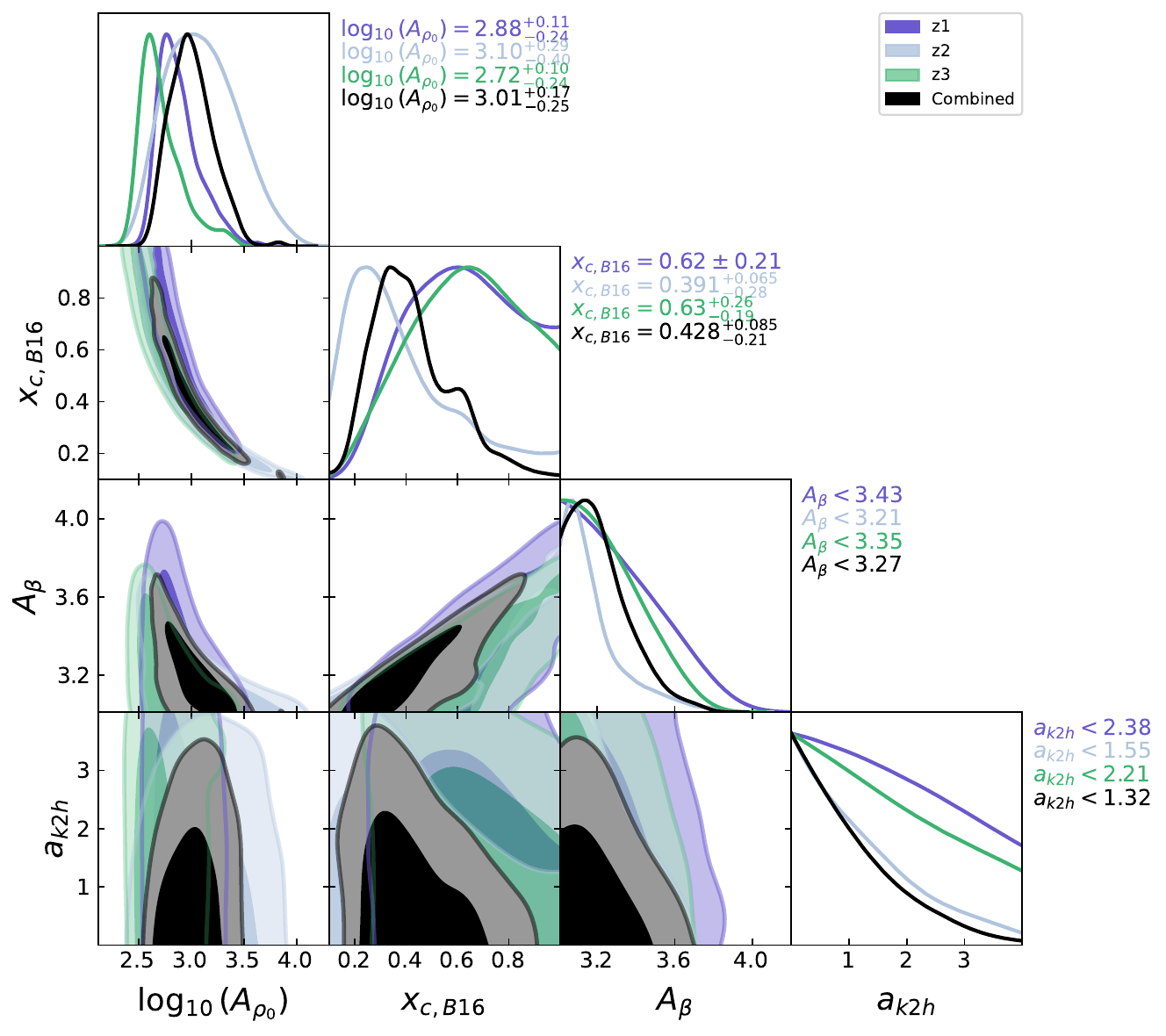}
    \caption{Marginalized posteriors for the four GNFW gas profile parameters in the HOD-marginalized analysis (numerical constraints in Table~\ref{table.marginalize_hod}); the counterpart of Fig.~\ref{fig:Corner}. Purple, light blue, and green contours correspond to the three individual redshift bins $0.4 < z < 0.6$, $0.6 < z < 0.8$, and $0.8 < z < 1.1$, and black contours to the combined sample. Compared with the fixed-HOD contours of Fig.~\ref{fig:Corner}, marginalizing over the HOD broadens the constraints on the amplitude $\log_{10}\rho_0$ and turns the outer slope $\beta_k$ into an upper limit in every bin.}
    \label{fig:marg_corner}
\end{figure}

Figure~\ref{fig:marg_corner} shows the marginalized-HOD posterior contours for the four GNFW gas parameters, the counterpart of Fig.~\ref{fig:Corner} for the fixed-HOD case. The contours broaden relative to the fixed-HOD analysis, most visibly in $\log_{10}\rho_0$ and along the $\log_{10}\rho_0$-$x_{c,k}$ degeneracy direction, and the outer slope $\beta_k$ is only bounded from above in every bin including the combined sample. The parameter medians remain consistent with the fixed-HOD values of Table~\ref{table.prior} within $1\sigma$, confirming that fixing the HOD does not bias the inferred gas profile; it only understates the amplitude uncertainty.

\bibliography{sample631}



\end{document}

%% file: authors.tex
\author{F.~J.~Qu}
\email{jq247@cantab.ac.uk}
\affiliation{Kavli Institute for Particle Astrophysics and Cosmology, Stanford University, 452 Lomita Mall, Stanford, CA, 94305, USA}
\affiliation{Department of Physics, Stanford University, 382 Via Pueblo Mall, Stanford, CA, 94305, USA}
\affiliation{SLAC National Accelerator Laboratory, 2575 Sand Hill Road, Menlo Park, California 94025, USA}

\author{B.~Ried~Guachalla}
\affiliation{Kavli Institute for Particle Astrophysics and Cosmology, Stanford University, 452 Lomita Mall, Stanford, CA, 94305, USA}
\affiliation{Department of Physics, Stanford University, 382 Via Pueblo Mall, Stanford, CA, 94305, USA}
\affiliation{SLAC National Accelerator Laboratory, 2575 Sand Hill Road, Menlo Park, California 94025, USA}

\author{E.~Schaan}
\affiliation{Kavli Institute for Particle Astrophysics and Cosmology, Stanford University, 452 Lomita Mall, Stanford, CA, 94305, USA}
\affiliation{Department of Physics, Stanford University, 382 Via Pueblo Mall, Stanford, CA, 94305, USA}
\affiliation{SLAC National Accelerator Laboratory, 2575 Sand Hill Road, Menlo Park, California 94025, USA}

\author{B.~Hadzhiyska}
\affiliation{Institute of Astronomy, Madingley Road, Cambridge CB3 0HA, UK}
\affiliation{Kavli Institute for Cosmology Cambridge, Madingley Road, Cambridge CB3 0HA, UK}

\author{S.~Ferraro}
\affiliation{Physics Division, Lawrence Berkeley National Laboratory, Berkeley, CA 94720, USA}
\affiliation{Berkeley Center for Cosmological Physics, Department of Physics, University of California, Berkeley, CA 94720, USA}


\author{J.~Aguilar}
\affiliation{Lawrence Berkeley National Laboratory, 1 Cyclotron Road, Berkeley, CA 94720, USA}

\author{S.~Ahlen}
\affiliation{Department of Physics, Boston University, 590 Commonwealth Avenue, Boston, MA 02215 USA}

\author{A.~Baleato Lizancos}
\affiliation{Berkeley Center for Cosmological Physics, Department of Physics, University of California, Berkeley, CA 94720, USA}
\affiliation{Physics Division, Lawrence Berkeley National Laboratory, Berkeley, CA 94720, USA}

\author{D.~Bianchi}
\affiliation{Dipartimento di Fisica ``Aldo Pontremoli'', Universit\`a degli Studi di Milano, Via Celoria 16, I-20133 Milano, Italy}
\affiliation{INAF-Osservatorio Astronomico di Brera, Via Brera 28, 20122 Milano, Italy}

\author{D.~Brooks}
\affiliation{Department of Physics \& Astronomy, University College London, Gower Street, London, WC1E 6BT, UK}

\author{R.~Canning}
\affiliation{Institute of Cosmology and Gravitation, University of Portsmouth, Dennis Sciama Building, Portsmouth, PO1 3FX, UK}

\author{F.~J.~Castander}
\affiliation{Institut d'Estudis Espacials de Catalunya (IEEC), c/ Esteve Terradas 1, Edifici RDIT, Campus PMT-UPC, 08860 Castelldefels, Spain}
\affiliation{Institute of Space Sciences, ICE-CSIC, Campus UAB, Carrer de Can Magrans s/n, 08913 Bellaterra, Barcelona, Spain}

\author{E.~Chaussidon}
\affiliation{Lawrence Berkeley National Laboratory, 1 Cyclotron Road, Berkeley, CA 94720, USA}

\author{T.~Claybaugh}
\affiliation{Lawrence Berkeley National Laboratory, 1 Cyclotron Road, Berkeley, CA 94720, USA}

\author{A.~Cuceu}
\affiliation{Lawrence Berkeley National Laboratory, 1 Cyclotron Road, Berkeley, CA 94720, USA}

\author{A.~de la Macorra}
\affiliation{Instituto de F\'{\i}sica, Universidad Nacional Aut\'{o}noma de M\'{e}xico,  Circuito de la Investigaci\'{o}n Cient\'{\i}fica, Ciudad Universitaria, Cd. de M\'{e}xico  C.~P.~04510,  M\'{e}xico}

\author{B. ~Dey}
\affiliation{Department of Astronomy \& Astrophysics, University of Toronto, Toronto, ON M5S 3H4, Canada}
\affiliation{Department of Physics \& Astronomy and Pittsburgh Particle Physics, Astrophysics, and Cosmology Center (PITT PACC), University of Pittsburgh, 3941 O'Hara Street, Pittsburgh, PA 15260, USA}

\author{P.~Doel}
\affiliation{Department of Physics \& Astronomy, University College London, Gower Street, London, WC1E 6BT, UK}

\author{A.~Font-Ribera}
\affiliation{Instituci\'{o} Catalana de Recerca i Estudis Avan\c{c}ats, Passeig de Llu\'{\i}s Companys, 23, 08010 Barcelona, Spain}
\affiliation{Institut de F\'{i}sica d'Altes Energies (IFAE), The Barcelona Institute of Science and Technology, Edifici Cn, Campus UAB, 08193, Bellaterra (Barcelona), Spain}

\author{J.~E.~Forero-Romero}
\affiliation{Departamento de F\'isica, Universidad de los Andes, Cra. 1 No. 18A-10, Edificio Ip, CP 111711, Bogot\'a, Colombia}
\affiliation{Observatorio Astron\'omico, Universidad de los Andes, Cra. 1 No. 18A-10, Edificio H, CP 111711 Bogot\'a, Colombia}

\author{E.~Gazta\~{n}aga}
\affiliation{Institut d'Estudis Espacials de Catalunya (IEEC), c/ Esteve Terradas 1, Edifici RDIT, Campus PMT-UPC, 08860 Castelldefels, Spain}
\affiliation{Institute of Cosmology and Gravitation, University of Portsmouth, Dennis Sciama Building, Portsmouth, PO1 3FX, UK}
\affiliation{Institute of Space Sciences, ICE-CSIC, Campus UAB, Carrer de Can Magrans s/n, 08913 Bellaterra, Barcelona, Spain}

\author{S. ~{Gontcho A Gontcho}}
\affiliation{University of Virginia, Department of Astronomy, Charlottesville, VA 22904, USA}

\author{G.~Gutierrez}
\affiliation{Fermi National Accelerator Laboratory, PO Box 500, Batavia, IL 60510, USA}

\author{H.~K.~Herrera-Alcantar}
\affiliation{Institut d'Astrophysique de Paris. 98 bis boulevard Arago. 75014 Paris, France}
\affiliation{IRFU, CEA, Universit\'{e} Paris-Saclay, F-91191 Gif-sur-Yvette, France}

\author{K.~Honscheid}
\affiliation{Center for Cosmology and AstroParticle Physics, The Ohio State University, 191 West Woodruff Avenue, Columbus, OH 43210, USA}
\affiliation{Department of Physics, The Ohio State University, 191 West Woodruff Avenue, Columbus, OH 43210, USA}
\affiliation{The Ohio State University, Columbus, 43210 OH, USA}

\author{C.~Howlett}
\affiliation{School of Mathematics and Physics, University of Queensland, Brisbane, QLD 4072, Australia}

\author{D.~Huterer}
\affiliation{Department of Physics, University of Michigan, 450 Church Street, Ann Arbor, MI 48109, USA}
\affiliation{University of Michigan, 500 S. State Street, Ann Arbor, MI 48109, USA}

\author{M.~Ishak}
\affiliation{Department of Physics, The University of Texas at Dallas, 800 W. Campbell Rd., Richardson, TX 75080, USA}

\author{R.~Kehoe}
\affiliation{Department of Physics, Southern Methodist University, 3215 Daniel Avenue, Dallas, TX 75275, USA}

\author{T.~Kisner}
\affiliation{Lawrence Berkeley National Laboratory, 1 Cyclotron Road, Berkeley, CA 94720, USA}

\author{A.~Kremin}
\affiliation{Lawrence Berkeley National Laboratory, 1 Cyclotron Road, Berkeley, CA 94720, USA}

\author{O.~Lahav}
\affiliation{Department of Physics \& Astronomy, University College London, Gower Street, London, WC1E 6BT, UK}

\author{M.~Landriau}
\affiliation{Lawrence Berkeley National Laboratory, 1 Cyclotron Road, Berkeley, CA 94720, USA}

\author{L.~Le~Guillou}
\affiliation{Sorbonne Universit\'{e}, CNRS/IN2P3, Laboratoire de Physique Nucl\'{e}aire et de Hautes Energies (LPNHE), FR-75005 Paris, France}

\author{M.~E.~Levi}
\affiliation{Lawrence Berkeley National Laboratory, 1 Cyclotron Road, Berkeley, CA 94720, USA}

\author{M.~Manera}
\affiliation{Departament de F\'{i}sica, Serra H\'{u}nter, Universitat Aut\`{o}noma de Barcelona, 08193 Bellaterra (Barcelona), Spain}
\affiliation{Institut de F\'{i}sica d'Altes Energies (IFAE), The Barcelona Institute of Science and Technology, Edifici Cn, Campus UAB, 08193, Bellaterra (Barcelona), Spain}

\author{A.~Meisner}
\affiliation{NSF NOIRLab, 950 N. Cherry Ave., Tucson, AZ 85719, USA}

\author{R.~Miquel}
\affiliation{Instituci\'{o} Catalana de Recerca i Estudis Avan\c{c}ats, Passeig de Llu\'{\i}s Companys, 23, 08010 Barcelona, Spain}
\affiliation{Institut de F\'{i}sica d'Altes Energies (IFAE), The Barcelona Institute of Science and Technology, Edifici Cn, Campus UAB, 08193, Bellaterra (Barcelona), Spain}

\author{S.~Nadathur}
\affiliation{Institute of Cosmology and Gravitation, University of Portsmouth, Dennis Sciama Building, Portsmouth, PO1 3FX, UK}

\author{J.~A.~Newman}
\affiliation{Department of Physics \& Astronomy and Pittsburgh Particle Physics, Astrophysics, and Cosmology Center (PITT PACC), University of Pittsburgh, 3941 O'Hara Street, Pittsburgh, PA 15260, USA}

\author{W.~J.~Percival}
\affiliation{Department of Physics and Astronomy, University of Waterloo, 200 University Ave W, Waterloo, ON N2L 3G1, Canada}
\affiliation{Perimeter Institute for Theoretical Physics, 31 Caroline St. North, Waterloo, ON N2L 2Y5, Canada}
\affiliation{Waterloo Centre for Astrophysics, University of Waterloo, 200 University Ave W, Waterloo, ON N2L 3G1, Canada}

\author{I.~P\'erez-R\`afols}
\affiliation{Departament de F\'isica, EEBE, Universitat Polit\`ecnica de Catalunya, c/Eduard Maristany 10, 08930 Barcelona, Spain}

\author{G.~Rossi}
\affiliation{Department of Physics and Astronomy, Sejong University, 209 Neungdong-ro, Gwangjin-gu, Seoul 05006, Republic of Korea}

\author{L.~Samushia}
\affiliation{Abastumani Astrophysical Observatory, Tbilisi, GE-0179, Georgia}
\affiliation{Department of Physics, Kansas State University, 116 Cardwell Hall, Manhattan, KS 66506, USA}
\affiliation{Faculty of Natural Sciences and Medicine, Ilia State University, 0194 Tbilisi, Georgia}

\author{E.~Sanchez}
\affiliation{CIEMAT, Avenida Complutense 40, E-28040 Madrid, Spain}

\author{E.~F.~Schlafly}
\affiliation{Space Telescope Science Institute, 3700 San Martin Drive, Baltimore, MD 21218, USA}

\author{D.~Schlegel}
\affiliation{Lawrence Berkeley National Laboratory, 1 Cyclotron Road, Berkeley, CA 94720, USA}

\author{M.~Schubnell}
\affiliation{Department of Physics, University of Michigan, 450 Church Street, Ann Arbor, MI 48109, USA}
\affiliation{University of Michigan, 500 S. State Street, Ann Arbor, MI 48109, USA}

\author{H.~Seo}
\affiliation{Department of Physics \& Astronomy, Ohio University, 139 University Terrace, Athens, OH 45701, USA}

\author{J.~Silber}
\affiliation{Lawrence Berkeley National Laboratory, 1 Cyclotron Road, Berkeley, CA 94720, USA}

\author{D.~Sprayberry}
\affiliation{NSF NOIRLab, 950 N. Cherry Ave., Tucson, AZ 85719, USA}

\author{G.~Tarl\'{e}}
\affiliation{University of Michigan, 500 S. State Street, Ann Arbor, MI 48109, USA}

\author{B.~A.~Weaver}
\affiliation{NSF NOIRLab, 950 N. Cherry Ave., Tucson, AZ 85719, USA}

\author{R.~Zhou}
\affiliation{Lawrence Berkeley National Laboratory, 1 Cyclotron Road, Berkeley, CA 94720, USA}

%% file: acknowledgements.tex
During the final stages of this work, we became aware of independent parallel efforts by \citep{harscouet2025kszeveryonepseudoclapproach}. 
We thank David Alonso and the authors of that work for useful discussion as both papers were progressing. 
We thank Blake Sherwin, Boris Bolliet, Risa Wechsler, Marcelo Alvarez, Abhishek Maniyar, Zeeshan Ahmed and attendees of the KITP workshop \textit{New Synergies in Multi-Probe Cosmology} at UC Santa Barbara for useful discussions, and Alex Krolewski for helpful suggestions during the internal review process.
This work received support from the U.S. Department of Energy under contract number DE-AC02-76SF00515 to SLAC National Accelerator Laboratory. S.F. is supported by Lawrence Berkeley National Laboratory and the Director, Office of Science, Office of High Energy Physics of the U.S. Department of Energy under Contract No.\ DE-AC02-05CH11231.

Computations were performed on the Niagara supercomputer at the SciNet HPC Consortium. SciNet is funded by Innovation, Science and Economic Development Canada; the Digital Research Alliance of Canada; the Ontario Research Fund: Research Excellence; and the University of Toronto. This research also used resources of the National Energy Research Scientific Computing Center (NERSC), a DOE Office of Science User Facility supported by the Office of Science of the U.S. Department of Energy under Contract No. DE-AC02-05CH11231.

This material is based upon work supported by the U.S. Department of Energy (DOE), Office of Science, Office of High-Energy Physics, under Contract No. DE–AC02–05CH11231, and by the National Energy Research Scientific Computing Center, a DOE Office of Science User Facility under the same contract. Additional support for DESI was provided by the U.S. National Science Foundation (NSF), Division of Astronomical Sciences under Contract No. AST-0950945 to the NSF’s National Optical-Infrared Astronomy Research Laboratory; the Science and Technology Facilities Council of the United Kingdom; the Gordon and Betty Moore Foundation; the Heising-Simons Foundation; the French Alternative Energies and Atomic Energy Commission (CEA); the National Council of Humanities, Science and Technology of Mexico (CONAHCYT); the Ministry of Science, Innovation and Universities of Spain (MICIU/AEI/10.13039/501100011033), and by the DESI Member Institutions: \url{https://www.desi.lbl.gov/collaborating-institutions}. Any opinions, findings, and conclusions or recommendations expressed in this material are those of the author(s) and do not necessarily reflect the views of the U. S. National Science Foundation, the U. S. Department of Energy, or any of the listed funding agencies.

The authors are honored to be permitted to conduct scientific research on I'oligam Du'ag (Kitt Peak), a mountain with particular significance to the Tohono O’odham Nation.

We acknowledge the use of public ACT data products made available through the National Energy Research Scientific Computing Center (NERSC), a U.S. Department of Energy Office of Science User Facility operated under Contract No. DE-AC02-05CH11231.

%% file: sample631.bib
@misc{guachalla2025backlightingextendedgashalos,
      title={Backlighting extended gas halos around luminous red galaxies: kinematic Sunyaev-Zel'dovich effect from DESI Y1 x ACT}, 
      author={Bernardita Ried Guachalla and Emmanuel Schaan and Boryana Hadzhiyska and Simone Ferraro and Jessica N. Aguilar and Steven Ahlen and Nicholas Battaglia and Davide Bianchi and Richard Bond and David Brooks and Todd Claybaugh and William R. Coulton and Axel de la Macorra and Mark J. Devlin and Arjun Dey and Peter Doel and Jo Dunkley and Kevin Fanning and Jaime Forero-Romero and Enrique Gazta{\~n}aga and Satya Gontcho A Gontcho and Gaston Gutierrez and Julien Guy and J. Colin Hill and Klaus Honscheid and Stephanie Juneau and Theodore Kisner and Anthony Kremin and Andrew Lambert and Martin Landriau and Laurent Le Guillou and Niall MacCrann and Marc Manera and Aaron Meisner and Ramon Miquel and Kavilan Moodley and John Moustakas and Tony Mroczkowski and Adam D. Myers and Michael D. Niemack and Gustavo Niz and Nathalie Palanque-Delabrouille and Will Percival and Ignasi P{\'e}rez-R{\`a}fols and Claire Poppett and Francisco Prada and Frank J. Qu and Graziano Rossi and Eusebio Sanchez and David Schlegel and Michael Schubnell and Hee-Jong Seo and Crist{\'o}bal Sif{\'o}n and David N. Spergel and David Sprayberry and Gregory Tarl{\'e} and Mariana Vargas-Maga{\~n}a and Eve M. Vavagiakis and Benjamin A. Weaver and Edward J. Wollack and Pauline Zarrouk},
      year={2025},
      eprint={2503.19870},
      archivePrefix={arXiv},
      primaryClass={astro-ph.GA},
      url={https://arxiv.org/abs/2503.19870}, 
}

@misc{hadzhiyska2025evidencelargebaryonicfeedback,
      title={Evidence for large baryonic feedback at low and intermediate redshifts from kinematic Sunyaev-Zel'dovich observations with ACT and DESI photometric galaxies}, 
      author={B. Hadzhiyska and S. Ferraro and B. Ried Guachalla and E. Schaan and J. Aguilar and N. Battaglia and J. R. Bond and D. Brooks and E. Calabrese and S. K. Choi and T. Claybaugh and W. R. Coulton and K. Dawson and M. Devlin and B. Dey and P. Doel and A. J. Duivenvoorden and J. Dunkley and G. S. Farren and A. Font-Ribera and J. E. Forero-Romero and P. A. Gallardo and E. Gazta{\~n}aga and S. Gontcho Gontcho and M. Gralla and L. Le Guillou and G. Gutierrez and J. Guy and J. C. Hill and R. Hlo{\v{z}}ek and K. Honscheid and S. Juneau and T. Kisner and A. Kremin and M. Landriau and R. H. Liu and T. Louis and N. MacCrann and A. de Macorra and M. Madhavacheril and M. Manera and A. Meisner and R. Miquel and K. Moodley and J. Moustakas and T. Mroczkowski and S. Naess and J. Newman and M. D. Niemack and G. Niz and L. Page and N. Palanque-Delabrouille and B. Partridge and W. J. Percival and F. Prada and F. J. Qu and G. Rossi and E. Sanchez and D. Schlegel and M. Schubnell and N. Sehgal and H. Seo and C. Sif{\'o}n and D. Spergel and D. Sprayberry and S. Staggs and G. Tarl{\'e} and C. Vargas and E. M. Vavagiakis and B. A. Weaver and E. J. Wollack and R. Zhou and H. Zou},
      year={2025},
      eprint={2407.07152},
      archivePrefix={arXiv},
      primaryClass={astro-ph.CO},
      url={https://arxiv.org/abs/2407.07152}, 
}

@article{PhysRevLett.88.211301,
  title = {Nonlinear Kinetic Sunyaev-Zeldovich Effect},
  author = {Ma, Chung-Pei and Fry, J. N.},
  journal = {Phys. Rev. Lett.},
  volume = {88},
  issue = {21},
  pages = {211301},
  numpages = {4},
  year = {2002},
  month = {May},
  publisher = {American Physical Society},
  doi = {10.1103/PhysRevLett.88.211301},
  url = {https://link.aps.org/doi/10.1103/PhysRevLett.88.211301}
}

@article{PhysRevD.109.103534,
  title = {Velocity reconstruction in the era of DESI and Rubin/LSST. II. Realistic samples on the light cone},
  author = {Hadzhiyska, Boryana and Ferraro, Simone and Ried Guachalla, Bernardita and Schaan, Emmanuel},
  journal = {Phys. Rev. D},
  volume = {109},
  issue = {10},
  pages = {103534},
  numpages = {17},
  year = {2024},
  month = {May},
  publisher = {American Physical Society},
  doi = {10.1103/PhysRevD.109.103534},
  url = {https://link.aps.org/doi/10.1103/PhysRevD.109.103534}
}

@article{McQuinn_2005,
   title={The Kinetic SunyaevZeldovich Effect from Reionization},
   volume={630},
   ISSN={1538-4357},
   url={http://dx.doi.org/10.1086/432049},
   DOI={10.1086/432049},
   number={2},
   journal={The Astrophysical Journal},
   publisher={American Astronomical Society},
   author={McQuinn, Matthew and Furlanetto, Steven R. and Hernquist, Lars and Zahn, Oliver and Zaldarriaga, Matias},
   year={2005},
   month=sep, pages={643--656} }

@article{Park_2016,
   title={THE IMPACT OF NONLINEAR STRUCTURE FORMATION ON THE POWER SPECTRUM OF TRANSVERSE MOMENTUM FLUCTUATIONS AND THE KINETIC SUNYAEV--ZELDOVICH EFFECT},
   volume={818},
   ISSN={1538-4357},
   url={http://dx.doi.org/10.3847/0004-637X/818/1/37},
   DOI={10.3847/0004-637x/818/1/37},
   number={1},
   journal={The Astrophysical Journal},
   publisher={American Astronomical Society},
   author={Park, Hyunbae and Komatsu, Eiichiro and Shapiro, Paul R. and Koda, Jun and Mao, Yi},
   year={2016},
   month=feb, pages={37} }

@article{Alonso_2019,
   title={A unified pseudo-C framework},
   volume={484},
   ISSN={1365-2966},
   url={http://dx.doi.org/10.1093/mnras/stz093},
   DOI={10.1093/mnras/stz093},
   number={3},
   journal={Monthly Notices of the Royal Astronomical Society},
   publisher={Oxford University Press (OUP)},
   author={Alonso, David and Sanchez, Javier and Slosar, An{\v{z}}e},
   year={2019},
   month=jan, pages={4127--4151} }

@article{Torrado_2021,
   title={Cobaya: code for Bayesian analysis of hierarchical physical models},
   volume={2021},
   ISSN={1475-7516},
   url={http://dx.doi.org/10.1088/1475-7516/2021/05/057},
   DOI={10.1088/1475-7516/2021/05/057},
   number={05},
   journal={Journal of Cosmology and Astroparticle Physics},
   publisher={IOP Publishing},
   author={Torrado, Jes{\'u}s and Lewis, Antony},
   year={2021},
   month=may, pages={057} }

@article{Bolliet_2023,
   title={Projected-field kinetic Sunyaev-Zeldovich Cross-correlations: halo model and forecasts},
   volume={2023},
   ISSN={1475-7516},
   url={http://dx.doi.org/10.1088/1475-7516/2023/03/039},
   DOI={10.1088/1475-7516/2023/03/039},
   number={03},
   journal={Journal of Cosmology and Astroparticle Physics},
   publisher={IOP Publishing},
   author={Bolliet, Boris and Colin Hill, J. and Ferraro, Simone and Kusiak, Aleksandra and Krolewski, Alex},
   year={2023},
   month=mar, pages={039} }

@article{Bolliet:2023eob,
    author = "Bolliet, B. and others",
    title = "{class\_sz I: Overview}",
    eprint = "2310.18482",
    archivePrefix = "arXiv",
    primaryClass = "astro-ph.IM",
    doi = "10.1051/epjconf/202429300008",
    journal = "EPJ Web Conf.",
    volume = "293",
    pages = "00008",
    year = "2024"
}

@article{Hartlap_2006,
   title={Why your model parameter confidences might be too optimistic.  Unbiased estimation of the inverse covariance matrix},
   volume={464},
   ISSN={1432-0746},
   url={http://dx.doi.org/10.1051/0004-6361:20066170},
   DOI={10.1051/0004-6361:20066170},
   number={1},
   journal={Astronomy and Astrophysics},
   publisher={EDP Sciences},
   author={Hartlap, J. and Simon, P. and Schneider, P.},
   year={2006},
   month=dec, pages={399--404} }

@article{Planck:2018vyg,
    author = "Aghanim, N. and others",
    collaboration = "Planck",
    title = "{Planck 2018 results. VI. Cosmological parameters}",
    eprint = "1807.06209",
    archivePrefix = "arXiv",
    primaryClass = "astro-ph.CO",
    doi = "10.1051/0004-6361/201833910",
    journal = "Astron. Astrophys.",
    volume = "641",
    pages = "A6",
    year = "2020",
    note = "[Erratum: Astron.Astrophys. 652, C4 (2021)]"
}

@misc{yuan2023desionepercentsurveyexploring,
      title={The DESI One-Percent Survey: Exploring the Halo Occupation Distribution of Luminous Red Galaxies and Quasi-Stellar Objects with AbacusSummit}, 
      author={Sihan Yuan and Hanyu Zhang and Ashley J. Ross and Jamie Donald-McCann and Boryana Hadzhiyska and Risa H. Wechsler and Zheng Zheng and Shadab Alam and Violeta Gonzalez-Perez and Jessica Nicole Aguilar and Steven Ahlen and Davide Bianchi and David Brooks and Axel de la Macorra and Kevin Fanning and Jaime E. Forero-Romero and Klaus Honscheid and Mustapha Ishak and Robert Kehoe and James Lasker and Martin Landriau and Marc Manera and Paul Martini and Aaron Meisner and Ramon Miquel and John Moustakas and Seshadri Nadathur and Jeffrey A. Newman and Jundan Nie and Will Percival and Claire Poppett and Antoine Rocher and Graziano Rossi and Eusebio Sanchez and Lado Samushia and Michael Schubnell and Hee-Jong Seo and Gregory Tarle and Benjamin Alan Weaver and Jiaxi Yu and Zhimin Zhou and Hu Zou},
      year={2023},
      eprint={2306.06314},
      archivePrefix={arXiv},
      primaryClass={astro-ph.CO},
      url={https://arxiv.org/abs/2306.06314}, 
}

@article{Bundy_2015,
doi = {10.1088/0067-0049/221/1/15},
url = {https://dx.doi.org/10.1088/0067-0049/221/1/15},
year = {2015},
month = {nov},
publisher = {The American Astronomical Society},
volume = {221},
number = {1},
pages = {15},
author = {Bundy, Kevin and Leauthaud, Alexie and Saito, Shun and Bolton, Adam and Lin, Yen-Ting and Maraston, Claudia and Nichol, Robert C. and Schneider, Donald P. and Thomas, Daniel and Wake, David A.},
title = {THE STRIPE 82 MASSIVE GALAXY PROJECT. I. CATALOG CONSTRUCTION},
journal = {The Astrophysical Journal Supplement Series},
}

@article{Zhou_2023,
doi = {10.3847/1538-3881/aca5fb},
url = {https://dx.doi.org/10.3847/1538-3881/aca5fb},
year = {2023},
month = {jan},
publisher = {The American Astronomical Society},
volume = {165},
number = {2},
pages = {58},
author = {Zhou, Rongpu and Dey, Biprateep and Newman, Jeffrey A. and Eisenstein, Daniel J. and Dawson, K. and Bailey, S. and Berti, A. and Guy, J. and Lan, Ting-Wen and Zou, H. and Aguilar, J. and Ahlen, S. and Alam, Shadab and Brooks, D. and de la Macorra, A. and Dey, A. and Dhungana, G. and Fanning, K. and Font-Ribera, A. and Gontcho, S. Gontcho A. and Honscheid, K. and Ishak, Mustapha and Kisner, T. and Kov{\'a}cs, A. and Kremin, A. and Landriau, M. and Levi, Michael E. and Magneville, C. and Manera, Marc and Martini, P. and Meisner, Aaron M. and Miquel, R. and Moustakas, J. and Myers, Adam D. and Nie, Jundan and Palanque-Delabrouille, N. and Percival, W. J. and Poppett, C. and Prada, F. and Raichoor, A. and Ross, A. J. and Schlafly, E. and Schlegel, D. and Schubnell, M. and Tarl{\'e}, Gregory and Weaver, B. A. and Wechsler, R. H. and Y{\'e}che, Christophe and Zhou, Zhimin},
title = {Target Selection and Validation of DESI Luminous Red Galaxies},
journal = {The Astronomical Journal}
}

@article{Duffy_2008,
   title={Dark matter halo concentrations in the Wilkinson Microwave Anisotropy Probe year 5 cosmology},
   volume={390},
   ISSN={1745-3925},
   url={http://dx.doi.org/10.1111/j.1745-3933.2008.00537.x},
   DOI={10.1111/j.1745-3933.2008.00537.x},
   number={1},
   journal={Monthly Notices of the Royal Astronomical Society: Letters},
   publisher={Oxford University Press (OUP)},
   author={Duffy, Alan R. and Schaye, Joop and Kay, Scott T. and Dalla Vecchia, Claudio},
   year={2008},
   month=oct, pages={L64--L68} }

@article{Diemer_2018,
   title={COLOSSUS: A Python Toolkit for Cosmology, Large-scale Structure, and Dark Matter Halos},
   volume={239},
   ISSN={1538-4365},
   url={http://dx.doi.org/10.3847/1538-4365/aaee8c},
   DOI={10.3847/1538-4365/aaee8c},
   number={2},
   journal={The Astrophysical Journal Supplement Series},
   publisher={American Astronomical Society},
   author={Diemer, Benedikt},
   year={2018},
   month=dec, pages={35} }

@article{Battaglia_2016,
   title={The tau of galaxy clusters},
   volume={2016},
   ISSN={1475-7516},
   url={http://dx.doi.org/10.1088/1475-7516/2016/08/058},
   DOI={10.1088/1475-7516/2016/08/058},
   number={08},
   journal={Journal of Cosmology and Astroparticle Physics},
   publisher={IOP Publishing},
   author={Battaglia, N.},
   year={2016},
   month=aug, pages={058--058} }

@article{PhysRevD.103.063514,
  title = {Atacama Cosmology Telescope: Modeling the gas thermodynamics in BOSS CMASS galaxies from kinematic and thermal Sunyaev-Zel'dovich measurements},
  author = {Amodeo, Stefania and Battaglia, Nicholas and Schaan, Emmanuel and Ferraro, Simone and Moser, Emily and Aiola, Simone and Austermann, Jason E. and Beall, James A. and Bean, Rachel and Becker, Daniel T. and Bond, Richard J. and Calabrese, Erminia and Calafut, Victoria and Choi, Steve K. and Denison, Edward V. and Devlin, Mark and Duff, Shannon M. and Duivenvoorden, Adriaan J. and Dunkley, Jo and D\"unner, Rolando and Gallardo, Patricio A. and Hall, Kirsten R. and Han, Dongwon and Hill, J. Colin and Hilton, Gene C. and Hilton, Matt and Hlo\ifmmode \check{z}\else \v{z}\fi{}ek, Ren\'ee and Hubmayr, Johannes and Huffenberger, Kevin M. and Hughes, John P. and Koopman, Brian J. and MacInnis, Amanda and McMahon, Jeff and Madhavacheril, Mathew S. and Moodley, Kavilan and Mroczkowski, Tony and Naess, Sigurd and Nati, Federico and Newburgh, Laura B. and Niemack, Michael D. and Page, Lyman A. and Partridge, Bruce and Schillaci, Alessandro and Sehgal, Neelima and Sif\'on, Crist\'obal and Spergel, David N. and Staggs, Suzanne and Storer, Emilie R. and Ullom, Joel N. and Vale, Leila R. and van Engelen, Alexander and Van Lanen, Jeff and Vavagiakis, Eve M. and Wollack, Edward J. and Xu, Zhilei},
  journal = {Phys. Rev. D},
  volume = {103},
  issue = {6},
  pages = {063514},
  numpages = {23},
  year = {2021},
  month = {Mar},
  publisher = {American Physical Society},
  doi = {10.1103/PhysRevD.103.063514},
  url = {https://link.aps.org/doi/10.1103/PhysRevD.103.063514}
}

@article{PhysRevD.109.103533,
  title = {Velocity reconstruction in the era of DESI and Rubin/LSST. I. Exploring spectroscopic, photometric, and hybrid samples},
  author = {Ried Guachalla, Bernardita and Schaan, Emmanuel and Hadzhiyska, Boryana and Ferraro, Simone},
  journal = {Phys. Rev. D},
  volume = {109},
  issue = {10},
  pages = {103533},
  numpages = {20},
  year = {2024},
  month = {May},
  publisher = {American Physical Society},
  doi = {10.1103/PhysRevD.109.103533},
  url = {https://link.aps.org/doi/10.1103/PhysRevD.109.103533}
}

@misc{lsstdarkenergysciencecollaboration2012largesynopticsurveytelescope,
      title={Large Synoptic Survey Telescope: Dark Energy Science Collaboration}, 
      author={LSST Dark Energy Science Collaboration},
      year={2012},
      eprint={1211.0310},
      archivePrefix={arXiv},
      primaryClass={astro-ph.CO},
      url={https://arxiv.org/abs/1211.0310}, 
}

@article{Ivezi__2019,
   title={LSST: From Science Drivers to Reference Design and Anticipated Data Products},
   volume={873},
   ISSN={1538-4357},
   url={http://dx.doi.org/10.3847/1538-4357/ab042c},
   DOI={10.3847/1538-4357/ab042c},
   number={2},
   journal={The Astrophysical Journal},
   publisher={American Astronomical Society},
   author={Ivezic, Zeljko and Kahn, Steven M. and Tyson, J. Anthony and Abel, Bob and Acosta, Emily and Allsman, Robyn and Alonso, David and AlSayyad, Yusra and Anderson, Scott F. and Andrew, John and P. Angel, James Roger and Angeli, George Z. and Ansari, Reza and Antilogus, Pierre and Araujo, Constanza and Armstrong, Robert and Arndt, Kirk T. and Astier, Pierre and Aubourg, {\'E}ric and Auza, Nicole and Axelrod, Tim S. and Bard, Deborah J. and Barr, Jeff D. and Barrau, Aurelian and Bartlett, James G. and Bauer, Amanda E. and Bauman, Brian J. and Baumont, Sylvain and Bechtol, Ellen and Bechtol, Keith and Becker, Andrew C. and Becla, Jacek and Beldica, Cristina and Bellavia, Steve and Bianco, Federica B. and Biswas, Rahul and Blanc, Guillaume and Blazek, Jonathan and Blandford, Roger D. and Bloom, Josh S. and Bogart, Joanne and Bond, Tim W. and Booth, Michael T. and Borgland, Anders W. and Borne, Kirk and Bosch, James F. and Boutigny, Dominique and Brackett, Craig A. and Bradshaw, Andrew and Brandt, William Nielsen and Brown, Michael E. and Bullock, James S. and Burchat, Patricia and Burke, David L. and Cagnoli, Gianpietro and Calabrese, Daniel and Callahan, Shawn and Callen, Alice L. and Carlin, Jeffrey L. and Carlson, Erin L. and Chandrasekharan, Srinivasan and Charles-Emerson, Glenaver and Chesley, Steve and Cheu, Elliott C. and Chiang, Hsin-Fang and Chiang, James and Chirino, Carol and Chow, Derek and Ciardi, David R. and Claver, Charles F. and Cohen-Tanugi, Johann and Cockrum, Joseph J. and Coles, Rebecca and Connolly, Andrew J. and Cook, Kem H. and Cooray, Asantha and Covey, Kevin R. and Cribbs, Chris and Cui, Wei and Cutri, Roc and Daly, Philip N. and Daniel, Scott F. and Daruich, Felipe and Daubard, Guillaume and Daues, Greg and Dawson, William and Delgado, Francisco and Dellapenna, Alfred and Peyster, Robert de and Val-Borro, Miguel de and Digel, Seth W. and Doherty, Peter and Dubois, Richard and Dubois-Felsmann, Gregory P. and Durech, Josef and Economou, Frossie and Eifler, Tim and Eracleous, Michael and Emmons, Benjamin L. and Neto, Angelo Fausti and Ferguson, Henry and Figueroa, Enrique and Fisher-Levine, Merlin and Focke, Warren and Foss, Michael D. and Frank, James and Freemon, Michael D. and Gangler, Emmanuel and Gawiser, Eric and Geary, John C. and Gee, Perry and Geha, Marla and Gessner, Charles J. B. and Gibson, Robert R. and Gilmore, D. Kirk and Glanzman, Thomas and Glick, William and Goldina, Tatiana and Goldstein, Daniel A. and Goodenow, Iain and Graham, Melissa L. and Gressler, William J. and Gris, Philippe and Guy, Leanne P. and Guyonnet, Augustin and Haller, Gunther and Harris, Ron and Hascall, Patrick A. and Haupt, Justine and Hernandez, Fabio and Herrmann, Sven and Hileman, Edward and Hoblitt, Joshua and Hodgson, John A. and Hogan, Craig and Howard, James D. and Huang, Dajun and Huffer, Michael E. and Ingraham, Patrick and Innes, Walter R. and Jacoby, Suzanne H. and Jain, Bhuvnesh and Jammes, Fabrice and Jee, M. James and Jenness, Tim and Jernigan, Garrett and Jevremovi, Darko and Johns, Kenneth and Johnson, Anthony S. and Johnson, Margaret W. G. and Jones, R. Lynne and Juramy-Gilles, Claire and Juri, Mario and Kalirai, Jason S. and Kallivayalil, Nitya J. and Kalmbach, Bryce and Kantor, Jeffrey P. and Karst, Pierre and Kasliwal, Mansi M. and Kelly, Heather and Kessler, Richard and Kinnison, Veronica and Kirkby, David and Knox, Lloyd and Kotov, Ivan V. and Krabbendam, Victor L. and Krughoff, K. Simon and Kub{\'a}nek, Petr and Kuczewski, John and Kulkarni, Shri and Ku, John and Kurita, Nadine R. and Lage, Craig S. and Lambert, Ron and Lange, Travis and Langton, J. Brian and Guillou, Laurent Le and Levine, Deborah and Liang, Ming and Lim, Kian-Tat and Lintott, Chris J. and Long, Kevin E. and Lopez, Margaux and Lotz, Paul J. and Lupton, Robert H. and Lust, Nate B. and MacArthur, Lauren A. and Mahabal, Ashish and Mandelbaum, Rachel and Markiewicz, Thomas W. and Marsh, Darren S. and Marshall, Philip J. and Marshall, Stuart and May, Morgan and McKercher, Robert and McQueen, Michelle and Meyers, Joshua and Migliore, Myriam and Miller, Michelle and Mills, David J. and Miraval, Connor and Moeyens, Joachim and Moolekamp, Fred E. and Monet, David G. and Moniez, Marc and Monkewitz, Serge and Montgomery, Christopher and Morrison, Christopher B. and Mueller, Fritz and Muller, Gary P. and Arancibia, Freddy Mu{\~n}oz and Neill, Douglas R. and Newbry, Scott P. and Nief, Jean-Yves and Nomerotski, Andrei and Nordby, Martin and OConnor, Paul and Oliver, John and Olivier, Scot S. and Olsen, Knut and OMullane, William and Ortiz, Sandra and Osier, Shawn and Owen, Russell E. and Pain, Reynald and Palecek, Paul E. and Parejko, John K. and Parsons, James B. and Pease, Nathan M. and Peterson, J. Matt and Peterson, John R. and Petravick, Donald L. and Petrick, M. E. Libby and Petry, Cathy E. and Pierfederici, Francesco and Pietrowicz, Stephen and Pike, Rob and Pinto, Philip A. and Plante, Raymond and Plate, Stephen and Plutchak, Joel P. and Price, Paul A. and Prouza, Michael and Radeka, Veljko and Rajagopal, Jayadev and Rasmussen, Andrew P. and Regnault, Nicolas and Reil, Kevin A. and Reiss, David J. and Reuter, Michael A. and Ridgway, Stephen T. and Riot, Vincent J. and Ritz, Steve and Robinson, Sean and Roby, William and Roodman, Aaron and Rosing, Wayne and Roucelle, Cecille and Rumore, Matthew R. and Russo, Stefano and Saha, Abhijit and Sassolas, Benoit and Schalk, Terry L. and Schellart, Pim and Schindler, Rafe H. and Schmidt, Samuel and Schneider, Donald P. and Schneider, Michael D. and Schoening, William and Schumacher, German and Schwamb, Megan E. and Sebag, Jacques and Selvy, Brian and Sembroski, Glenn H. and Seppala, Lynn G. and Serio, Andrew and Serrano, Eduardo and Shaw, Richard A. and Shipsey, Ian and Sick, Jonathan and Silvestri, Nicole and Slater, Colin T. and Smith, J. Allyn and Smith, R. Chris and Sobhani, Shahram and Soldahl, Christine and Storrie-Lombardi, Lisa and Stover, Edward and Strauss, Michael A. and Street, Rachel A. and Stubbs, Christopher W. and Sullivan, Ian S. and Sweeney, Donald and Swinbank, John D. and Szalay, Alexander and Takacs, Peter and Tether, Stephen A. and Thaler, Jon J. and Thayer, John Gregg and Thomas, Sandrine and Thornton, Adam J. and Thukral, Vaikunth and Tice, Jeffrey and Trilling, David E. and Turri, Max and Berg, Richard Van and Berk, Daniel Vanden and Vetter, Kurt and Virieux, Francoise and Vucina, Tomislav and Wahl, William and Walkowicz, Lucianne and Walsh, Brian and Walter, Christopher W. and Wang, Daniel L. and Wang, Shin-Yawn and Warner, Michael and Wiecha, Oliver and Willman, Beth and Winters, Scott E. and Wittman, David and Wolff, Sidney C. and Wood-Vasey, W. Michael and Wu, Xiuqin and Xin, Bo and Yoachim, Peter and Zhan, Hu},
   year={2019},
   month=mar, pages={111} }

@ARTICLE{2022A&A...662A.112E,
       author = {{Euclid Collaboration} and others},
        title = "{Euclid preparation. I. The Euclid Wide Survey}",
      journal = {\aap},
         year = 2022,
        month = jun,
       volume = {662},
          eid = {A112},
        pages = {A112},
          doi = {10.1051/0004-6361/202141938},
archivePrefix = {arXiv},
       eprint = {2108.01201},
 primaryClass = {astro-ph.CO},
       adsurl = {https://ui.adsabs.harvard.edu/abs/2022A&A...662A.112E}
}

@misc{spergel2015widefieldinfrarredsurveytelescopeastrophysics,
      title={Wide-Field InfrarRed Survey Telescope-Astrophysics Focused Telescope Assets WFIRST-AFTA 2015 Report}, 
      author={D. Spergel and N. Gehrels and C. Baltay and D. Bennett and J. Breckinridge and M. Donahue and A. Dressler and B. S. Gaudi and T. Greene and O. Guyon and C. Hirata and J. Kalirai and N. J. Kasdin and B. Macintosh and W. Moos and S. Perlmutter and M. Postman and B. Rauscher and J. Rhodes and Y. Wang and D. Weinberg and D. Benford and M. Hudson and W. -S. Jeong and Y. Mellier and W. Traub and T. Yamada and P. Capak and J. Colbert and D. Masters and M. Penny and D. Savransky and D. Stern and N. Zimmerman and R. Barry and L. Bartusek and K. Carpenter and E. Cheng and D. Content and F. Dekens and R. Demers and K. Grady and C. Jackson and G. Kuan and J. Kruk and M. Melton and B. Nemati and B. Parvin and I. Poberezhskiy and C. Peddie and J. Ruffa and J. K. Wallace and A. Whipple and E. Wollack and F. Zhao},
      year={2015},
      eprint={1503.03757},
      archivePrefix={arXiv},
      primaryClass={astro-ph.IM},
      url={https://arxiv.org/abs/1503.03757}, 
}

@ARTICLE{2006ApJ...650..560C,
       author = {{Cen}, Renyue and {Ostriker}, Jeremiah P.},
        title = "{Where Are the Baryons? II. Feedback Effects}",
      journal = {\apj},
         year = 2006,
        month = oct,
       volume = {650},
       number = {2},
        pages = {560-572},
          doi = {10.1086/506505},
archivePrefix = {arXiv},
       eprint = {astro-ph/0601008},
 primaryClass = {astro-ph},
       adsurl = {https://ui.adsabs.harvard.edu/abs/2006ApJ...650..560C}
}

@article{Birkinshaw_1999,
   title={The Sunyaev--Zeldovich effect},
   volume={310},
   ISSN={0370-1573},
   url={http://dx.doi.org/10.1016/S0370-1573(98)00080-5},
   DOI={10.1016/s0370-1573(98)00080-5},
   number={2--3},
   journal={Physics Reports},
   publisher={Elsevier BV},
   author={Birkinshaw, M},
   year={1999},
   month=mar, pages={97--195} }

@article{Mroczkowski_2019,
   title={Astrophysics with the Spatially and Spectrally Resolved Sunyaev-Zeldovich Effects: A Millimetre/Submillimetre Probe of the Warm and Hot Universe},
   volume={215},
   ISSN={1572-9672},
   url={http://dx.doi.org/10.1007/s11214-019-0581-2},
   DOI={10.1007/s11214-019-0581-2},
   number={1},
   journal={Space Science Reviews},
   publisher={Springer Science and Business Media LLC},
   author={Mroczkowski, Tony and Nagai, Daisuke and Basu, Kaustuv and Chluba, Jens and Sayers, Jack and Adam, R{\'e}mi and Churazov, Eugene and Crites, Abigail and Di Mascolo, Luca and Eckert, Dominique and Macias-Perez, Juan and Mayet, Fr{\'e}d{\'e}ric and Perotto, Laurence and Pointecouteau, Etienne and Romero, Charles and Ruppin, Florian and Scannapieco, Evan and ZuHone, John},
   year={2019},
   month=feb }

@article{PhysRevLett.109.041101,
  title = {Evidence of Galaxy Cluster Motions with the Kinematic Sunyaev-Zel'dovich Effect},
  author = {Hand, Nick and Addison, Graeme E. and Aubourg, Eric and Battaglia, Nick and Battistelli, Elia S. and Bizyaev, Dmitry and Bond, J. Richard and Brewington, Howard and Brinkmann, Jon and Brown, Benjamin R. and Das, Sudeep and Dawson, Kyle S. and Devlin, Mark J. and Dunkley, Joanna and Dunner, Rolando and Eisenstein, Daniel J. and Fowler, Joseph W. and Gralla, Megan B. and Hajian, Amir and Halpern, Mark and Hilton, Matt and Hincks, Adam D. and Hlozek, Ren\'ee and Hughes, John P. and Infante, Leopoldo and Irwin, Kent D. and Kosowsky, Arthur and Lin, Yen-Ting and Malanushenko, Elena and Malanushenko, Viktor and Marriage, Tobias A. and Marsden, Danica and Menanteau, Felipe and Moodley, Kavilan and Niemack, Michael D. and Nolta, Michael R. and Oravetz, Daniel and Page, Lyman A. and Palanque-Delabrouille, Nathalie and Pan, Kaike and Reese, Erik D. and Schlegel, David J. and Schneider, Donald P. and Sehgal, Neelima and Shelden, Alaina and Sievers, Jon and Sif\'on, Crist\'obal and Simmons, Audrey and Snedden, Stephanie and Spergel, David N. and Staggs, Suzanne T. and Swetz, Daniel S. and Switzer, Eric R. and Trac, Hy and Weaver, Benjamin A. and Wollack, Edward J. and Yeche, Christophe and Zunckel, Caroline},
  journal = {Phys. Rev. Lett.},
  volume = {109},
  issue = {4},
  pages = {041101},
  numpages = {6},
  year = {2012},
  month = {Jul},
  publisher = {American Physical Society},
  doi = {10.1103/PhysRevLett.109.041101},
  url = {https://link.aps.org/doi/10.1103/PhysRevLett.109.041101}
}

@article{PhysRevLett.115.191301,
  title = {Evidence of the Missing Baryons from the Kinematic Sunyaev-Zeldovich Effect in Planck Data},
  author = {Hern\'andez-Monteagudo, Carlos and Ma, Yin-Zhe and Kitaura, Francisco S. and Wang, Wenting and G\'enova-Santos, Ricardo and Mac\'{\i}as-P\'erez, Juan and Herranz, Diego},
  journal = {Phys. Rev. Lett.},
  volume = {115},
  issue = {19},
  pages = {191301},
  numpages = {5},
  year = {2015},
  month = {Nov},
  publisher = {American Physical Society},
  doi = {10.1103/PhysRevLett.115.191301},
  url = {https://link.aps.org/doi/10.1103/PhysRevLett.115.191301}
}

@article{Bernardis_2017,
   title={Detection of the pairwise kinematic Sunyaev-Zeldovich effect with BOSS DR11 and the Atacama Cosmology Telescope},
   volume={2017},
   ISSN={1475-7516},
   url={http://dx.doi.org/10.1088/1475-7516/2017/03/008},
   DOI={10.1088/1475-7516/2017/03/008},
   number={03},
   journal={Journal of Cosmology and Astroparticle Physics},
   publisher={IOP Publishing},
   author={Bernardis, F. De and Aiola, S. and Vavagiakis, E.M. and Battaglia, N. and Niemack, M.D. and Beall, J. and Becker, D.T. and Bond, J.R. and Calabrese, E. and Cho, H. and Coughlin, K. and Datta, R. and Devlin, M. and Dunkley, J. and Dunner, R. and Ferraro, S. and Fox, A. and Gallardo, P.A. and Halpern, M. and Hand, N. and Hasselfield, M. and Henderson, S.W. and Hill, J.C. and Hilton, G.C. and Hilton, M. and Hincks, A.D. and Hlozek, R. and Hubmayr, J. and Huffenberger, K. and Hughes, J.P. and Irwin, K.D. and Koopman, B.J. and Kosowsky, A. and Li, D. and Louis, T. and Lungu, M. and Madhavacheril, M.S. and Maurin, L. and McMahon, J. and Moodley, K. and Naess, S. and Nati, F. and Newburgh, L. and Nibarger, J.P. and Page, L.A. and Partridge, B. and Schaan, E. and Schmitt, B. L. and Sehgal, N. and Sievers, J. and Simon, S.M. and Spergel, D.N. and Staggs, S.T. and Stevens, J.R. and Thornton, R.J. and Engelen, A. van and Lanen, J. Van and Wollack, E.J.},
   year={2017},
   month=mar, pages={008--008} }

@article{Soergel_2016,
   title={Detection of the kinematic Sunyaev--Zeldovich effect with DES Year 1 and SPT},
   volume={461},
   ISSN={1365-2966},
   url={http://dx.doi.org/10.1093/mnras/stw1455},
   DOI={10.1093/mnras/stw1455},
   number={3},
   journal={Monthly Notices of the Royal Astronomical Society},
   publisher={Oxford University Press (OUP)},
   author={Soergel, B. and Flender, S. and Story, K. T. and Bleem, L. and Giannantonio, T. and Efstathiou, G. and Rykoff, E. and Benson, B. A. and Crawford, T. and Dodelson, S. and Habib, S. and Heitmann, K. and Holder, G. and Jain, B. and Rozo, E. and Saro, A. and Weller, J. and Abdalla, F. B. and Allam, S. and Annis, J. and Armstrong, R. and Benoit-L{\'e}vy, A. and Bernstein, G. M. and Carlstrom, J. E. and Carnero Rosell, A. and Carrasco Kind, M. and Castander, F. J. and Chiu, I. and Chown, R. and Crocce, M. and Cunha, C. E. and DAndrea, C. B. and da Costa, L. N. and de Haan, T. and Desai, S. and Diehl, H. T. and Dietrich, J. P. and Doel, P. and Estrada, J. and Evrard, A. E. and Flaugher, B. and Fosalba, P. and Frieman, J. and Gaztanaga, E. and Gruen, D. and Gruendl, R. A. and Holzapfel, W. L. and Honscheid, K. and James, D. J. and Keisler, R. and Kuehn, K. and Kuropatkin, N. and Lahav, O. and Lima, M. and Marshall, J. L. and McDonald, M. and Melchior, P. and Miller, C. J. and Miquel, R. and Nord, B. and Ogando, R. and Omori, Y. and Plazas, A. A. and Rapetti, D. and Reichardt, C. L. and Romer, A. K. and Roodman, A. and Saliwanchik, B. R. and Sanchez, E. and Schubnell, M. and Sevilla-Noarbe, I. and Sheldon, E. and Smith, R. C. and Soares-Santos, M. and Sobreira, F. and Stark, A. and Suchyta, E. and Swanson, M. E. C. and Tarle, G. and Thomas, D. and Vieira, J. D. and Walker, A. R. and Whitehorn, N.},
   year={2016},
   month=jun, pages={3172--3193} }

@article{Sugiyama_2018,
   title={A direct measure of free electron gas via the kinematic Sunyaev--Zeldovich effect in Fourier-space analysis},
   volume={475},
   ISSN={1365-2966},
   url={http://dx.doi.org/10.1093/mnras/stx3362},
   DOI={10.1093/mnras/stx3362},
   number={3},
   journal={Monthly Notices of the Royal Astronomical Society},
   publisher={Oxford University Press (OUP)},
   author={Sugiyama, Naonori S and Okumura, Teppei and Spergel, David N},
   year={2018},
   month=jan, pages={3764--3785} }

@article{PhysRevD.104.043502,
  title = {The Atacama Cosmology Telescope: Detection of the pairwise kinematic Sunyaev-Zel'dovich effect with SDSS DR15 galaxies},
  author = {Calafut, V. and Gallardo, P. A. and Vavagiakis, E. M. and Amodeo, S. and Aiola, S. and Austermann, J. E. and Battaglia, N. and Battistelli, E. S. and Beall, J. A. and Bean, R. and Bond, J. R. and Calabrese, E. and Choi, S. K. and Cothard, N. F. and Devlin, M. J. and Duell, C. J. and Duff, S. M. and Duivenvoorden, A. J. and Dunkley, J. and Dunner, R. and Ferraro, S. and Guan, Y. and Hill, J. C. and Hilton, G. C. and Hilton, M. and Hlo\ifmmode \check{z}\else \v{z}\fi{}ek, R. and Huber, Z. B. and Hubmayr, J. and Huffenberger, K. M. and Hughes, J. P. and Koopman, B. J. and Kosowsky, A. and Li, Y. and Lokken, M. and Madhavacheril, M. and McMahon, J. and Moodley, K. and Naess, S. and Nati, F. and Newburgh, L. B. and Niemack, M. D. and Page, L. A. and Partridge, B. and Schaan, E. and Schillaci, A. and Sif\'on, C. and Spergel, D. N. and Staggs, S. T. and Ullom, J. N. and Vale, L. R. and Van Engelen, A. and Van Lanen, J. and Wollack, E. J. and Xu, Z.},
  journal = {Phys. Rev. D},
  volume = {104},
  issue = {4},
  pages = {043502},
  numpages = {16},
  year = {2021},
  month = {Aug},
  publisher = {American Physical Society},
  doi = {10.1103/PhysRevD.104.043502},
  url = {https://link.aps.org/doi/10.1103/PhysRevD.104.043502}
}

@article{Liu:2025zqo,
    author = "Liu, R. Henry and others",
    title = "{Measurements of the thermal Sunyaev-Zel{\textquoteright}dovich effect with ACT and DESI luminous red galaxies}",
    eprint = "2502.08850",
    archivePrefix = "arXiv",
    primaryClass = "astro-ph.CO",
    reportNumber = "FERMILAB-PUB-25-0094-PPD",
    doi = "10.1103/jqn8-19gx",
    journal = "Phys. Rev. D",
    volume = "112",
    number = "8",
    pages = "083561",
    year = "2025"
}

@misc{li2024detectionpairwisekineticsunyaevzeldovich,
      title={Detection of pairwise kinetic Sunyaev-Zel'dovich effect with DESI galaxy groups and Planck in Fourier space}, 
      author={Shaohong Li and Yi Zheng and Ziyang Chen and Haojie Xu and Xiaohu Yang},
      year={2024},
      eprint={2401.03507},
      archivePrefix={arXiv},
      primaryClass={astro-ph.CO},
      url={https://arxiv.org/abs/2401.03507}, 
}

@misc{Hadzhiyska:2025egz,
    author = {Hadzhiyska, B. and others},
    title = {{Probing cosmic velocities with the pairwise kinematic Sunyaev-Zel'dovich signal in DESI Bright Galaxy Sample DR1 and ACT DR6}},
    eprint = {2510.14135},
    archivePrefix = {arXiv},
    primaryClass = {astro-ph.CO},
    reportNumber = {FERMILAB-PUB-25-0758-PPD},
    year = {2025}
}

@article{PhysRevD.103.063513,
  title = {Atacama Cosmology Telescope: Combined kinematic and thermal Sunyaev-Zel'dovich measurements from BOSS CMASS and LOWZ halos},
  author = {Schaan, Emmanuel and Ferraro, Simone and Amodeo, Stefania and Battaglia, Nicholas and Aiola, Simone and Austermann, Jason E. and Beall, James A. and Bean, Rachel and Becker, Daniel T. and Bond, Richard J. and Calabrese, Erminia and Calafut, Victoria and Choi, Steve K. and Denison, Edward V. and Devlin, Mark J. and Duff, Shannon M. and Duivenvoorden, Adriaan J. and Dunkley, Jo and D\"unner, Rolando and Gallardo, Patricio A. and Guan, Yilun and Han, Dongwon and Hill, J. Colin and Hilton, Gene C. and Hilton, Matt and Hlo\ifmmode \check{z}\else \v{z}\fi{}ek, Ren\'ee and Hubmayr, Johannes and Huffenberger, Kevin M. and Hughes, John P. and Koopman, Brian J. and MacInnis, Amanda and McMahon, Jeff and Madhavacheril, Mathew S. and Moodley, Kavilan and Mroczkowski, Tony and Naess, Sigurd and Nati, Federico and Newburgh, Laura B. and Niemack, Michael D. and Page, Lyman A. and Partridge, Bruce and Salatino, Maria and Sehgal, Neelima and Schillaci, Alessandro and Sif\'on, Crist\'obal and Smith, Kendrick M. and Spergel, David N. and Staggs, Suzanne and Storer, Emilie R. and Trac, Hy and Ullom, Joel N. and Van Lanen, Jeff and Vale, Leila R. and van Engelen, Alexander and Maga\~na, Mariana Vargas and Vavagiakis, Eve M. and Wollack, Edward J. and Xu, Zhilei},
  collaboration = {Atacama Cosmology Telescope Collaboration},
  journal = {Phys. Rev. D},
  volume = {103},
  issue = {6},
  pages = {063513},
  numpages = {26},
  year = {2021},
  month = {Mar},
  publisher = {American Physical Society},
  doi = {10.1103/PhysRevD.103.063513},
  url = {https://link.aps.org/doi/10.1103/PhysRevD.103.063513}
}

@article{PhysRevD.93.082002,
  title = {Evidence for the kinematic Sunyaev-Zel'dovich effect with the Atacama Cosmology Telescope and velocity reconstruction from the Baryon Oscillation Spectroscopic Survey},
  author = {Schaan, Emmanuel and Ferraro, Simone and Vargas-Maga\~na, Mariana and Smith, Kendrick M. and Ho, Shirley and Aiola, Simone and Battaglia, Nicholas and Bond, J. Richard and De Bernardis, Francesco and Calabrese, Erminia and Cho, Hsiao-Mei and Devlin, Mark J. and Dunkley, Joanna and Gallardo, Patricio A. and Hasselfield, Matthew and Henderson, Shawn and Hill, J. Colin and Hincks, Adam D. and Hlozek, Ren\'ee and Hubmayr, Johannes and Hughes, John P. and Irwin, Kent D. and Koopman, Brian and Kosowsky, Arthur and Li, Dale and Louis, Thibaut and Lungu, Marius and Madhavacheril, Mathew and Maurin, Lo\"{\i}c and McMahon, Jeffrey John and Moodley, Kavilan and Naess, Sigurd and Nati, Federico and Newburgh, Laura and Niemack, Michael D. and Page, Lyman A. and Pappas, Christine G. and Partridge, Bruce and Schmitt, Benjamin L. and Sehgal, Neelima and Sherwin, Blake D. and Sievers, Jonathan L. and Spergel, David N. and Staggs, Suzanne T. and van Engelen, Alexander and Wollack, Edward J.},
  collaboration = {ACTPol Collaboration},
  journal = {Phys. Rev. D},
  volume = {93},
  issue = {8},
  pages = {082002},
  numpages = {8},
  year = {2016},
  month = {Apr},
  publisher = {American Physical Society},
  doi = {10.1103/PhysRevD.93.082002},
  url = {https://link.aps.org/doi/10.1103/PhysRevD.93.082002}
}

@article{PhysRevD.108.023516,
  title = {Kinematic Sunyaev-Zel'dovich effect with ACT, DES, and BOSS: A novel hybrid estimator},
  author = {Mallaby-Kay, M. and Amodeo, S. and Hill, J. C. and Aguena, M. and Allam, S. and Alves, O. and Annis, J. and Battaglia, N. and Battistelli, E. S. and Baxter, E. J. and Bechtol, K. and Becker, M. R. and Bertin, E. and Bond, J. R. and Brooks, D. and Calabrese, E. and Carnero Rosell, A. and Carrasco Kind, M. and Carretero, J. and Choi, A. and Crocce, M. and da Costa, L. N. and Pereira, M. E. S. and De Vicente, J. and Desai, S. and Dietrich, J. P. and Doel, P. and Doux, C. and Drlica-Wagner, A. and Dunkley, J. and Elvin-Poole, J. and Everett, S. and Ferraro, S. and Ferrero, I. and Frieman, J. and Gallardo, P. A. and Garc\'{\i}a-Bellido, J. and Giannini, G. and Gruen, D. and Gruendl, R. A. and Gutierrez, G. and Hinton, S. R. and Hollowood, D. L. and James, D. J. and Kosowsky, A. and Kuehn, K. and Lokken, M. and Louis, T. and Marshall, J. L. and McMahon, J. and Mena-Fern\'andez, J. and Menanteau, F. and Miquel, R. and Moodley, K. and Mroczkowski, T. and Naess, S. and Niemack, M. D. and Ogando, R. L. C. and Page, L. and Pandey, S. and Pieres, A. and Plazas Malag\'on, A. A. and Raveri, M. and Rodriguez-Monroy, M. and Rykoff, E. S. and Samuroff, S. and Sanchez, E. and Schaan, E. and Sevilla-Noarbe, I. and Sheldon, E. and Sif\'on, C. and Smith, M. and Soares-Santos, M. and Sobreira, F. and Suchyta, E. and Tarle, G. and To, C. and Vargas, C. and Vavagiakis, E. M. and Weaverdyck, N. and Weller, J. and Wiseman, P. and Yanny, B.},
  journal = {Phys. Rev. D},
  volume = {108},
  issue = {2},
  pages = {023516},
  numpages = {19},
  year = {2023},
  month = {Jul},
  publisher = {American Physical Society},
  doi = {10.1103/PhysRevD.108.023516},
  url = {https://link.aps.org/doi/10.1103/PhysRevD.108.023516}
}

@article{PhysRevD.109.063530,
  title = {Atacama Cosmology Telescope: High-resolution component-separated maps across one third of the sky},
  author = {Coulton, William and Madhavacheril, Mathew S. and Duivenvoorden, Adriaan J. and Hill, J. Colin and Abril-Cabezas, Irene and Ade, Peter A. R. and Aiola, Simone and Alford, Tommy and Amiri, Mandana and Amodeo, Stefania and An, Rui and Atkins, Zachary and Austermann, Jason E. and Battaglia, Nicholas and Battistelli, Elia Stefano and Beall, James A. and Bean, Rachel and Beringue, Benjamin and Bhandarkar, Tanay and Biermann, Emily and Bolliet, Boris and Bond, J. Richard and Cai, Hongbo and Calabrese, Erminia and Calafut, Victoria and Capalbo, Valentina and Carrero, Felipe and Chesmore, Grace E. and Cho, Hsiao-mei and Choi, Steve K. and Clark, Susan E. and Rosado, Rodrigo C\'ordova and Cothard, Nicholas F. and Coughlin, Kevin and Crowley, Kevin T. and Devlin, Mark J. and Dicker, Simon and Doze, Peter and Duell, Cody J. and Duff, Shannon M. and Dunkley, Jo and D\"unner, Rolando and Fanfani, Valentina and Fankhanel, Max and Farren, Gerrit and Ferraro, Simone and Freundt, Rodrigo and Fuzia, Brittany and Gallardo, Patricio A. and Garrido, Xavier and Givans, Jahmour and Gluscevic, Vera and Golec, Joseph E. and Guan, Yilun and Halpern, Mark and Han, Dongwon and Hasselfield, Matthew and Healy, Erin and Henderson, Shawn and Hensley, Brandon and Herv\'{\i}as-Caimapo, Carlos and Hilton, Gene C. and Hilton, Matt and Hincks, Adam D. and Hlo\ifmmode \check{z}\else \v{z}\fi{}ek, Ren\'ee and Ho, Shuay-Pwu Patty and Huber, Zachary B. and Hubmayr, Johannes and Huffenberger, Kevin M. and Hughes, John P. and Irwin, Kent and Isopi, Giovanni and Jense, Hidde T. and Keller, Ben and Kim, Joshua and Knowles, Kenda and Koopman, Brian J. and Kosowsky, Arthur and Kramer, Darby and Kusiak, Aleksandra and La Posta, Adrien and Lakey, Victoria and Lee, Eunseong and Li, Zack and Li, Yaqiong and Limon, Michele and Lokken, Martine and Louis, Thibaut and Lungu, Marius and MacCrann, Niall and MacInnis, Amanda and Maldonado, Diego and Maldonado, Felipe and Mallaby-Kay, Maya and Marques, Gabriela A. and van Marrewijk, Joshiwa and McCarthy, Fiona and McMahon, Jeff and Mehta, Yogesh and Menanteau, Felipe and Moodley, Kavilan and Morris, Thomas W. and Mroczkowski, Tony and Naess, Sigurd and Namikawa, Toshiya and Nati, Federico and Newburgh, Laura and Nicola, Andrina and Niemack, Michael D. and Nolta, Michael R. and Orlowski-Scherer, John and Page, Lyman A. and Pandey, Shivam and Partridge, Bruce and Prince, Heather and Puddu, Roberto and Qu, Frank J. and Radiconi, Federico and Robertson, Naomi and Rojas, Felipe and Sakuma, Tai and Salatino, Maria and Schaan, Emmanuel and Schmitt, Benjamin L. and Sehgal, Neelima and Shaikh, Shabbir and Sherwin, Blake D. and Sierra, Carlos and Sievers, Jon and Sif\'on, Crist\'obal and Simon, Sara and Sonka, Rita and Spergel, David N. and Staggs, Suzanne T. and Storer, Emilie and Switzer, Eric R. and Tampier, Niklas and Thornton, Robert and Trac, Hy and Treu, Jesse and Tucker, Carole and Ullom, Joel and Vale, Leila R. and Van Engelen, Alexander and Van Lanen, Jeff and Vargas, Cristian and Vavagiakis, Eve M. and Wagoner, Kasey and Wang, Yuhan and Wenzl, Lukas and Wollack, Edward J. and Xu, Zhilei and Zago, Fernando and Zheng, Kaiwen},
  journal = {Phys. Rev. D},
  volume = {109},
  issue = {6},
  pages = {063530},
  numpages = {32},
  year = {2024},
  month = {Mar},
  publisher = {American Physical Society},
  doi = {10.1103/PhysRevD.109.063530},
  url = {https://link.aps.org/doi/10.1103/PhysRevD.109.063530}
}

@misc{Sunseri:2025hhj,
    author = "Sunseri, James and Amon, Alexandra and Dunkley, Jo and Battaglia, Nicholas and Ferraro, Simone and Hadzhiyska, Boryana and Ried Guachalla, Bernadita and Schaan, Emmanuel",
    title = "{Disentangling the Halo: Joint Model for Measurements of the Kinetic Sunyaev-Zeldovich Effect and Galaxy-Galaxy Lensing}",
    eprint = "2505.20413",
    archivePrefix = "arXiv",
    primaryClass = "astro-ph.CO",
    month = "5",
    year = "2025"
}

@misc{Bigwood:2025kur,
    author = "Bigwood, Leah and Yamamoto, Masaya and Siegel, Jared and Amon, Alexandra and McCarthy, Ian G. and Dave, Romeel and Salcido, Jaime and Schaller, Matthieu and Schaye, Joop and Yang, Tianyi",
    title = "{The kinetic Sunyaev Zeldovich effect as a benchmark for AGN feedback models in hydrodynamical simulations: insights from DESI + ACT}",
    eprint = "2510.15822",
    archivePrefix = "arXiv",
    primaryClass = "astro-ph.CO",
    month = "10",
    year = "2025"
}

@misc{Siegel:2025ivd,
    author = "Siegel, Jared and Bigwood, Leah and Amon, Alexandra and McCullough, Jamie and Yamamoto, Masaya and McCarthy, Ian G. and Schaller, Matthieu and Schneider, Aurel and Schaye, Joop",
    title = "{The suppression of the matter power spectrum: strong feedback from X-ray gas mass fractions, kSZ effect profiles, and galaxy-galaxy lensing}",
    eprint = "2512.02954",
    archivePrefix = "arXiv",
    primaryClass = "astro-ph.CO",
    month = "12",
    year = "2025"
}

@article{DES:2024iny,
    author = "Bigwood, L. and others",
    collaboration = "DES",
    title = "{Weak lensing combined with the kinetic Sunyaev{\textendash}Zel{\textquoteright}dovich effect: a study of baryonic feedback}",
    eprint = "2404.06098",
    archivePrefix = "arXiv",
    primaryClass = "astro-ph.CO",
    reportNumber = "DES-2024-0827, FERMILAB-PUB-24-0130-PPD",
    doi = "10.1093/mnras/stae2100",
    journal = "Mon. Not. Roy. Astron. Soc.",
    volume = "534",
    number = "1",
    pages = "655--682",
    year = "2024"
}

@ARTICLE{2023AJ....165...58Z,
       author = {{Zhou}, Rongpu and {Dey}, Biprateep and {Newman}, Jeffrey A. and {Eisenstein}, Daniel J. and {Dawson}, K. and {Bailey}, S. and {Berti}, A. and {Guy}, J. and {Lan}, Ting-Wen and {Zou}, H. and {Aguilar}, J. and {Ahlen}, S. and {Alam}, Shadab and {Brooks}, D. and {de la Macorra}, A. and {Dey}, A. and {Dhungana}, G. and {Fanning}, K. and {Font-Ribera}, A. and {Gontcho}, S. Gontcho A. and {Honscheid}, K. and {Ishak}, Mustapha and {Kisner}, T. and {Kov{\'a}cs}, A. and {Kremin}, A. and {Landriau}, M. and {Levi}, Michael E. and {Magneville}, C. and {Manera}, Marc and {Martini}, P. and {Meisner}, Aaron M. and {Miquel}, R. and {Moustakas}, J. and {Myers}, Adam D. and {Nie}, Jundan and {Palanque-Delabrouille}, N. and {Percival}, W.~J. and {Poppett}, C. and {Prada}, F. and {Raichoor}, A. and {Ross}, A.~J. and {Schlafly}, E. and {Schlegel}, D. and {Schubnell}, M. and {Tarl{\'e}}, Gregory and {Weaver}, B.~A. and {Wechsler}, R.~H. and {Y{\'e}che}, Christophe and {Zhou}, Zhimin},
        title = "{Target Selection and Validation of DESI Luminous Red Galaxies}",
      journal = {\aj},
         year = 2023,
        month = feb,
       volume = {165},
       number = {2},
          eid = {58},
        pages = {58},
          doi = {10.3847/1538-3881/aca5fb},
archivePrefix = {arXiv},
       eprint = {2208.08515},
 primaryClass = {astro-ph.CO},
       adsurl = {https://ui.adsabs.harvard.edu/abs/2023AJ....165...58Z}
}

@ARTICLE{1986ApJ...306L..51O,
       author = {{Ostriker}, J.~P. and {Vishniac}, E.~T.},
        title = "{Generation of Microwave Background Fluctuations from Nonlinear Perturbations at the ERA of Galaxy Formation}",
      journal = {\apjl},
         year = 1986,
        month = jul,
       volume = {306},
        pages = {L51},
          doi = {10.1086/184704},
       adsurl = {https://ui.adsabs.harvard.edu/abs/1986ApJ...306L..51O}
}

@article{Nelson_2015,
   title={The illustris simulation: Public data release},
   volume={13},
   ISSN={2213-1337},
   url={http://dx.doi.org/10.1016/j.ascom.2015.09.003},
   DOI={10.1016/j.ascom.2015.09.003},
   journal={Astronomy and Computing},
   publisher={Elsevier BV},
   author={Nelson, D. and Pillepich, A. and Genel, S. and Vogelsberger, M. and Springel, V. and Torrey, P. and Rodriguez-Gomez, V. and Sijacki, D. and Snyder, G.F. and Griffen, B. and Marinacci, F. and Blecha, L. and Sales, L. and Xu, D. and Hernquist, L.},
   year={2015},
   month=nov, pages={12--37} }

@misc{nelson2021illustristngsimulationspublicdata,
      title={The IllustrisTNG Simulations: Public Data Release}, 
      author={Dylan Nelson and Volker Springel and Annalisa Pillepich and Vicente Rodriguez-Gomez and Paul Torrey and Shy Genel and Mark Vogelsberger and Ruediger Pakmor and Federico Marinacci and Rainer Weinberger and Luke Kelley and Mark Lovell and Benedikt Diemer and Lars Hernquist},
      year={2021},
      eprint={1812.05609},
      archivePrefix={arXiv},
      primaryClass={astro-ph.GA},
      url={https://arxiv.org/abs/1812.05609}, 
}

@article{PhysRevLett.117.051301,
  title = {Kinematic Sunyaev-Zel'dovich Effect with Projected Fields: A Novel Probe of the Baryon Distribution with Planck, WMAP, and WISE Data},
  author = {Hill, J. Colin and Ferraro, Simone and Battaglia, Nick and Liu, Jia and Spergel, David N.},
  journal = {Phys. Rev. Lett.},
  volume = {117},
  issue = {5},
  pages = {051301},
  numpages = {6},
  year = {2016},
  month = {Jul},
  publisher = {American Physical Society},
  doi = {10.1103/PhysRevLett.117.051301},
  url = {https://link.aps.org/doi/10.1103/PhysRevLett.117.051301}
}

@article{PhysRevD.94.123526,
  title = {Kinematic Sunyaev-Zel'dovich effect with projected fields. II. Prospects, challenges, and comparison with simulations},
  author = {Ferraro, Simone and Hill, J. Colin and Battaglia, Nick and Liu, Jia and Spergel, David N.},
  journal = {Phys. Rev. D},
  volume = {94},
  issue = {12},
  pages = {123526},
  numpages = {14},
  year = {2016},
  month = {Dec},
  publisher = {American Physical Society},
  doi = {10.1103/PhysRevD.94.123526},
  url = {https://link.aps.org/doi/10.1103/PhysRevD.94.123526}
}

@article{PhysRevD.104.043518,
  title = {Constraining the baryon abundance with the kinematic Sunyaev-Zel'dovich effect: Projected-field detection using $Planck$, $WMAP$, and $unWISE$},
  author = {Kusiak, Aleksandra and Bolliet, Boris and Ferraro, Simone and Hill, J. Colin and Krolewski, Alex},
  journal = {Phys. Rev. D},
  volume = {104},
  issue = {4},
  pages = {043518},
  numpages = {28},
  year = {2021},
  month = {Aug},
  publisher = {American Physical Society},
  doi = {10.1103/PhysRevD.104.043518},
  url = {https://link.aps.org/doi/10.1103/PhysRevD.104.043518}
}

@misc{sunseri2025disentanglinghalojointmodel,
      title={Disentangling the Halo: Joint Model for Measurements of the Kinetic Sunyaev-Zeldovich Effect and Galaxy-Galaxy Lensing}, 
      author={James Sunseri and Alexandra Amon and Jo Dunkley and Nicholas Battaglia and Simone Ferraro and Boryana Hadzhiyska and Bernardita Ried Guachalla and Emmanuel Schaan},
      year={2025},
      eprint={2505.20413},
      archivePrefix={arXiv},
      primaryClass={astro-ph.CO},
      url={https://arxiv.org/abs/2505.20413}, 
}

@article{McKay1979,
  author = {McKay, M. D. and Beckman, R. J. and Conover, W. J.},
  title = {A Comparison of Three Methods for Selecting Values of Input Variables in the Analysis of Output from a Computer Code},
  journal = {Technometrics},
  volume = {21},
  number = {2},
  pages = {239--245},
  year = {1979},
  month = {May},
  publisher = {Taylor \& Francis, Ltd.},
  doi = {10.2307/1268522},
  url = {https://www.jstor.org/stable/1268522}
}

@article{karamanis2022accelerating,
    title={Accelerating astronomical and cosmological inference with preconditioned Monte Carlo},
    author={Karamanis, Minas and Beutler, Florian and Peacock, John A and Nabergoj, David and Seljak, Uro{\v{s}}},
    journal={Monthly Notices of the Royal Astronomical Society},
    volume={516},
    number={2},
    pages={1644--1653},
    year={2022},
    publisher={Oxford University Press}
}

@article{karamanis2022pocomc,
    title={pocoMC: A Python package for accelerated Bayesian inference in astronomy and cosmology},
    author={Karamanis, Minas and Nabergoj, David and Beutler, Florian and Peacock, John A and Seljak, Uros},
    journal={arXiv preprint arXiv:2207.05660},
    year={2022}
}

@article{scikit-learn,
  title={Scikit-learn: Machine Learning in {P}ython},
  author={Pedregosa, F. and Varoquaux, G. and Gramfort, A. and Michel, V.
          and Thirion, B. and Grisel, O. and Blondel, M. and Prettenhofer, P.
          and Weiss, R. and Dubourg, V. and Vanderplas, J. and Passos, A. and
          Cournapeau, D. and Brucher, M. and Perrot, M. and Duchesnay, E.},
  journal={Journal of Machine Learning Research},
  volume={12},
  pages={2825--2830},
  year={2011}
}

@misc{Lague:2025txe,
    author = {Lagu{\"e}, Alex and Madhavacheril, Mathew S. and Borrow, Josh and Smith, Kendrick M. and Chen, Xinyi and Schaller, Matthieu and Schaye, Joop},
    title = "{Inferring the Impacts of Baryonic Feedback from Kinetic Sunyaev-Zeldovich Cross-Correlations}",
    eprint = "2511.20595",
    archivePrefix = "arXiv",
    primaryClass = "astro-ph.CO",
    month = "11",
    year = "2025"
}

@misc{mccarthy2025flamingocombiningkineticsz,
      title={FLAMINGO: combining kinetic SZ effect and galaxy-galaxy lensing measurements to gauge the impact of feedback on large-scale structure}, 
      author={Ian G. McCarthy and Alexandra Amon and Joop Schaye and Emmanuel Schaan and Raul E. Angulo and Jaime Salcido and Matthieu Schaller and Leah Bigwood and Willem Elbers and Roi Kugel and John C. Helly and Victor J. Forouhar Moreno and Carlos S. Frenk and Robert J. McGibbon and Lurdes Ondaro-Mallea and Marcel P. van Daalen},
      year={2025},
      eprint={2410.19905},
      archivePrefix={arXiv},
      primaryClass={astro-ph.CO},
      url={https://arxiv.org/abs/2410.19905}, 
}

@misc{salcido2025implicationsfeedbacksolutionss8,
      title={Implications of feedback solutions to the $S_8$ tension for the baryon fractions of galaxy groups and clusters}, 
      author={Jaime Salcido and Ian G. McCarthy},
      year={2025},
      eprint={2409.05716},
      archivePrefix={arXiv},
      primaryClass={astro-ph.CO},
      url={https://arxiv.org/abs/2409.05716}, 
}

@misc{harscouet2025kszeveryonepseudoclapproach,
      title={kSZ for everyone: the pseudo-Cl approach to stacking}, 
      author={Lea Harscouet and Kevin Wolz and Amy Wayland and David Alonso and Boryana Hadzhiyska},
      year={2025},
      eprint={2512.14625},
      archivePrefix={arXiv},
      primaryClass={astro-ph.CO},
      url={https://arxiv.org/abs/2512.14625}, 
}

@misc{HadzhiyskaBGSELG,
      title={Precision Kinematic Sunyaev--Zel'dovich Measurements Across Halo Mass and Redshift with DESI DR2 and ACT DR6: Part II. Bright Galaxy Survey and Emission-Line Galaxies}, 
      author={Boryana Hadzhiyska and Simone Ferraro and Frank J. Qu and Bernardita Ried Guachalla and Emmanuel Schaan and others},
      year={2026},
      eprint={26XX.XXXX},
      archivePrefix={in prep, arXiv},
      primaryClass={astro-ph.CO},
}

@article{Hadzhiyska:2025mvt,
    author = "Hadzhiyska, Boryana and Ferraro, Simone and Farren, Gerrit S. and Sailer, Noah and Zhou, Rongpu",
    title = "{Missing baryons recovered: A measurement of the gas fraction in galaxies and groups with the kinematic Sunyaev-Zel{\textquoteright}dovich effect and CMB lensing}",
    eprint = "2507.14136",
    archivePrefix = "arXiv",
    primaryClass = "astro-ph.CO",
    doi = "10.1103/mdhz-fgj8",
    journal = "Phys. Rev. D",
    volume = "112",
    number = "12",
    pages = "123507",
    year = "2025"
}

@ARTICLE{DESI2022.KP1.Instr,
       author = {{DESI Collaboration} and {Abareshi}, B. and {Aguilar}, J. and {Ahlen}, S. and {Alam}, Shadab and {Alexander}, David M. and {Alfarsy}, R. and {Allen}, L. and {Allende Prieto}, C. and {Alves}, O. and {Ameel}, J. and {Armengaud}, E. and {Asorey}, J. and {Aviles}, Alejandro and {Bailey}, S. and {Balaguera-Antol{\'\i}nez}, A. and {Ballester}, O. and {Baltay}, C. and {Bault}, A. and {Beltran}, S.~F. and {Benavides}, B. and {BenZvi}, S. and {Berti}, A. and {Besuner}, R. and {Beutler}, Florian and {Bianchi}, D. and {Blake}, C. and {Blanc}, P. and {Blum}, R. and {Bolton}, A. and {Bose}, S. and {Bramall}, D. and {Brieden}, S. and {Brodzeller}, A. and {Brooks}, D. and {Brownewell}, C. and {Buckley-Geer}, E. and {Cahn}, R.~N. and {Cai}, Z. and {Canning}, R. and {Capasso}, R. and {Carnero Rosell}, A. and {Carton}, P. and {Casas}, R. and {Castander}, F.~J. and {Cervantes-Cota}, J.~L. and {Chabanier}, S. and {Chaussidon}, E. and {Chuang}, C. and {Circosta}, C. and {Cole}, S. and {Cooper}, A.~P. and {da Costa}, L. and {Cousinou}, M. -C. and {Cuceu}, A. and {Davis}, T.~M. and {Dawson}, K. and {de la Cruz-Noriega}, R. and {de la Macorra}, A. and {de Mattia}, A. and {Della Costa}, J. and {Demmer}, P. and {Derwent}, M. and {Dey}, A. and {Dey}, B. and {Dhungana}, G. and {Ding}, Z. and {Dobson}, C. and {Doel}, P. and {Donald-McCann}, J. and {Donaldson}, J. and {Douglass}, K. and {Duan}, Y. and {Dunlop}, P. and {Edelstein}, J. and {Eftekharzadeh}, S. and {Eisenstein}, D.~J. and {Enriquez-Vargas}, M. and {Escoffier}, S. and {Evatt}, M. and {Fagrelius}, P. and {Fan}, X. and {Fanning}, K. and {Fawcett}, V.~A. and {Ferraro}, S. and {Ereza}, J. and {Flaugher}, B. and {Font-Ribera}, A. and {Forero-Romero}, J.~E. and {Frenk}, C.~S. and {Fromenteau}, S. and {G{\"a}nsicke}, B.~T. and {Garcia-Quintero}, C. and {Garrison}, L. and {Gazta{\~n}aga}, E. and {Gerardi}, F. and {Gil-Mar{\'\i}n}, H. and {Gontcho a Gontcho}, S. and {Gonzalez-Morales}, Alma X. and {Gonzalez-de-Rivera}, G. and {Gonzalez-Perez}, V. and {Gordon}, C. and {Graur}, O. and {Green}, D. and {Grove}, C. and {Gruen}, D. and {Gutierrez}, G. and {Guy}, J. and {Hahn}, C. and {Harris}, S. and {Herrera}, D. and {Herrera-Alcantar}, Hiram K. and {Honscheid}, K. and {Howlett}, C. and {Huterer}, D. and {Ir{\v{s}}i{\v{c}}}, V. and {Ishak}, M. and {Jelinsky}, P. and {Jiang}, L. and {Jimenez}, J. and {Jing}, Y.~P. and {Joyce}, R. and {Jullo}, E. and {Juneau}, S. and {Kara{\c{c}}ayl{\i}}, N.~G. and {Karamanis}, M. and {Karcher}, A. and {Karim}, T. and {Kehoe}, R. and {Kent}, S. and {Kirkby}, D. and {Kisner}, T. and {Kitaura}, F. and {Koposov}, S.~E. and {Kov{\'a}cs}, A. and {Kremin}, A. and {Krolewski}, Alex and {L'Huillier}, B. and {Lahav}, O. and {Lambert}, A. and {Lamman}, C. and {Lan}, Ting-Wen and {Landriau}, M. and {Lane}, S. and {Lang}, D. and {Lange}, J.~U. and {Lasker}, J. and {Le Guillou}, L. and {Leauthaud}, A. and {Le Van Suu}, A. and {Levi}, Michael E. and {Li}, T.~S. and {Magneville}, C. and {Manera}, M. and {Manser}, Christopher J. and {Marshall}, B. and {Martini}, Paul and {McCollam}, W. and {McDonald}, P. and {Meisner}, Aaron M. and {Mena-Fern{\'a}ndez}, J. and {Meneses-Rizo}, J. and {Mezcua}, M. and {Miller}, T. and {Miquel}, R. and {Montero-Camacho}, P. and {Moon}, J. and {Moustakas}, J. and {Mueller}, E. and {Mu{\~n}oz-Guti{\'e}rrez}, Andrea and {Myers}, Adam D. and {Nadathur}, S. and {Najita}, J. and {Napolitano}, L. and {Neilsen}, E. and {Newman}, Jeffrey A. and {Nie}, J.~D. and {Ning}, Y. and {Niz}, G. and {Norberg}, P. and {Noriega}, Hern{\'a}n E. and {O'Brien}, T. and {Obuljen}, A. and {Palanque-Delabrouille}, N. and {Palmese}, A. and {Zhiwei}, P. and {Pappalardo}, D. and {PENG}, X. and {Percival}, W.~J. and {Perruchot}, S. and {Pogge}, R. and {Poppett}, C. and {Porredon}, A. and {Prada}, F. and {Prochaska}, J. and {Pucha}, R. and {P{\'e}rez-Fern{\'a}ndez}, A. and {P{\'e}rez-R{\`a}fols}, I. and {Rabinowitz}, D. and {Raichoor}, A. and {Ramirez-Solano}, S. and {Ram{\'\i}rez-P{\'e}rez}, C{\'e}sar and {Ravoux}, C. and {Reil}, K. and {Rezaie}, M. and {Rocher}, A. and {Rockosi}, C. and {Roe}, N.~A. and {Roodman}, A. and {Ross}, A.~J. and {Rossi}, G. and {Ruggeri}, R. and {Ruhlmann-Kleider}, V. and {Sabiu}, C.~G. and {Safonova}, S. and {Said}, K. and {Saintonge}, A. and {Salas Catonga}, Javier and {Samushia}, L. and {Sanchez}, E. and {Saulder}, C. and {Schaan}, E. and {Schlafly}, E. and {Schlegel}, D. and {Schmoll}, J. and {Scholte}, D. and {Schubnell}, M. and {Secroun}, A. and {Seo}, H. and {Serrano}, S. and {Sharples}, Ray M. and {Sholl}, Michael J. and {Silber}, Joseph Harry and {Silva}, D.~R. and {Sirk}, M. and {Siudek}, M. and {Smith}, A. and {Sprayberry}, D. and {Staten}, R. and {Stupak}, B. and {Tan}, T. and {Tarl{\'e}}, Gregory and {Tie}, Suk Sien and {Tojeiro}, R. and {Ure{\~n}a-L{\'o}pez}, L.~A. and {Valdes}, F. and {Valenzuela}, O. and {Valluri}, M. and {Vargas-Maga{\~n}a}, M. and {Verde}, L. and {Walther}, M. and {Wang}, B. and {Wang}, M.~S. and {Weaver}, B.~A. and {Weaverdyck}, C. and {Wechsler}, R. and {Wilson}, Michael J. and {Yang}, J. and {Yu}, Y. and {Yuan}, S. and {Y{\`e}che}, Christophe and {Zhang}, H. and {Zhang}, K. and {Zhao}, Cheng and {Zhou}, Rongpu and {Zhou}, Zhimin and {Zou}, H. and {Zou}, J. and {Zou}, S. and {Zu}, Y. and {DESI Collaboration}},
        title = "{Overview of the Instrumentation for the Dark Energy Spectroscopic Instrument}",
      journal = {\aj},
         year = 2022,
        month = nov,
       volume = {164},
       number = {5},
          eid = {207},
        pages = {207},
          doi = {10.3847/1538-3881/ac882b},
archivePrefix = {arXiv},
       eprint = {2205.10939},
 primaryClass = {astro-ph.IM},
       adsurl = {https://ui.adsabs.harvard.edu/abs/2022AJ....164..207A}
}

@ARTICLE{Corrector.Miller.2023,
       author = {{Miller}, Timothy N. and {Doel}, Peter and {Gutierrez}, Gaston and {Besuner}, Robert and {Brooks}, David and {Gallo}, Giuseppe and {Heetderks}, Henry and {Jelinsky}, Patrick and {Kent}, Stephen M. and {Lampton}, Michael and {Levi}, Michael E. and {Liang}, Ming and {Meisner}, Aaron and {Sholl}, Michael J. and {Silber}, Joseph Harry and {Sprayberry}, David and {Aguilar}, Jessica Nicole and {de la Macorra}, Axel and {Eisenstein}, Daniel and {Fanning}, Kevin and {Font-Ribera}, Andreu and {Gazta{\~n}aga}, Enrique and {Gontcho A Gontcho}, Satya and {Honscheid}, Klaus and {Jimenez}, Jorge and {Joyce}, Dick and {Kehoe}, Robert and {Kisner}, Theodore and {Kremin}, Anthony and {Landriau}, Martin and {Le Guillou}, Laurent and {Magneville}, Christophe and {Martini}, Paul and {Miquel}, Ramon and {Moustakas}, John and {Nie}, Jundan and {Percival}, Will and {Poppett}, Claire and {Prada}, Francisco and {Rossi}, Graziano and {Schlegel}, David and {Schubnell}, Michael and {Seo}, Hee-Jong and {Sharples}, Ray and {Tarl{\'e}}, Gregory and {Vargas-Maga{\~n}a}, Mariana and {Zhou}, Zhimin and {the DESI Collaboration}},
        title = "{The Optical Corrector for the Dark Energy Spectroscopic Instrument}",
      journal = {\aj},
         year = 2024,
        month = aug,
       volume = {168},
       number = {2},
          eid = {95},
        pages = {95},
          doi = {10.3847/1538-3881/ad45fe},
archivePrefix = {arXiv},
       eprint = {2306.06310},
 primaryClass = {astro-ph.IM},
       adsurl = {https://ui.adsabs.harvard.edu/abs/2024AJ....168...95M}
}

@ARTICLE{FiberSystem.Poppett.2024,
       author = {{Poppett}, Claire and {Tyas}, Luke and {Aguilar}, J. and {Bebek}, Christopher and {Bramall}, D. and {Claybaugh}, T. and {Edelstein}, J. and {Fagrelius}, P. and {Heetderks}, H. and {Jelinsky}, P. and {Jelinsky}, S. and {Lafever}, Robin and {Lambert}, A. and {Lampton}, M. and {Levi}, Michael E. and {Martini}, P. and {Rockosi}, C. and {Schmoll}, J. and {Sharples}, Ray M. and {Sirk}, Martin and {Wishnow}, Edward and {Yu}, Jiaxi and {Ahlen}, S. and {Bault}, A. and {BenZvi}, S. and {Brooks}, D. and {Cole}, S. and {de la Macorra}, A. and {Dey}, Arjun and {Doel}, P. and {Fanning}, K. and {Font-Ribera}, A. and {Forero-Romero}, J.~E. and {Gazta{\~n}aga}, E. and {Gontcho A Gontcho}, S. and {Gonzalez-Morales}, A.~X. and {Hahn}, C. and {Honscheid}, K. and {Jimenez}, J. and {Juneau}, S. and {Kirkby}, D. and {Kremin}, A. and {Landriau}, M. and {Le Guillou}, L. and {Manera}, M. and {Meisner}, A. and {Miquel}, R. and {Moustakas}, J. and {Mueller}, E. and {Mu{\~n}oz-Guti{\'e}rrez}, A. and {Myers}, A.~D. and {Nie}, J. and {Niz}, G. and {Palanque-Delabrouille}, N. and {Percival}, W.~J. and {Prada}, F. and {Rabinowitz}, D. and {Rezaie}, M. and {Rossi}, G. and {Sanchez}, E. and {Schlafly}, Edward F. and {Schlegel}, D. and {Schubnell}, M. and {Seo}, H. and {Sprayberry}, D. and {Tarl{\'e}}, G. and {Vargas-Maga{\~n}a}, M. and {Weaver}, B.~A. and {Zhou}, R.},
        title = "{Overview of the Fiber System for the Dark Energy Spectroscopic Instrument}",
      journal = {\aj},
         year = 2024,
        month = dec,
       volume = {168},
       number = {6},
          eid = {245},
        pages = {245},
          doi = {10.3847/1538-3881/ad76a4},
       adsurl = {https://ui.adsabs.harvard.edu/abs/2024AJ....168..245P}
}

@ARTICLE{DESI2016b.Instr,
       author = {{DESI Collaboration} and {Aghamousa}, Amir and {Aguilar}, Jessica and {Ahlen}, Steve and {Alam}, Shadab and {Allen}, Lori E. and {Allende Prieto}, Carlos and {Annis}, James and {Bailey}, Stephen and {Balland}, Christophe and {Ballester}, Otger and {Baltay}, Charles and {Beaufore}, Lucas and {Bebek}, Chris and {Beers}, Timothy C. and {Bell}, Eric F. and {Bernal}, Jos{\'e} Luis and {Besuner}, Robert and {Beutler}, Florian and {Blake}, Chris and {Bleuler}, Hannes and {Blomqvist}, Michael and {Blum}, Robert and {Bolton}, Adam S. and {Briceno}, Cesar and {Brooks}, David and {Brownstein}, Joel R. and {Buckley-Geer}, Elizabeth and {Burden}, Angela and {Burtin}, Etienne and {Busca}, Nicolas G. and {Cahn}, Robert N. and {Cai}, Yan-Chuan and {Cardiel-Sas}, Laia and {Carlberg}, Raymond G. and {Carton}, Pierre-Henri and {Casas}, Ricard and {Castander}, Francisco J. and {Cervantes-Cota}, Jorge L. and {Claybaugh}, Todd M. and {Close}, Madeline and {Coker}, Carl T. and {Cole}, Shaun and {Comparat}, Johan and {Cooper}, Andrew P. and {Cousinou}, M. -C. and {Crocce}, Martin and {Cuby}, Jean-Gabriel and {Cunningham}, Daniel P. and {Davis}, Tamara M. and {Dawson}, Kyle S. and {de la Macorra}, Axel and {De Vicente}, Juan and {Delubac}, Timoth{\'e}e and {Derwent}, Mark and {Dey}, Arjun and {Dhungana}, Govinda and {Ding}, Zhejie and {Doel}, Peter and {Duan}, Yutong T. and {Ealet}, Anne and {Edelstein}, Jerry and {Eftekharzadeh}, Sarah and {Eisenstein}, Daniel J. and {Elliott}, Ann and {Escoffier}, St{\'e}phanie and {Evatt}, Matthew and {Fagrelius}, Parker and {Fan}, Xiaohui and {Fanning}, Kevin and {Farahi}, Arya and {Farihi}, Jay and {Favole}, Ginevra and {Feng}, Yu and {Fernandez}, Enrique and {Findlay}, Joseph R. and {Finkbeiner}, Douglas P. and {Fitzpatrick}, Michael J. and {Flaugher}, Brenna and {Flender}, Samuel and {Font-Ribera}, Andreu and {Forero-Romero}, Jaime E. and {Fosalba}, Pablo and {Frenk}, Carlos S. and {Fumagalli}, Michele and {Gaensicke}, Boris T. and {Gallo}, Giuseppe and {Garcia-Bellido}, Juan and {Gaztanaga}, Enrique and {Pietro Gentile Fusillo}, Nicola and {Gerard}, Terry and {Gershkovich}, Irena and {Giannantonio}, Tommaso and {Gillet}, Denis and {Gonzalez-de-Rivera}, Guillermo and {Gonzalez-Perez}, Violeta and {Gott}, Shelby and {Graur}, Or and {Gutierrez}, Gaston and {Guy}, Julien and {Habib}, Salman and {Heetderks}, Henry and {Heetderks}, Ian and {Heitmann}, Katrin and {Hellwing}, Wojciech A. and {Herrera}, David A. and {Ho}, Shirley and {Holland}, Stephen and {Honscheid}, Klaus and {Huff}, Eric and {Hutchinson}, Timothy A. and {Huterer}, Dragan and {Hwang}, Ho Seong and {Illa Laguna}, Joseph Maria and {Ishikawa}, Yuzo and {Jacobs}, Dianna and {Jeffrey}, Niall and {Jelinsky}, Patrick and {Jennings}, Elise and {Jiang}, Linhua and {Jimenez}, Jorge and {Johnson}, Jennifer and {Joyce}, Richard and {Jullo}, Eric and {Juneau}, St{\'e}phanie and {Kama}, Sami and {Karcher}, Armin and {Karkar}, Sonia and {Kehoe}, Robert and {Kennamer}, Noble and {Kent}, Stephen and {Kilbinger}, Martin and {Kim}, Alex G. and {Kirkby}, David and {Kisner}, Theodore and {Kitanidis}, Ellie and {Kneib}, Jean-Paul and {Koposov}, Sergey and {Kovacs}, Eve and {Koyama}, Kazuya and {Kremin}, Anthony and {Kron}, Richard and {Kronig}, Luzius and {Kueter-Young}, Andrea and {Lacey}, Cedric G. and {Lafever}, Robin and {Lahav}, Ofer and {Lambert}, Andrew and {Lampton}, Michael and {Landriau}, Martin and {Lang}, Dustin and {Lauer}, Tod R. and {Le Goff}, Jean-Marc and {Le Guillou}, Laurent and {Le Van Suu}, Auguste and {Lee}, Jae Hyeon and {Lee}, Su-Jeong and {Leitner}, Daniela and {Lesser}, Michael and {Levi}, Michael E. and {L'Huillier}, Benjamin and {Li}, Baojiu and {Liang}, Ming and {Lin}, Huan and {Linder}, Eric and {Loebman}, Sarah R. and {Luki{\'c}}, Zarija and {Ma}, Jun and {MacCrann}, Niall and {Magneville}, Christophe and {Makarem}, Laleh and {Manera}, Marc and {Manser}, Christopher J. and {Marshall}, Robert and {Martini}, Paul and {Massey}, Richard and {Matheson}, Thomas and {McCauley}, Jeremy and {McDonald}, Patrick and {McGreer}, Ian D. and {Meisner}, Aaron and {Metcalfe}, Nigel and {Miller}, Timothy N. and {Miquel}, Ramon and {Moustakas}, John and {Myers}, Adam and {Naik}, Milind and {Newman}, Jeffrey A. and {Nichol}, Robert C. and {Nicola}, Andrina and {Nicolati da Costa}, Luiz and {Nie}, Jundan and {Niz}, Gustavo and {Norberg}, Peder and {Nord}, Brian and {Norman}, Dara and {Nugent}, Peter and {O'Brien}, Thomas and {Oh}, Minji and {Olsen}, Knut A.~G.},
        title = "{The DESI Experiment Part II: Instrument Design}",
      journal = {arXiv e-prints},
         year = 2016,
        month = oct,
          eid = {arXiv:1611.00037},
        pages = {arXiv:1611.00037},
          doi = {10.48550/arXiv.1611.00037},
archivePrefix = {arXiv},
       eprint = {1611.00037},
 primaryClass = {astro-ph.IM},
       adsurl = {https://ui.adsabs.harvard.edu/abs/2016arXiv161100037D}
}

@ARTICLE{Spectro.Pipeline.Guy.2023,
       author = {{Guy}, J. and {Bailey}, S. and {Kremin}, A. and {Alam}, Shadab and {Alexander}, D.~M. and {Allende Prieto}, C. and {BenZvi}, S. and {Bolton}, A.~S. and {Brooks}, D. and {Chaussidon}, E. and {Cooper}, A.~P. and {Dawson}, K. and {de la Macorra}, A. and {Dey}, A. and {Dey}, Biprateep and {Dhungana}, G. and {Eisenstein}, D.~J. and {Font-Ribera}, A. and {Forero-Romero}, J.~E. and {Gazta{\~n}aga}, E. and {Gontcho A Gontcho}, S. and {Green}, D. and {Honscheid}, K. and {Ishak}, M. and {Kehoe}, R. and {Kirkby}, D. and {Kisner}, T. and {Koposov}, Sergey E. and {Lan}, Ting-Wen and {Landriau}, M. and {Le Guillou}, L. and {Levi}, Michael E. and {Magneville}, C. and {Manser}, Christopher J. and {Martini}, P. and {Meisner}, Aaron M. and {Miquel}, R. and {Moustakas}, J. and {Myers}, Adam D. and {Newman}, Jeffrey A. and {Nie}, Jundan and {Palanque-Delabrouille}, N. and {Percival}, W.~J. and {Poppett}, C. and {Prada}, F. and {Raichoor}, A. and {Ravoux}, C. and {Ross}, A.~J. and {Schlafly}, E.~F. and {Schlegel}, D. and {Schubnell}, M. and {Sharples}, Ray M. and {Tarl{\'e}}, Gregory and {Weaver}, B.~A. and {Y{\'e}che}, Christophe and {Zhou}, Rongpu and {Zhou}, Zhimin and {Zou}, H.},
        title = "{The Spectroscopic Data Processing Pipeline for the Dark Energy Spectroscopic Instrument}",
      journal = {\aj},
         year = 2023,
        month = apr,
       volume = {165},
       number = {4},
          eid = {144},
        pages = {144},
          doi = {10.3847/1538-3881/acb212},
archivePrefix = {arXiv},
       eprint = {2209.14482},
 primaryClass = {astro-ph.IM},
       adsurl = {https://ui.adsabs.harvard.edu/abs/2023AJ....165..144G}
}

@ARTICLE{SurveyOps.Schlafly.2023,
       author = {{Schlafly}, Edward F. and {Kirkby}, David and {Schlegel}, David J. and {Myers}, Adam D. and {Raichoor}, Anand and {Dawson}, Kyle and {Aguilar}, Jessica and {Allende Prieto}, Carlos and {Bailey}, Stephen and {BenZvi}, Segev and {Bermejo-Climent}, Jose and {Brooks}, David and {de la Macorra}, Axel and {Dey}, Arjun and {Doel}, Peter and {Fanning}, Kevin and {Font-Ribera}, Andreu and {Forero-Romero}, Jaime E. and {Garc{\'\i}a-Bellido}, Juan and {Gontcho A Gontcho}, Satya and {Guy}, Julien and {Hahn}, ChangHoon and {Honscheid}, Klaus and {Ishak}, Mustapha and {Juneau}, St{\'e}phanie and {Kehoe}, Robert and {Kisner}, Theodore and {Kremin}, Anthony and {Landriau}, Martin and {Lang}, Dustin A. and {Lasker}, James and {Levi}, Michael E. and {Magneville}, Christophe and {Manser}, Christopher J. and {Martini}, Paul and {Meisner}, Aaron M. and {Miquel}, Ramon and {Moustakas}, John and {Newman}, Jeffrey A. and {Nie}, Jundan and {Palanque-Delabrouille}, Nathalie. and {Percival}, Will J. and {Poppett}, Claire and {Rockosi}, Constance and {Ross}, Ashley J. and {Rossi}, Graziano and {Tarl{\'e}}, Gregory and {Weaver}, Benjamin A. and {Y{\`e}che}, Christophe and {Zhou}, Rongpu and {DESI Collaboration}},
        title = "{Survey Operations for the Dark Energy Spectroscopic Instrument}",
      journal = {\aj},
         year = 2023,
        month = dec,
       volume = {166},
       number = {6},
          eid = {259},
        pages = {259},
          doi = {10.3847/1538-3881/ad0832},
archivePrefix = {arXiv},
       eprint = {2306.06309},
 primaryClass = {astro-ph.CO},
       adsurl = {https://ui.adsabs.harvard.edu/abs/2023AJ....166..259S}
}

@ARTICLE{DESI2024.I.DR1,
       author = {{DESI Collaboration} and {Abdul-Karim}, M  and {Adame}, A.~G. and {Aguado}, D. and {Aguilar}, J. and {Ahlen}, S. and {Alam}, S. and {Aldering}, G. and {Alexander}, D.~M. and {Alfarsy}, R. and {Allen}, L. and {Allende Prieto}, C. and {Alves}, O. and {Anand}, A. and {Andrade}, U. and {Armengaud}, E. and {Avila}, S. and {Aviles}, A. and {Awan}, H. and {Bailey}, S. and {Baleato Lizancos}, A. and {Ballester}, O. and {Bault}, A. and {Bautista}, J. and {BenZvi}, S. and {Beraldo e Silva}, L. and {Bermejo-Climent}, J.~R. and {Beutler}, F. and {Bianchi}, D. and {Blake}, C. and {Blum}, R. and {Bolton}, A.~S. and {Bonici}, M. and {Brieden}, S. and {Brodzeller}, A. and {Brooks}, D. and {Buckley-Geer}, E. and {Burtin}, E. and {Canning}, R. and {Carnero Rosell}, A. and {Carr}, A. and {Carrilho}, P. and {Casas}, L. and {Castander}, F.~J. and {Cereskaite}, R. and {Cervantes-Cota}, J.~L. and {Chaussidon}, E. and {Chaves-Montero}, J. and {Chen}, S. and {Chen}, X. and {Claybaugh}, T. and {Cole}, S. and {Cooper}, A.~P. and {Cousinou}, M. -C. and {Cuceu}, A. and {Davis}, T.~M. and {Dawson}, K.~S. and {de Belsunce}, R. and {de la Cruz}, R. and {de la Macorra}, A. and {de Mattia}, A. and {Deiosso}, N. and {Della Costa}, J. and {Demina}, R. and {Demirbozan}, U. and {DeRose}, J. and {Dey}, A. and {Dey}, B. and {Ding}, J. and {Ding}, Z. and {Doel}, P. and {Douglass}, K. and {Dowicz}, M. and {Ebina}, H. and {Edelstein}, J. and {Eisenstein}, D.~J. and {Elbers}, W. and {Emas}, N. and {Escoffier}, S. and {Fagrelius}, P. and {Fan}, X. and {Fanning}, K. and {Fawcett}, V.~A. and {Fern{\'a}ndez-Garc{\'\i}a}, E. and {Ferraro}, S. and {Findlay}, N. and {Font-Ribera}, A. and {Forero-Romero}, J.~E. and {Forero-S{\'a}nchez}, D. and {Frenk}, C.~S. and {G{\"a}nsicke}, B.~T. and {Galbany}, L. and {Garc{\'\i}a-Bellido}, J. and {Garcia-Quintero}, C. and {Garrison}, L.~H. and {Gazta{\~n}aga}, E. and {Gil-Mar{\'\i}n}, H. and {Gnedin}, O.~Y. and {Gontcho}, S. Gontcho A and {Gonzalez-Morales}, A.~X. and {Gonzalez-Perez}, V. and {Gordon}, C. and {Graur}, O. and {Green}, D. and {Gruen}, D. and {Gsponer}, R. and {Guandalin}, C. and {Gutierrez}, G. and {Guy}, J. and {Hahn}, C. and {Han}, J.~J. and {Han}, J. and {He}, S. and {Herrera-Alcantar}, H.~K. and {Honscheid}, K. and {Hou}, J. and {Howlett}, C. and {Huterer}, D. and {Ir{\v{s}}i{\v{c}}}, V. and {Ishak}, M. and {Jacques}, A. and {Jimenez}, J. and {Jing}, Y.~P. and {Joachimi}, B. and {Joudaki}, S. and {Joyce}, R. and {Jullo}, E. and {Juneau}, S. and {Kara{\c{c}}ayl{\i}}, N.~G. and {Karim}, T. and {Kehoe}, R. and {Kent}, S. and {Khederlarian}, A. and {Kirkby}, D. and {Kisner}, T. and {Kitaura}, F. -S. and {Kizhuprakkat}, N. and {Kong}, H. and {Koposov}, S.~E. and {Kremin}, A. and {Krolewski}, A. and {Lahav}, O. and {Lai}, Y. and {Lamman}, C. and {Lan}, T. -W. and {Landriau}, M. and {Lang}, D. and {Lange}, J.~U. and {Lasker}, J. and {Le Goff}, J.~M. and {Le Guillou}, L. and {Leauthaud}, A. and {Levi}, M.~E. and {Li}, S. and {Li}, T.~S. and {Lodha}, K. and {Lokken}, M. and {Luo}, Y. and {Magneville}, C. and {Manera}, M. and {Manser}, C.~J. and {Margala}, D. and {Martini}, P. and {Maus}, M. and {McCullough}, J. and {McDonald}, P. and {Medina}, G.~E. and {Medina-Varela}, L. and {Meisner}, A. and {Mena-Fern{\'a}ndez}, J. and {Menegas}, A. and {Mezcua}, M. and {Miquel}, R. and {Montero-Camacho}, P. and {Moon}, J. and {Moustakas}, J. and {Mu{\~n}oz-Guti{\'e}rrez}, A. and {Mu{\~n}oz-Santos}, D. and {Myers}, A.~D. and {Myles}, J. and {Nadathur}, S. and {Najita}, J. and {Napolitano}, L. and {Newman}, J.~A. and {Nikakhtar}, F. and {Nikutta}, R. and {Niz}, G. and {Noriega}, H.~E. and {Padmanabhan}, N. and {Paillas}, E. and {Palanque-Delabrouille}, N. and {Palmese}, A. and {Pan}, J. and {Pan}, Z. and {Parkinson}, D. and {Peacock}, J.~A.  and {Percival}, W.~J. and {P{\'e}rez-Fern{\'a}ndez}, A. and {P{\'e}rez-R{\`a}fols}, I. and {Peterson}, P.},
        title = "{Data Release 1 of the Dark Energy Spectroscopic Instrument}",
      journal = {arXiv e-prints},
         year = 2025,
        month = mar,
          eid = {arXiv:2503.14745},
        pages = {arXiv:2503.14745},
          doi = {10.48550/arXiv.2503.14745},
archivePrefix = {arXiv},
       eprint = {2503.14745},
 primaryClass = {astro-ph.CO},
       adsurl = {https://ui.adsabs.harvard.edu/abs/2025arXiv250314745D}
}

@ARTICLE{DESI2024.VII.KP7B,
       author = {{DESI Collaboration} and {Adame}, A.~G. and {Aguilar}, J. and {Ahlen}, S. and {Alam}, S. and {Alexander}, D.~M. and {Allende Prieto}, C. and {Alvarez}, M. and {Alves}, O. and {Anand}, A. and {Andrade}, U. and {Armengaud}, E. and {Avila}, S. and {Aviles}, A. and {Awan}, H. and {Bahr-Kalus}, B. and {Bailey}, S. and {Baltay}, C. and {Bault}, A. and {Behera}, J. and {BenZvi}, S. and {Beutler}, F. and {Bianchi}, D. and {Blake}, C. and {Blum}, R. and {Bonici}, M. and {Brieden}, S. and {Brodzeller}, A. and {Brooks}, D. and {Buckley-Geer}, E. and {Burtin}, E. and {Calderon}, R. and {Canning}, R. and {Carnero Rosell}, A. and {Cereskaite}, R. and {Cervantes-Cota}, J.~L. and {Chabanier}, S. and {Chaussidon}, E. and {Chaves-Montero}, J. and {Chebat}, D. and {Chen}, S. and {Chen}, X. and {Claybaugh}, T. and {Cole}, S. and {Cuceu}, A. and {Davis}, T.~M. and {Dawson}, K. and {de la Macorra}, A. and {de Mattia}, A. and {Deiosso}, N. and {Dey}, A. and {Dey}, B. and {Ding}, Z. and {Doel}, P. and {Edelstein}, J. and {Eftekharzadeh}, S. and {Eisenstein}, D.~J. and {Elbers}, W. and {Elliott}, A. and {Fagrelius}, P. and {Fanning}, K. and {Ferraro}, S. and {Ereza}, J. and {Findlay}, N. and {Flaugher}, B. and {Font-Ribera}, A. and {Forero-S{\'a}nchez}, D. and {Forero-Romero}, J.~E. and {Frenk}, C.~S. and {Garcia-Quintero}, C. and {Garrison}, L.~H. and {Gazta{\~n}aga}, E. and {Gil-Mar{\'\i}n}, H. and {Gontcho}, S. Gontcho A. and {Gonzalez-Morales}, A.~X. and {Gonzalez-Perez}, V. and {Gordon}, C. and {Green}, D. and {Gruen}, D. and {Gsponer}, R. and {Gutierrez}, G. and {Guy}, J. and {Hadzhiyska}, B. and {Hahn}, C. and {Hanif}, M.~M.~S. and {Herrera-Alcantar}, H.~K. and {Honscheid}, K. and {Howlett}, C. and {Huterer}, D. and {Ir{\v{s}}i{\v{c}}}, V. and {Ishak}, M. and {Joyce}, R. and {Juneau}, S. and {Kara{\c{c}}ayl{\i}}, N.~G. and {Kehoe}, R. and {Kent}, S. and {Kirkby}, D. and {Kong}, H. and {Koposov}, S.~E. and {Kremin}, A. and {Krolewski}, A. and {Lahav}, O. and {Lai}, Y. and {Lan}, T. -W. and {Landriau}, M. and {Lang}, D. and {Lasker}, J. and {Le Goff}, J.~M. and {Le Guillou}, L. and {Leauthaud}, A. and {Levi}, M.~E. and {Li}, T.~S. and {Lodha}, K. and {Magneville}, C. and {Manera}, M. and {Margala}, D. and {Martini}, P. and {Matthewson}, W. and {Maus}, M. and {McDonald}, P. and {Medina-Varela}, L. and {Meisner}, A. and {Mena-Fern{\'a}ndez}, J. and {Miquel}, R. and {Moon}, J. and {Moore}, S. and {Moustakas}, J. and {Mudur}, N. and {Mueller}, E. and {Mu{\~n}oz-Guti{\'e}rrez}, A. and {Myers}, A.~D. and {Nadathur}, S. and {Napolitano}, L. and {Neveux}, R. and {Newman}, J.~A. and {Nguyen}, N.~M. and {Nie}, J. and {Niz}, G. and {Noriega}, H.~E. and {Padmanabhan}, N. and {Paillas}, E. and {Palanque-Delabrouille}, N. and {Pan}, J. and {Penmetsa}, S. and {Percival}, W.~J. and {Pieri}, M.~M. and {Pinon}, M. and {Poppett}, C. and {Porredon}, A. and {Prada}, F. and {P{\'e}rez-Fern{\'a}ndez}, A. and {P{\'e}rez-R{\`a}fols}, I. and {Rabinowitz}, D. and {Raichoor}, A. and {Ram{\'\i}rez-P{\'e}rez}, C. and {Ramirez-Solano}, S. and {Rashkovetskyi}, M. and {Ravoux}, C. and {Rezaie}, M. and {Rich}, J. and {Rocher}, A. and {Rockosi}, C. and {Roe}, N.~A. and {Rosado-Marin}, A. and {Ross}, A.~J. and {Rossi}, G. and {Ruggeri}, R. and {Ruhlmann-Kleider}, V. and {Samushia}, L. and {Sanchez}, E. and {Saulder}, C. and {Schlafly}, E.~F. and {Schlegel}, D. and {Schubnell}, M. and {Seo}, H. and {Shafieloo}, A. and {Sharples}, R. and {Silber}, J. and {Slosar}, A. and {Smith}, A. and {Sprayberry}, D. and {Tan}, T. and {Tarl{\'e}}, G. and {Taylor}, P. and {Trusov}, S. and {Vaisakh}, R. and {Valcin}, D. and {Valdes}, F. and {Valogiannis}, G. and {Vargas-Maga{\~n}a}, M. and {Verde}, L. and {Walther}, M. and {Wang}, B. and {Wang}, M.~S. and {Weaver}, B.~A. and {Weaverdyck}, N. and {Wechsler}, R.~H. and {Weinberg}, D.~H. and {White}, M. and {Wilson}, M.~J. and {Yi}, L.},
        title = "{DESI 2024 VII: cosmological constraints from the full-shape modeling of clustering measurements}",
      journal = {\jcap},
         year = 2025,
        month = jul,
       volume = {2025},
       number = {7},
          eid = {028},
        pages = {028},
          doi = {10.1088/1475-7516/2025/07/028},
archivePrefix = {arXiv},
       eprint = {2411.12022},
 primaryClass = {astro-ph.CO},
       adsurl = {https://ui.adsabs.harvard.edu/abs/2025JCAP...07..028A}
}

@ARTICLE{DESI.DR2.DR2,
  title = {DESI DR2 results. II. Measurements of baryon acoustic oscillations and cosmological constraints},
  author = {Abdul Karim, M. and Aguilar, J. and Ahlen, S. and Alam, S. and Allen, L. and Prieto, C. Allende and Alves, O. and Anand, A. and Andrade, U. and Armengaud, E. and Aviles, A. and Bailey, S. and Baltay, C. and Bansal, P. and Bault, A. and Behera, J. and BenZvi, S. and Bianchi, D. and Blake, C. and Brieden, S. and Brodzeller, A. and Brooks, D. and Buckley-Geer, E. and Burtin, E. and Calderon, R. and Canning, R. and Rosell, A. Carnero and Carrilho, P. and Casas, L. and Castander, F. J. and Charles, M. and Chaussidon, E. and Chaves-Montero, J. and Chebat, D. and Chen, X. and Claybaugh, T. and Cole, S. and Cooper, A. P. and Cuceu, A. and Dawson, K. S. and de la Macorra, A. and de Mattia, A. and Deiosso, N. and Della Costa, J. and Demina, R. and Dey, A. and Dey, B. and Ding, Z. and Doel, P. and Edelstein, J. and Eisenstein, D. J. and Elbers, W. and Fagrelius, P. and Fanning, K. and Fern{\'a}ndez-Garc{\'i}a, E. and Ferraro, S. and Font-Ribera, A. and Forero-Romero, J. E. and Frenk, C. S. and Garcia-Quintero, C. and Garrison, L. H. and Gazta{\~n}aga, E. and Gil-Mar{\'i}n, H. and Gontcho, S. Gontcho A. and Gonzalez, D. and Gonzalez-Morales, A. X. and Gordon, C. and Green, D. and Gutierrez, G. and Guy, J. and Hadzhiyska, B. and Hahn, C. and He, S. and Herbold, M. and Herrera-Alcantar, H. K. and Ho, M.-F. and Honscheid, K. and Howlett, C. and Huterer, D. and Ishak, M. and Juneau, S. and Kamble, N. V. and Kara{\c{c}}ayl{\i}, N. G. and Kehoe, R. and Kent, S. and Kim, A. G. and Kirkby, D. and Kisner, T. and Koposov, S. E. and Kremin, A. and Krolewski, A. and Lahav, O. and Lamman, C. and Landriau, M. and Lang, D. and Lasker, J. and Le Goff, J. M. and Le Guillou, L. and Leauthaud, A. and Levi, M. E. and Li, Q. and Li, T. S. and Lodha, K. and Lokken, M. and Lozano-Rodr{\'i}guez, F. and Magneville, C. and Manera, M. and Martini, P. and Matthewson, W. L. and Meisner, A. and Mena-Fern{\'a}ndez, J. and Menegas, A. and Mergulh{\~a}o, T. and Miquel, R. and Moustakas, J. and Mu{\~n}oz-Guti{\'e}rrez, A. and Mu{\~n}oz-Santos, D. and Myers, A. D. and Nadathur, S. and Naidoo, K. and Napolitano, L. and Newman, J. A. and Niz, G. and Noriega, H. E. and Paillas, E. and Palanque-Delabrouille, N. and Pan, J. and Peacock, J. A. and Ibanez, M. P. and Percival, W. J. and P{\'e}rez-Fern{\'a}ndez, A. and P{\'e}rez-R{\`a}fols, I. and Pieri, M. M. and Poppett, C. and Prada, F. and Rabinowitz, D. and Raichoor, A. and Ram{\'i}rez-P{\'e}rez, C. and Rashkovetskyi, M. and Ravoux, C. and Rich, J. and Rocher, A. and Rockosi, C. and Rohlf, J. and Rom{\'a}n-Herrera, J. O. and Ross, A. J. and Rossi, G. and Ruggeri, R. and Ruhlmann-Kleider, V. and Samushia, L. and Sanchez, E. and Sanders, N. and Schlegel, D. and Schubnell, M. and Seo, H. and Shafieloo, A. and Sharples, R. and Silber, J. and Sinigaglia, F. and Sprayberry, D. and Tan, T. and Tarl{\'e}, G. and Taylor, P. and Turner, W. and Ure{\~n}a-L{\'o}pez, L. A. and Vaisakh, R. and Valdes, F. and Valogiannis, G. and Vargas-Maga{\~n}a, M. and Verde, L. and Walther, M. and Weaver, B. A. and Weinberg, D. H. and White, M. and Wolfson, M. and Y{\`e}che, C. and Yu, J. and Zaborowski, E. A. and Zarrouk, P. and Zhai, Z. and Zhang, H. and Zhao, C. and Zhao, G. B. and Zhou, R. and Zou, H.},
  collaboration = {DESI Collaboration},
  journal = {Phys. Rev. D},
  volume = {112},
  issue = {8},
  pages = {083515},
  numpages = {40},
  year = {2025},
  month = {Oct},
  publisher = {American Physical Society},
  doi = {10.1103/tr6y-kpc6},
  url = {https://link.aps.org/doi/10.1103/tr6y-kpc6},
  archivePrefix = {arXiv},
  eprint = {2503.14738},
  primaryClass = {astro-ph.CO}
}

@article{DESI2023.KP1.SV,
  author = {{DESI Collaboration} and Adame, A. G. and Aguilar, J. and others},
  title = {Validation of the Scientific Program for the Dark Energy Spectroscopic Instrument},
  journal = {AJ},
  volume = {167},
  pages = {62},
  year = {2024},
  doi = {10.3847/1538-3881/ad0b08},
  archivePrefix = {arXiv},
  eprint = {2306.06307},
  primaryClass = {astro-ph.CO}
}

@misc{data_zenodo,
  author       = {Qu, Frank J. and Ried Guachalla, Bernardita and Schaan, Emmanuel and Hadzhiyska, Boryana and Ferraro, Simone},
  title        = {Data products for: Precision Kinematic Sunyaev-Zel'dovich Measurements with DESI DR2 and ACT DR6: Part I. Luminous Red Galaxies},
  year         = {2025},
  publisher    = {Zenodo},
  doi          = {10.5281/zenodo.18334134},
  url          = {https://zenodo.org/uploads/18334134}
}

@article{Knox_1995,
   title={Determination of inflationary observables by cosmic microwave background anisotropy experiments},
   volume={52},
   ISSN={0556-2821},
   url={http://dx.doi.org/10.1103/PhysRevD.52.4307},
   DOI={10.1103/physrevd.52.4307},
   number={8},
   journal={Physical Review D},
   publisher={American Physical Society (APS)},
   author={Knox, Lloyd},
   year={1995},
   month=oct, pages={4307–4318} }

@article{Paillas_2025,
   title={Optimal reconstruction of baryon acoustic oscillations for DESI 2024},
   volume={2025},
   ISSN={1475-7516},
   url={http://dx.doi.org/10.1088/1475-7516/2025/01/142},
   DOI={10.1088/1475-7516/2025/01/142},
   number={01},
   journal={Journal of Cosmology and Astroparticle Physics},
   publisher={IOP Publishing},
   author={Paillas, E. and Ding, Z. and Chen, X. and Seo, H. and Padmanabhan, N. and de Mattia, A. and Ross, A.J. and Nadathur, S. and Howlett, C. and Aguilar, J. and Ahlen, S. and Alves, O. and Andrade, U. and Brooks, D. and Buckley-Geer, E. and Burtin, E. and Chen, S. and Claybaugh, T. and Cole, S. and Dawson, K. and de la Macorra, A. and Dey, Arjun and Doel, P. and Fanning, K. and Ferraro, S. and Forero-Romero, J.E. and Garcia-Quintero, C. and Gaztañaga, E. and Gil-Marín, H. and Gontcho, S.Gontcho A. and Gutierrez, G. and Hahn, C. and Hanif, M.M.S. and Honscheid, K. and Ishak, M. and Kehoe, R. and Kremin, A. and Landriau, M. and Le Guillou, L. and Levi, M.E. and Manera, M. and Martini, P. and Medina-Varela, L. and Meisner, A. and Mena-Fernández, J. and Miquel, R. and Moustakas, J. and Mueller, E. and Muñoz-Gutiérrez, A. and Myers, A.D. and Newman, J.A. and Nie, J. and Niz, G. and Palanque-Delabrouille, N. and Percival, W.J. and Poppett, C. and Prada, F. and Pérez-Fernández, A. and Rashkovetskyi, M. and Rezaie, M. and Rosado-Marin, A. and Rossi, G. and Ruggeri, R. and Sanchez, E. and Saulder, C. and Schlafly, E.F. and Schlegel, D. and Schubnell, M. and Sprayberry, D. and Tarlé, G. and Valcin, D. and Vargas-Magaña, M. and Yu, J. and Yuan, S. and Zhou, R. and Zou, H.},
   year={2025},
   month=jan, pages={142} }

@misc{chen2024extensiveanalysisreconstructionalgorithms,
      title={Extensive analysis of reconstruction algorithms for DESI 2024 baryon acoustic oscillations}, 
      author={X. Chen and Z. Ding and E. Paillas and S. Nadathur and H. Seo and S. Chen and N. Padmanabhan and M. White and A. de Mattia and P. McDonald and A. J. Ross and A. Variu and A. Carnero Rosell and B. Hadzhiyska and M. M. S Hanif and D. Forero-Sánchez and S. Ahlen and O. Alves and U. Andrade and S. BenZvi and D. Bianchi and D. Brooks and E. Chaussidon and T. Claybaugh and A. de la Macorra and Biprateep Dey and K. Fanning and S. Ferraro and A. Font-Ribera and J. E. Forero-Romero and C. Garcia-Quintero and E. Gaztañaga and S. Gontcho A Gontcho and G. Gutierrez and C. Hahn and K. Honscheid and S. Juneau and R. Kehoe and D. Kirkby and T. Kisner and A. Kremin and M. E. Levi and A. Meisner and J. Mena-Fernández and R. Miquel and J. Moustakas and A. Muñoz-Gutiérrez and F. Nikakhtar and N. Palanque-Delabrouille and W. J. Percival and F. Prada and I. Pérez-Ràfols and M. Rashkovetskyi and G. Rossi and R. Ruggeri and E. Sanchez and C. Saulder and D. Schlegel and M. Schubnell and A. Smith and D. Sprayberry and G. Tarlé and D. Valcin and M. Vargas-Magaña and B. A. Weaver and S. Yuan and R. Zhou},
      year={2024},
      eprint={2411.19738},
      archivePrefix={arXiv},
      primaryClass={astro-ph.CO},
      url={https://arxiv.org/abs/2411.19738}, 
}

@ARTICLE{1994ApJ...421L...1N,
       author = {{Nusser}, Adi and {Davis}, Marc},
        title = "{On the Prediction of Velocity Fields from Redshift Space Galaxy Samples}",
      journal = {\apjl},
         year = 1994,
        month = jan,
       volume = {421},
        pages = {L1},
          doi = {10.1086/187172},
archivePrefix = {arXiv},
       eprint = {astro-ph/9309009},
 primaryClass = {astro-ph},
       adsurl = {https://ui.adsabs.harvard.edu/abs/1994ApJ...421L...1N}
}

@article{
DESI2024.III.KP4,
   title={DESI 2024 III: baryon acoustic oscillations from galaxies and quasars},
   volume={2025},
   ISSN={1475-7516},
   url={http://dx.doi.org/10.1088/1475-7516/2025/04/012},
   DOI={10.1088/1475-7516/2025/04/012},
   number={04},
   journal={Journal of Cosmology and Astroparticle Physics},
   publisher={IOP Publishing},
   author={Adame, A.G. and Aguilar, J. and Ahlen, S. and Alam, S. and Alexander, D.M. and Alvarez, M. and Alves, O. and Anand, A. and Andrade, U. and Armengaud, E. and Avila, S. and Aviles, A. and Awan, H. and Bailey, S. and Baltay, C. and Bault, A. and Behera, J. and BenZvi, S. and Beutler, F. and Bianchi, D. and Blake, C. and Blum, R. and Brieden, S. and Brodzeller, A. and Brooks, D. and Buckley-Geer, E. and Burtin, E. and Calderon, R. and Canning, R. and Carnero Rosell, A. and Cereskaite, R. and Cervantes-Cota, J.L. and Chabanier, S. and Chaussidon, E. and Chaves-Montero, J. and Chen, S. and Chen, X. and Claybaugh, T. and Cole, S. and Cuceu, A. and Davis, T.M. and Dawson, K. and de la Macorra, A. and de Mattia, A. and Deiosso, N. and Dey, A. and Dey, B. and Ding, Z. and Doel, P. and Edelstein, J. and Eftekharzadeh, S. and Eisenstein, D.J. and Elliott, A. and Fagrelius, P. and Fanning, K. and Ferraro, S. and Ereza, J. and Findlay, N. and Flaugher, B. and Font-Ribera, A. and Forero-Sánchez, D. and Forero-Romero, J.E. and Garcia-Quintero, C. and Gaztañaga, E. and Gil-Marín, H. and Gontcho, S.Gontcho A. and Gonzalez-Morales, A.X. and Gonzalez-Perez, V. and Gordon, C. and Green, D. and Gruen, D. and Gsponer, R. and Gutierrez, G. and Guy, J. and Hadzhiyska, B. and Hahn, C. and Hanif, M.M.S. and Herrera-Alcantar, H.K. and Honscheid, K. and Howlett, C. and Huterer, D. and Iršič, V. and Ishak, M. and Juneau, S. and Karaçaylı, N.G. and Kehoe, R. and Kent, S. and Kirkby, D. and Kong, H. and Kremin, A. and Krolewski, A. and Lai, Y. and Lan, T.-W. and Landriau, M. and Lang, D. and Lasker, J. and Le Goff, J.M. and Le Guillou, L. and Leauthaud, A. and Levi, M.E. and Li, T.S. and Linder, E. and Lodha, K. and Magneville, C. and Manera, M. and Margala, D. and Martini, P. and Maus, M. and McDonald, P. and Medina-Varela, L. and Meisner, A. and Mena-Fernández, J. and Miquel, R. and Moon, J. and Moore, S. and Moustakas, J. and Mueller, E. and Muñoz-Gutiérrez, A. and Myers, A.D. and Nadathur, S. and Napolitano, L. and Neveux, R. and Newman, J.A. and Nguyen, N.M. and Nie, J. and Niz, G. and Noriega, H.E. and Padmanabhan, N. and Paillas, E. and Palanque-Delabrouille, N. and Pan, J. and Penmetsa, S. and Percival, W.J. and Pieri, M.M. and Pinon, M. and Poppett, C. and Porredon, A. and Prada, F. and Pérez-Fernández, A. and Pérez-Ràfols, I. and Rabinowitz, D. and Raichoor, A. and Ramírez-Pérez, C. and Ramirez-Solano, S. and Rashkovetskyi, M. and Ravoux, C. and Rezaie, M. and Rich, J. and Rocher, A. and Rockosi, C. and Roe, N.A. and Rosado-Marin, A. and Ross, A.J. and Rossi, G. and Ruggeri, R. and Ruhlmann-Kleider, V. and Samushia, L. and Sanchez, E. and Saulder, C. and Schlafly, E.F. and Schlegel, D. and Schubnell, M. and Seo, H. and Sharples, R. and Silber, J. and Slosar, A. and Smith, A. and Sprayberry, D. and Swanson, J. and Tan, T. and Tarlé, G. and Trusov, S. and Vaisakh, R. and Valcin, D. and Valdes, F. and Vargas-Magaña, M. and Verde, L. and Walther, M. and Wang, B. and Wang, M.S. and Weaver, B.A. and Weaverdyck, N. and Wechsler, R.H. and Weinberg, D.H. and White, M. and Wilson, M.J. and Yu, J. and Yu, Y. and Yuan, S. and Yèche, C. and Zaborowski, E.A. and Zarrouk, P. and Zhang, H. and Zhao, C. and Zhao, R. and Zhou, R. and Zou, H.},
   year={2025},
   month=apr, pages={012} }

@ARTICLE{2020A&A...641A...6P,
       author = {{Planck Collaboration} and {Aghanim}, N. and {Akrami}, Y. and {Ashdown}, M. and {Aumont}, J. and {Baccigalupi}, C. and {Ballardini}, M. and {Banday}, A.~J. and {Barreiro}, R.~B. and {Bartolo}, N. and {Basak}, S. and {Battye}, R. and {Benabed}, K. and {Bernard}, J.-P. and {Bersanelli}, M. and {Bielewicz}, P. and {Bock}, J.~J. and {Bond}, J.~R. and {Borrill}, J. and {Bouchet}, F.~R. and {Boulanger}, F. and {Bucher}, M. and {Burigana}, C. and {Butler}, R.~C. and {Calabrese}, E. and {Cardoso}, J.-F. and {Carron}, J. and {Challinor}, A. and {Chiang}, H.~C. and {Chluba}, J. and {Colombo}, L.~P.~L. and {Combet}, C. and {Contreras}, D. and {Crill}, B.~P. and {Cuttaia}, F. and {de Bernardis}, P. and {de Zotti}, G. and {Delabrouille}, J. and {Delouis}, J.-M. and {Di Valentino}, E. and {Diego}, J.~M. and {Dor{\'e}}, O. and {Douspis}, M. and {Ducout}, A. and {Dupac}, X. and {Dusini}, S. and {Efstathiou}, G. and {Elsner}, F. and {En{\ss}lin}, T.~A. and {Eriksen}, H.~K. and {Fantaye}, Y. and {Farhang}, M. and {Fergusson}, J. and {Fernandez-Cobos}, R. and {Finelli}, F. and {Forastieri}, F. and {Frailis}, M. and {Fraisse}, A.~A. and {Franceschi}, E. and {Frolov}, A. and {Galeotta}, S. and {Galli}, S. and {Ganga}, K. and {G{\'e}nova-Santos}, R.~T. and {Gerbino}, M. and {Ghosh}, T. and {Gonz{\'a}lez-Nuevo}, J. and {G{\'o}rski}, K.~M. and {Gratton}, S. and {Gruppuso}, A. and {Gudmundsson}, J.~E. and {Hamann}, J. and {Handley}, W. and {Hansen}, F.~K. and {Herranz}, D. and {Hildebrandt}, S.~R. and {Hivon}, E. and {Huang}, Z. and {Jaffe}, A.~H. and {Jones}, W.~C. and {Karakci}, A. and {Keih{\"a}nen}, E. and {Keskitalo}, R. and {Kiiveri}, K. and {Kim}, J. and {Kisner}, T.~S. and {Knox}, L. and {Krachmalnicoff}, N. and {Kunz}, M. and {Kurki-Suonio}, H. and {Lagache}, G. and {Lamarre}, J.-M. and {Lasenby}, A. and {Lattanzi}, M. and {Lawrence}, C.~R. and {Le Jeune}, M. and {Lemos}, P. and {Lesgourgues}, J. and {Levrier}, F. and {Lewis}, A. and {Liguori}, M. and {Lilje}, P.~B. and {Lilley}, M. and {Lindholm}, V. and {L{\'o}pez-Caniego}, M. and {Lubin}, P.~M. and {Ma}, Y.-Z. and {Mac{\'\i}as-P{\'e}rez}, J.~F. and {Maggio}, G. and {Maino}, D. and {Mandolesi}, N. and {Mangilli}, A. and {Marcos-Caballero}, A. and {Maris}, M. and {Martin}, P.~G. and {Martinelli}, M. and {Mart{\'\i}nez-Gonz{\'a}lez}, E. and {Matarrese}, S. and {Mauri}, N. and {McEwen}, J.~D. and {Meinhold}, P.~R. and {Melchiorri}, A. and {Mennella}, A. and {Migliaccio}, M. and {Millea}, M. and {Mitra}, S. and {Miville-Desch{\^e}nes}, M.-A. and {Molinari}, D. and {Montier}, L. and {Morgante}, G. and {Moss}, A. and {Natoli}, P. and {N{\o}rgaard-Nielsen}, H.~U. and {Pagano}, L. and {Paoletti}, D. and {Partridge}, B. and {Patanchon}, G. and {Peiris}, H.~V. and {Perrotta}, F. and {Pettorino}, V. and {Piacentini}, F. and {Polastri}, L. and {Polenta}, G. and {Puget}, J.-L. and {Rachen}, J.~P. and {Reinecke}, M. and {Remazeilles}, M. and {Renzi}, A. and {Rocha}, G. and {Rosset}, C. and {Roudier}, G. and {Rubi{\~n}o-Mart{\'\i}n}, J.~A. and {Ruiz-Granados}, B. and {Salvati}, L. and {Sandri}, M. and {Savelainen}, M. and {Scott}, D. and {Shellard}, E.~P.~S. and {Sirignano}, C. and {Sirri}, G. and {Spencer}, L.~D. and {Sunyaev}, R. and {Suur-Uski}, A.-S. and {Tauber}, J.~A. and {Tavagnacco}, D. and {Tenti}, M. and {Toffolatti}, L. and {Tomasi}, M. and {Trombetti}, T. and {Valenziano}, L. and {Valiviita}, J. and {Van Tent}, B. and {Vibert}, L. and {Vielva}, P. and {Villa}, F. and {Vittorio}, N. and {Wandelt}, B.~D. and {Wehus}, I.~K. and {White}, M. and {White}, S.~D.~M. and {Zacchei}, A. and {Zonca}, A.},
        title = "{Planck 2018 results. VI. Cosmological parameters}",
      journal = {\aap},
         year = 2020,
        month = sep,
       volume = {641},
          eid = {A6},
        pages = {A6},
          doi = {10.1051/0004-6361/201833910},
archivePrefix = {arXiv},
       eprint = {1807.06209},
 primaryClass = {astro-ph.CO},
       adsurl = {https://ui.adsabs.harvard.edu/abs/2020A&A...641A...6P}
}

@article{Baleato_Lizancos_2025,
   title={Selecting samples of galaxies with fewer Fingers-of-God},
   volume={2025},
   ISSN={1475-7516},
   url={http://dx.doi.org/10.1088/1475-7516/2025/07/014},
   DOI={10.1088/1475-7516/2025/07/014},
   number={07},
   journal={Journal of Cosmology and Astroparticle Physics},
   publisher={IOP Publishing},
   author={Baleato Lizancos, Antón and Seljak, Uroš and Karamanis, Minas and Bonici, Marco and Ferraro, Simone},
   year={2025},
   month=jul, pages={014} }

@misc{hotinli2025velocityreconstructionkszmeasuring,
      title={Velocity Reconstruction from KSZ: Measuring $f_{NL}$ with ACT and DESILS}, 
      author={Selim C. Hotinli and Kendrick M. Smith and Simone Ferraro},
      year={2025},
      eprint={2506.21657},
      archivePrefix={arXiv},
      primaryClass={astro-ph.CO},
      url={https://arxiv.org/abs/2506.21657}, 
}

@misc{lai2025kszvelocityreconstructionact,
      title={KSZ Velocity Reconstruction with ACT and DESI-LS using a Tomographic QML Power Spectrum Estimator}, 
      author={Anderson C. M. Lai and Yurii Kvasiuk and Moritz Münchmeyer},
      year={2025},
      eprint={2506.21684},
      archivePrefix={arXiv},
      primaryClass={astro-ph.CO},
      url={https://arxiv.org/abs/2506.21684}, 
}

@misc{lague2024constraintslocalprimordialnongaussianity,
      title={Constraints on local primordial non-Gaussianity with 3d Velocity Reconstruction from the Kinetic Sunyaev-Zeldovich Effect}, 
      author={Alex Laguë and Mathew S. Madhavacheril and Kendrick M. Smith and Simone Ferraro and Emmanuel Schaan},
      year={2024},
      eprint={2411.08240},
      archivePrefix={arXiv},
      primaryClass={astro-ph.CO},
      url={https://arxiv.org/abs/2411.08240}, 
}

@misc{mccarthy2025atacamacosmologytelescopecrosscorrelation,
      title={The Atacama Cosmology Telescope: Cross-correlation of kSZ and continuity equation velocity reconstruction with photometric DESI LRGs}, 
      author={Fiona McCarthy and Boryana Hadzhiyska and J. Richard Bond and William R. Coulton and Jo Dunkley and Carmen Embil Villagra and Matthew C. Johnson and Kavilan Moodley and Toshiya Namikawa and Bernardita Ried Guachalla and Blake D. Sherwin and Cristóbal Sifón and Alexander van Engelen and Eve M. Vavagiakis and Edward J. Wollack},
      year={2025},
      eprint={2511.15701},
      archivePrefix={arXiv},
      primaryClass={astro-ph.CO},
      url={https://arxiv.org/abs/2511.15701}, 
}

@misc{mccarthy2024atacamacosmologytelescopelargescale,
      title={The Atacama Cosmology Telescope: Large-scale velocity reconstruction with the kinematic Sunyaev--Zel'dovich effect and DESI LRGs}, 
      author={Fiona McCarthy and Nicholas Battaglia and Rachel Bean and J. Richard Bond and Hongbo Cai and Erminia Calabrese and William R. Coulton and Mark J. Devlin and Jo Dunkley and Simone Ferraro and Vera Gluscevic and Yilun Guan and J. Colin Hill and Matthew C. Johnson and Aleksandra Kusiak and Alex Laguë and Niall MacCrann and Mathew S. Madhavacheril and Kavilan Moodley and Sigurd Naess and Frank J. Qu and Bernardita Ried Guachalla and Neelima Sehgal and Blake D. Sherwin and Cristóbal Sifón and Kendrick M. Smith and Suzanne T. Staggs and Alexander van Engelen and Eve M. Vavagiakis and Edward J. Wollack},
      year={2024},
      eprint={2410.06229},
      archivePrefix={arXiv},
      primaryClass={astro-ph.CO},
      url={https://arxiv.org/abs/2410.06229}, 
}

@article{Zheng_2007,
   title={Galaxy Evolution from Halo Occupation Distribution Modeling of DEEP2 and SDSS Galaxy Clustering},
   volume={667},
   ISSN={1538-4357},
   url={http://dx.doi.org/10.1086/521074},
   DOI={10.1086/521074},
   number={2},
   journal={The Astrophysical Journal},
   publisher={American Astronomical Society},
   author={Zheng, Zheng and Coil, Alison L. and Zehavi, Idit},
   year={2007},
   month=oct, pages={760–779} }

@article{Zhang_2026,
   title={AbacusHF-v2: HOD Mock Catalogs from the AbacusSummit Simulations},
   volume={},
   ISSN={},
   url={},
   DOI={},
   number={},
   journal={in prep.},
   publisher={},
   author={Zhang, Hanyu and others},
   year={2026},
   month={}, pages={} }

@article{Schneider_2022,
   title={Constraining baryonic feedback and cosmology with weak-lensing, X-ray, and kinematic Sunyaev–Zeldovich observations},
   volume={514},
   ISSN={1365-2966},
   url={http://dx.doi.org/10.1093/mnras/stac1493},
   DOI={10.1093/mnras/stac1493},
   number={3},
   journal={Monthly Notices of the Royal Astronomical Society},
   publisher={Oxford University Press (OUP)},
   author={Schneider, Aurel and Giri, Sambit K and Amodeo, Stefania and Refregier, Alexandre},
   year={2022},
   month=jun, pages={3802–3814} }

@misc{bigwood2025kineticsunyaevzeldovicheffect,
      title={The kinetic Sunyaev Zeldovich effect as a benchmark for AGN feedback models in hydrodynamical simulations: insights from DESI + ACT}, 
      author={Leah Bigwood and Masaya Yamamoto and Jared Siegel and Alexandra Amon and Ian G. McCarthy and Romeel Dave and Jaime Salcido and Matthieu Schaller and Joop Schaye and Tianyi Yang},
      year={2025},
      eprint={2510.15822},
      archivePrefix={arXiv},
      primaryClass={astro-ph.CO},
      url={https://arxiv.org/abs/2510.15822}, 
}

@misc{kovac2025baryonificationiiconstrainingfeedback,
      title={Baryonification II: Constraining feedback with X-ray and kinematic Sunyaev-Zel'dovich observations}, 
      author={Michael Kovač and Andrina Nicola and Jozef Bucko and Aurel Schneider and Robert Reischke and Sambit K. Giri and Romain Teyssier and Matthieu Schaller and Joop Schaye},
      year={2025},
      eprint={2507.07991},
      archivePrefix={arXiv},
      primaryClass={astro-ph.CO},
      url={https://arxiv.org/abs/2507.07991}, 
}

@article{Kravtsov_2018,
   title={Stellar Mass—Halo Mass Relation and Star Formation Efficiency in High-Mass Halos},
   volume={44},
   ISSN={1562-6873},
   url={http://dx.doi.org/10.1134/S1063773717120015},
   DOI={10.1134/s1063773717120015},
   number={1},
   journal={Astronomy Letters},
   publisher={Pleiades Publishing Ltd},
   author={Kravtsov, A. V. and Vikhlinin, A. A. and Meshcheryakov, A. V.},
   year={2018},
   month=jan, pages={8–34} }

@misc{act_0,
      title={The Atacama Cosmology Telescope: {DR6} Maps}, 
      author={Sigurd Naess and Yilun Guan and Adriaan J. Duivenvoorden and Matthew Hasselfield and Yuhan Wang et al},
      year={2025},
      eprint={2503.14451},
      archivePrefix={arXiv},
      primaryClass={astro-ph.CO},
      url={https://arxiv.org/abs/2503.14451}, 
}

@ARTICLE{hilton2021,
       author = {{Hilton}, M. and {Sif{\'o}n}, C. and {Naess}, S. and {Madhavacheril}, M. and {Oguri}, M. and {Rozo}, E. and {Rykoff}, E. and {Abbott}, T.~M.~C. and {Adhikari}, S. and {Aguena}, M. and {Aiola}, S. and {Allam}, S. and {Amodeo}, S. and {Amon}, A. and {Annis}, J. and {Ansarinejad}, B. and {Aros-Bunster}, C. and {Austermann}, J.~E. and {Avila}, S. and {Bacon}, D. and {Battaglia}, N. and {Beall}, J.~A. and {Becker}, D.~T. and {Bernstein}, G.~M. and {Bertin}, E. and {Bhandarkar}, T. and {Bhargava}, S. and {Bond}, J.~R. and {Brooks}, D. and {Burke}, D.~L. and {Calabrese}, E. and {Carrasco Kind}, M. and {Carretero}, J. and {Choi}, S.~K. and {Choi}, A. and {Conselice}, C. and {da Costa}, L.~N. and {Costanzi}, M. and {Crichton}, D. and {Crowley}, K.~T. and {D{\"u}nner}, R. and {Denison}, E.~V. and {Devlin}, M.~J. and {Dicker}, S.~R. and {Diehl}, H.~T. and {Dietrich}, J.~P. and {Doel}, P. and {Duff}, S.~M. and {Duivenvoorden}, A.~J. and {Dunkley}, J. and {Everett}, S. and {Ferraro}, S. and {Ferrero}, I. and {Fert{\'e}}, A. and {Flaugher}, B. and {Frieman}, J. and {Gallardo}, P.~A. and {Garc{\'\i}a-Bellido}, J. and {Gaztanaga}, E. and {Gerdes}, D.~W. and {Giles}, P. and {Golec}, J.~E. and {Gralla}, M.~B. and {Grandis}, S. and {Gruen}, D. and {Gruendl}, R.~A. and {Gschwend}, J. and {Gutierrez}, G. and {Han}, D. and {Hartley}, W.~G. and {Hasselfield}, M. and {Hill}, J.~C. and {Hilton}, G.~C. and {Hincks}, A.~D. and {Hinton}, S.~R. and {Ho}, S.-P.~P. and {Honscheid}, K. and {Hoyle}, B. and {Hubmayr}, J. and {Huffenberger}, K.~M. and {Hughes}, J.~P. and {Jaelani}, A.~T. and {Jain}, B. and {James}, D.~J. and {Jeltema}, T. and {Kent}, S. and {Knowles}, K. and {Koopman}, B.~J. and {Kuehn}, K. and {Lahav}, O. and {Lima}, M. and {Lin}, Y.-T. and {Lokken}, M. and {Loubser}, S.~I. and {MacCrann}, N. and {Maia}, M.~A.~G. and {Marriage}, T.~A. and {Martin}, J. and {McMahon}, J. and {Melchior}, P. and {Menanteau}, F. and {Miquel}, R. and {Miyatake}, H. and {Moodley}, K. and {Morgan}, R. and {Mroczkowski}, T. and {Nati}, F. and {Newburgh}, L.~B. and {Niemack}, M.~D. and {Nishizawa}, A.~J. and {Ogando}, R.~L.~C. and {Orlowski-Scherer}, J. and {Page}, L.~A. and {Palmese}, A. and {Partridge}, B. and {Paz-Chinch{\'o}n}, F. and {Phakathi}, P. and {Plazas}, A.~A. and {Robertson}, N.~C. and {Romer}, A.~K. and {Carnero Rosell}, A. and {Salatino}, M. and {Sanchez}, E. and {Schaan}, E. and {Schillaci}, A. and {Sehgal}, N. and {Serrano}, S. and {Shin}, T. and {Simon}, S.~M. and {Smith}, M. and {Soares-Santos}, M. and {Spergel}, D.~N. and {Staggs}, S.~T. and {Storer}, E.~R. and {Suchyta}, E. and {Swanson}, M.~E.~C. and {Tarle}, G. and {Thomas}, D. and {To}, C. and {Trac}, H. and {Ullom}, J.~N. and {Vale}, L.~R. and {Van Lanen}, J. and {Vavagiakis}, E.~M. and {De Vicente}, J. and {Wilkinson}, R.~D. and {Wollack}, E.~J. and {Xu}, Z. and {Zhang}, Y.},
        title = "{The Atacama Cosmology Telescope: A Catalog of >4000 Sunyaev-Zel{\textquoteright}dovich Galaxy Clusters}",
      journal = {\apjs},
         year = 2021,
        month = mar,
       volume = {253},
       number = {1},
          eid = {3},
        pages = {3},
          doi = {10.3847/1538-4365/abd023},
archivePrefix = {arXiv},
       eprint = {2009.11043},
 primaryClass = {astro-ph.CO},
       adsurl = {https://ui.adsabs.harvard.edu/abs/2021ApJS..253....3H}
}

@ARTICLE{2016ApJS..227...21T,
       author = {{Thornton}, R.~J. and {Ade}, P.~A.~R. and {Aiola}, S. and {Angil{\`e}}, F.~E. and {Amiri}, M. and {Beall}, J.~A. and {Becker}, D.~T. and {Cho}, H.-M. and {Choi}, S.~K. and {Corlies}, P. and {Coughlin}, K.~P. and {Datta}, R. and {Devlin}, M.~J. and {Dicker}, S.~R. and {D{\"u}nner}, R. and {Fowler}, J.~W. and {Fox}, A.~E. and {Gallardo}, P.~A. and {Gao}, J. and {Grace}, E. and {Halpern}, M. and {Hasselfield}, M. and {Henderson}, S.~W. and {Hilton}, G.~C. and {Hincks}, A.~D. and {Ho}, S.~P. and {Hubmayr}, J. and {Irwin}, K.~D. and {Klein}, J. and {Koopman}, B. and {Li}, Dale and {Louis}, T. and {Lungu}, M. and {Maurin}, L. and {McMahon}, J. and {Munson}, C.~D. and {Naess}, S. and {Nati}, F. and {Newburgh}, L. and {Nibarger}, J. and {Niemack}, M.~D. and {Niraula}, P. and {Nolta}, M.~R. and {Page}, L.~A. and {Pappas}, C.~G. and {Schillaci}, A. and {Schmitt}, B.~L. and {Sehgal}, N. and {Sievers}, J.~L. and {Simon}, S.~M. and {Staggs}, S.~T. and {Tucker}, C. and {Uehara}, M. and {van Lanen}, J. and {Ward}, J.~T. and {Wollack}, E.~J.},
        title = "{The Atacama Cosmology Telescope: The Polarization-sensitive ACTPol Instrument}",
      journal = {\apjs},
         year = 2016,
        month = dec,
       volume = {227},
       number = {2},
          eid = {21},
        pages = {21},
          doi = {10.3847/1538-4365/227/2/21},
archivePrefix = {arXiv},
       eprint = {1605.06569},
 primaryClass = {astro-ph.IM},
       adsurl = {https://ui.adsabs.harvard.edu/abs/2016ApJS..227...21T}
}

@misc{tazi2023intuitionframeworkapplyinggps,
      title={Beyond Intuition, a Framework for Applying GPs to Real-World Data}, 
      author={Kenza Tazi and Jihao Andreas Lin and Ross Viljoen and Alex Gardner and ST John and Hong Ge and Richard E. Turner},
      year={2023},
      eprint={2307.03093},
      archivePrefix={arXiv},
      primaryClass={cs.LG},
      url={https://arxiv.org/abs/2307.03093}, 
}

@misc{ross2024constructionlargescalestructurecatalogs,
      title={The Construction of Large-scale Structure Catalogs for the Dark Energy Spectroscopic Instrument}, 
      author={A. J. Ross and J. Aguilar and S. Ahlen and S. Alam and A. Anand and S. Bailey and D. Bianchi and S. Brieden and D. Brooks and E. Burtin and A. Carnero Rosell and E. Chaussidon and T. Claybaugh and S. Cole and K. Dawson and A. de la Macorra and A. de Mattia and Arjun Dey and Biprateep Dey and P. Doel and K. Fanning and S. Ferraro and J. Ereza and A. Font-Ribera and J. E. Forero-Romero and E. Gaztañaga and H. Gil-Marín and S. Gontcho A Gontcho and A. X. Gonzalez-Morales and J. Guy and C. Hahn and S. Heydenreich and K. Honscheid and C. Howlett and M. Ishak and T. Karim and D. Kirkby and T. Kisner and H. Kong and A. Kremin and A. Krolewski and A. Lambert and M. Landriau and J. Lasker and L. Le Guillou and M. E. Levi and M. Manera and P. Martini and P. McDonald and A. Meisner and R. Miquel and J. Moon and J. Moustakas and A. Muñoz-Gutiérrez and A. D. Myers and S. Nadathur and L. Napolitano and J. A. Newman and J. Nie and G. Niz and N. Palanque-Delabrouille and W. J. Percival and C. Poppett and F. Prada and A. Raichoor and C. Ravoux and M. Rezaie and A. Rosado-Marin and G. Rossi and L. Samushia and E. Sanchez and E. F. Schlafly and D. Schlegel and H. Seo and A. Smith and D. Sprayberry and G. Tarlé and D. Valcin and M. Vargas-Magaña and B. A. Weaver and M. Wilson and J. Yu and P. Zarrouk and C. Zhao and R. Zhou and H. Zou},
      year={2024},
      eprint={2405.16593},
      archivePrefix={arXiv},
      primaryClass={astro-ph.CO},
      url={https://arxiv.org/abs/2405.16593}, 
}

@misc{andrade2025validationdesidr2measurements,
      title={Validation of the DESI DR2 Measurements of Baryon Acoustic Oscillations from Galaxies and Quasars}, 
      author={U. Andrade and E. Paillas and J. Mena-Fernández and Q. Li and A. J. Ross and S. Nadathur and M. Rashkovetskyi and A. Pérez-Fernández and H. Seo and N. Sanders and O. Alves and X. Chen and N. Deiosso and A. de Mattia and M. White and M. Abdul-Karim and S. Ahlen and E. Armengaud and A. Aviles and D. Bianchi and S. Brieden and A. Brodzeller and D. Brooks and E. Burtin and R. Calderon and R. Canning and A. Carnero Rosell and L. Casas and F. J. Castander and M. Charles and E. Chaussidon and J. Chaves-Montero and T. Claybaugh and S. Cole and A. Cuceu and K. S. Dawson and A. de la Macorra and J. Della Costa and A. Dey and B. Dey and Z. Ding and P. Doel and D. J. Eisenstein and W. Elbers and E. Fernández-García and S. Ferraro and A. Font-Ribera and J. E. Forero-Romero and C. Garcia-Quintero and L. H. Garrison and E. Gaztañaga and H. Gil-Marín and S. Gontcho A Gontcho and A. X. Gonzalez-Morales and C. Gordon and G. Gutierrez and J. Guy and C. Hahn and S. He and H. K. Herrera-Alcantar and K. Honscheid and C. Howlett and D. Huterer and M. Ishak and S. Juneau and R. Kehoe and D. Kirkby and T. Kisner and A. Kremin and O. Lahav and C. Lamman and M. Landriau and L. Le Guillou and A. Leauthaud and M. E. Levi and C. Magneville and M. Manera and P. Martini and W. Matthewson and A. Meisner and R. Miquel and J. Moustakas and A. Muñoz-Gutiérrez and D. Muñoz-Santos and A. D. Myers and L. Napolitano and J. A. Newman and H. E. Noriega and N. Palanque-Delabrouille and J. Pan and W. J. Percival and I. Pérez-Ràfols and C. Poppett and F. Prada and A. Raichoor and C. Ramírez-Pérez and C. Ravoux and G. Rossi and R. Ruggeri and L. Samushia and E. Sanchez and D. Schlegel and M. Schubnell and F. Sinigaglia and D. Sprayberry and T. Tan and G. Tarlé and P. Taylor and W. Turner and R. Vaisakh and M. Vargas-Magaña and M. Walther and B. A. Weaver and M. Wolfson and J. Yu and C. Yèche and P. Zarrouk and R. Zhou and H. Zou},
      year={2025},
      eprint={2503.14742},
      archivePrefix={arXiv},
      primaryClass={astro-ph.CO},
      url={https://arxiv.org/abs/2503.14742}, 
}

@article{Abdul_Karim_2025,
   title={DESI DR2 results. II. Measurements of baryon acoustic oscillations and cosmological constraints},
   volume={112},
   ISSN={2470-0029},
   url={http://dx.doi.org/10.1103/tr6y-kpc6},
   DOI={10.1103/tr6y-kpc6},
   number={8},
   journal={Physical Review D},
   publisher={American Physical Society (APS)},
   author={Abdul Karim, M. and DESI},
   year={2025},
   month=oct }
